\documentstyle[eqsecnum,epsf,aps,prb,multicol]{revtex}
\begin{document} 
\draft

\title{Exact SO(8) Symmetry in the Weakly-Interacting Two-Leg Ladder}
\author{Hsiu-Hau Lin$^{1}$, Leon Balents$^{2}$ 
and Matthew P. A. Fisher$^{2}$} 
\address{$^1$ Department of Physics, University of
            California, Santa Barbara, CA 93106-9530\\
         $^2$ Institute for Theoretical Physics, University of 
California,
             Santa Barbara, CA 93106-4030}
\date{\today} 
\maketitle

\begin{abstract}

We revisit the problem of interacting electrons hopping on a two-leg 
ladder.  A perturbative renormalization group analysis reveals that at 
half-filling the model scales onto an exactly soluble Gross-Neveu 
model for {\it arbitrary} finite-ranged interactions, provided they 
are sufficiently weak.  The Gross-Neveu model has an enormous global 
$SO(8)$ symmetry, manifest in terms of eight real Fermion fields 
which, however, are highly non-local in terms of the electron 
operators.  For generic repulsive interactions, the two-leg ladder 
exhibits a Mott insulating phase at half-filling with d-wave pairing 
correlations.  Integrability of the Gross-Neveu model is employed to 
extract the {\it exact} energies, degeneracies and quantum numbers of 
{\it all} the low energy excited states, which fall into degenerate 
$SO(8)$ multiplets.  One $SO(8)$ vector includes two charged Cooper 
pair excitations, a neutral $s=1$ triplet of magnons, and three other 
neutral $s=0$ particle-hole excitations.  A {\it triality} symmetry 
relates these eight two-particle excitations to {\it two} other 
degenerate octets which are comprised of {\it single}-electron like 
excitations.  In addition to these 24 degenerate ``particle" states 
costing an energy (mass) $m$ to create, there is a 28 dimensional 
antisymmetric tensor multiplet of ``bound" states with energy 
$\sqrt{3}m$.  Doping away from half-filling liberates the Cooper pairs 
leading to quasi-long-range d-wave pair field correlations, but 
maintaining a gap to spin and single-electron excitations.  For very 
low doping levels, integrability allows one to extract {\it exact} 
values for these energy gaps.  Enlarging the space of interactions to 
include attractive interactions reveals that there are {\it four} 
robust phases possible for the weak coupling two-leg ladder.  While 
each of the four phases has a (different) $SO(8)$ symmetry, they are 
shown to all share a common $SO(5)$ symmetry - the one recently 
proposed by Zhang as a unifying feature of magnetism and 
superconductivity in the cuprates.

\end{abstract}

\begin{multicols}{2}

\section{Introduction}

Since the discovery of the cuprate superconductors\cite{Bednorz86}\ 
there has been renewed interest in the behavior of weakly doped Mott
insulators.\cite{Anderson87,Dagotto92,Rice93}\ There are two broad
classes of Mott insulators, distinguished by the presence or absence
of magnetic order.  More commonly spin rotational invariance is
spontaneously broken, and long-range magnetic order, typically
antiferromagnetic, is realized.\cite{foot:LRO}\ There are then low
energy spin excitations, the spin 1 magnons.  Alternatively, in a
spin-liquid Mott insulator there are no broken symmetries, the
magnetic order is short-ranged and there is a gap to all spin
excitations : a spin-gap.

In the cuprates the Mott insulator is antiferromagnetically ordered, 
but upon doping with holes the antiferromagnetism is rapidly 
destroyed, and above a certain level superconductivity occurs.  Below 
optimal doping levels, there are experimental signs of a spin gap 
opening at temperatures well above the transition into the 
superconducting phase.\cite{Levi96,Batlogg96,Ong96}\
The apparent connection between a spin-gap and 
superconductivity has been a source of motivation to search for Mott 
insulators of the spin-liquid variety.

Although spin-liquids are notoriously difficult to achieve in
two-dimensions,\cite{Schulz94}\ it was realized that
quasi-one-dimensional ladders would be more promising.  Particular
attention has focussed on the two-leg
ladder.\cite{Dagotto96}\
%\cite{Schulz86,Dagotto92,Rice93,White94,Affleck94,Dagotto96}\ 
At half-filling in the Mott insulator, the spin excitations can be
described by a Heisenberg antiferromagnet, and due to the tendency for
singlet bond formation across the rungs of the ladder, spin-liquid
behavior is expected.\cite{Dagotto92,Rice93,Schulz86,White94}\ In the
past several years there have been extensive analyses of two-leg
ladders, particularly the Hubbard\cite{Noack94,Noack95} and t-J
models,\cite{Dagotto92,Hayward96,Sano96,Troyer96}\ both at
half-filling and with doping.  Based on numerical methods, including
Monte Carlo and density matrix renormalization
group,\cite{Dagotto92,Rice93}\ as well as analytic approaches at weak
coupling,\cite{Balents96a,Lin97,Fabrizio93,Schulz96a,Schulz96b,Nagaosa95,Finkelstein93,Kuroki94}\ 
the basic behavior is established.  At half-filling there is a
spin-liquid phase with a spin-gap.  Upon doping, the spin-gap
survives, although smaller in magnitude, and the system exhibits
quasi-long-range superconducting pairing correlations, with
approximate d-wave symmetry.  This behavior is reminiscent of that
seen in the underdoped cuprate superconductors.

There are a number of experimental systems which can be described in 
terms of coupled two-leg ladders, which exhibit a spin-gap in the 
insulating compound.\cite{Azuma94,Ishida94,Kojima95}\
These materials are often very difficult to 
dope.  In one case, doping has apparently been achieved, and 
under a pressure of 3GPa  superconductivity is observed below 
12K.\cite{Uehara96,Isobe97}\  
Carbon nanotubes\cite{Ebbesen96a}\ constitute
another novel material which can be modelled in terms of a two-leg
ladder.\cite{Balents97,Krotov97,Lin97u1}\  Specifically, the low 
energy electronic excitations propagating down a single-walled 
nanotube can be mapped onto a two-leg ladder
model with very {\it weak} interactions, inversely proportional to the
tube radius.

An obvious advantage of such low-dimensional correlated electron 
systems is (relative) theoretical simplicity.  Indeed, in 
one-dimension many correlated electron models, including the Hubbard 
model, are exactly soluble.\cite{Emery79}\  
Unfortunately, the Mott insulating phases 
of these one-dimensional models typically have gapless 
spin-excitations, and upon doping do not exhibit pairing.  To date, we 
are unaware of any exactly soluble two-leg ladder models which exhibit 
a gapped spin-liquid ground state.

In this paper, we revisit models of interacting electron hopping on a
two-leg ladder, focusing on the behavior near half-filling.  For {\it
  generic} short-range potentials, we derive a perturbative
renormalization group valid for weak interactions, much smaller than
the bandwidth.\cite{Balents96a,Lin97}\ Remarkably, at half-filling the
renormalization group transformation scales the system towards a
special model with enormous symmetry - the $SO(8)$ Gross-Neveu (GN)
model.\cite{Gross74}\ Scaling onto the GN model occurs {\it
  independent} of the initial interaction parameters, provided they
are weak and predominantly repulsive.
%\cite{Lin97u5}\  
Thus, for weakly 
interacting two-leg ladders at half-filling {\it universal} low energy 
properties are expected.  Specifically, all properties on energy 
scales of order a characteristic GN {\sl mass} (gap) $m$ and distance 
scales longer than or of order $v/m$ (where $v$ is the Fermi velocity) 
are universal and determined by the GN model.  In terms of microscopic 
parameters, the GN mass is of order $m \sim te^{-t/U}$, where $t$ is 
the 1d bandwidth and $U$ is a typical interaction strength, but is 
more profitably treated, along with $v$, as a phenomenological 
parameter.  The universality predicted by the renormalization group 
can be profitably exploited because the $SO(8)$ GN model is {\sl 
integrable},\cite{Zamolodchikov79,Shankar78,Karowski81}\ 
so that many of these universal properties can be 
computed exactly.  To our knowledge, this is the first integrable 
model for a Mott-insulating spin liquid.  It describes a state we call 
the {\sl D-Mott} phase, because the Mott-insulator has short-range 
pairing correlations with approximate d-wave symmetry.  We now 
summarize the results obtained from the $SO(8)$ GN field theory.

The primary input from integrability is the complete excitation
spectrum.\cite{Zamolodchikov79,Shankar78,Karowski81,Dashen75}\ The
excitations of the GN model are comprised of ``particles'' (i.e.
sharp excitations with a single-valued energy-momentum relation)
organized into $SO(8)$ multiplets, as well as continuum scattering
states of these particles.  As expected for a Mott-insulating
spin-liquid with no broken symmetries, each of these excitations is
separated from the ground state by a non-zero gap.  The lowest-lying
particles come in three octets, all with mass $m$, i.e.  dispersing as
$\epsilon_{1}(q) = \sqrt{m^{2}+q^{2}}$, where $q$ is the deviation of
the particle's momentum from its minimum energy value.  One {\sl
  vector} multiplet (conveniently denoted formally by a vector of
Majorana fermions $\eta_{\scriptscriptstyle A}$, $A=1\ldots 8$)
consists entirely of collective two-particle excitations: two charge
$\pm 2e$ ``Cooper pairs'' around zero momentum, a triplet of spin-one
``magnons'' around momentum $(\pi,\pi)$, and three neutral spin-zero
``charge-density-wave'' (or particle-hole pair) excitations.  $SO(8)$
transformations rotate the components of the vector into one another,
unifying the pair, magnon, and charge-density-wave excitations.
Indeed, the $SO(5)$ subgroup rotating only the first five components
of this vector is exactly the symmetry proposed recently by
Zhang\cite{Zhang97}\ to unify antiferromagnetism and superconductivity
in the cuprates.  This vector octet, referred to as ``fundamental''
fermions in the field-theory literature, is related by a remarkable
{\sl triality} symmetry\cite{Shankar80,Shankar81}\ (present in the
$SO(N)$ GN model only for $N=8$) to two other mass $m$ octets: spinor
and isospinor multiplets, called the even and odd kinks.  These
sixteen particles have the quantum numbers of individual
quasi-electrons and quasi-holes.  The triality symmetry thus goes
beyond the $SO(8)$ algebra to relate single-particle and two-particle
properties in a fundamental way.\cite{Shankar80,Shankar81}\ This
relation also implies that pairing is present even in the
Mott-insulator: the minimum energy to add a pair of electrons (as a
member of the $SO(8)$ vector multiplet) is $m$, reduced by a {\sl
  binding energy} of $m$ from the cost of $2m$ needed to add two
quasi-electrons far apart.  At energies above the 24 mass $m$ states,
there exists an antisymmetric tensor multiplet of 28 particles with
mass $\sqrt{3}m$.  Each can be viewed as bound states of two different
fundamental fermions (or equivalently, two even or two odd kinks).  In
this way their quantum numbers can be easily deduced by simple
addition.  The tensor states contribute additional sharp
(delta-function) peaks to various spectral functions, providing, for
instance, the continuation of the magnon branch near momentum $(0,0)$.
For convenience, the quantum numbers (charge, spin, and momentum) of
the vector and tensor excitations are tabulated in Tables~1 and ~2.
Finally, continuum scattering states enter the spectrum above the
energy $2m$.

Combining the excitation spectrum of the GN model with the 
non-interacting spectrum and some additional arguments, we have also 
constructed schematic forms for several correlation functions of 
interest.  In particular, in Sec.~\ref{sec:correlators}\ we give 
detailed predictions and plots of the single particle spectral 
function (measurable by photoemission), the spin spectral function 
(measurable by inelastic neutron scattering), and the optical 
conductivity.  Integrability implies, for instance, sharp magnon peaks 
in the spin structure factor at ${\bf k} = (\pi,\pi)$, $(0,0)$, and 
$(\pm(k_{{\scriptscriptstyle F}1}-k_{{\scriptscriptstyle F}2}),\pi)$
 with minimum energy $m$, $\sqrt{3}m$ and 
$\sqrt{3}m$ respectively (here $k_{{\scriptscriptstyle F}1}$ 
and $k_{{\scriptscriptstyle F}2}$ are the Fermi 
momenta of the non-interacting system).  Complete details can be found 
in Sec.~\ref{sec:correlators}.  The optical conductivity has three 
principal features: a Drude peak around zero frequency, with 
exponentially small weight ($\sim e^{-m/T}$) at low temperature; an 
``exciton'' peak around $\omega = \sqrt{3}m$, exponentially narrow at 
low temperatures; and a continuum for 
$\omega \gtrsim 2m$, due to unbound quasi-particle quasi-hole pairs.  
See Sec.~\ref{sec:correlators}\ for more details and a figure.

Our next calculations concern the relation of these results to a 
recent study of microscopically $SO(5)$ invariant ladder models by 
Scalapino, Zhang and Hanke (SZH).\cite{Scalapino97u}\  
These authors consider the strong coupling limit 
of a certain locally-interacting two-leg ladder model designed to exhibit 
exact $SO(5)$ symmetry.  Their model has an on-site interaction
$|U| \gg t$, an 
intra-rung interaction $|V| \gg t$, and a 
magnetic rung-exchange interaction $J$, related to one another by the
$SO(5)$ symmetry.  In the 
$U$--$V$ plane they derive a strong-coupling phase diagram, including 
the case of attractive interactions with $U$ and $V$ negative.  We
have analyzed general $SO(5)$ invariant two-leg ladder 
models in the opposite limit of {\sl weak interactions}, deriving as 
a special case the corresponding weak-coupling phase diagram for their 
model.  In fact, although we have not explored the full 9-dimensional 
space completely, for all bare couplings we have considered, including 
attractive interactions that {\sl break} $SO(5)$ symmetry explicitly, 
the RG scales the system into the $SO(5)$ subspace.
%\cite{Lin97u5}\  
When the 
interactions are predominantly repulsive, the $SO(5)$ system falls 
into the basin of attraction of the D-Mott phase, and the above 
results apply.  As negative interactions are introduced, four other 
phases emerge: an {\sl S-Mott} spin-liquid, with short-range 
approximate s-wave pairing symmetry, a charge-density-wave (CDW) state with 
long-range positional order at $(\pi,\pi)$, a spin-Peierls phase with 
kinetic energy modulated at $(\pi,\pi)$, and a Luttinger liquid (C2S2, 
in the nomenclature of Ref.~\onlinecite{Balents96a}) 
phase continuously connected to the non-interacting system.  The first 
two of these also occur in the strong-coupling limit, though their 
positions in the phase diagram (Fig. 10) are modified.  
The phase diagrams at weak and strong coupling differ in non-trivial 
ways, implying a rather complex evolution of the system with 
increasing $U$ and $V$.  In weak coupling, all four non-trivial phases 
have distinct asymptotic $SO(8)$ symmetries, enhanced from the common 
bare $SO(5)$.  Furthermore, critical points describing the transitions 
between the various phases can also be identified.  In particular, the 
D-Mott to S-Mott and CDW to spin-Peierls critical points are $c=1$ 
conformal field theories (single mode Luttinger liquids), which in 
weak-coupling are acccompanied by a decoupled massive $SO(6)$ sector.  
The S-Mott to CDW and D-Mott to spin-Peierls transitions are Ising 
critical theories ($c=1/2$), with decoupled massive $SO(7)$ sectors in 
weak-coupling.  There is also a multi-critical point describing a 
direct transition from the D-Mott to CDW or from the S-Mott to 
spin-Peierls phases, which is simply
a product of the $c=1$ 
and $c=1/2$ critical points.

Our final results concern the effects of doping a small density of 
holes (or electrons) into the D-Mott spin-liquid phase at
half-filling.  For very small hole 
concentrations, the modifications of the Fermi velocities by band 
curvature effects can be ignored, and the doping incorporated simply 
by including a chemical potential term coupled to 
the total charge $Q$ in the GN model; $H_{\mu} = H -\mu Q$.  An analogous procedure is 
employed by Zhang\cite{Zhang97}
in his study of the $SO(5)$ non-linear sigma model.  
Because the charge $Q$ is a global $SO(8)$ generator, integrability of 
the GN model is preserved, and furthermore many of the $SO(8)$ quantum 
numbers can still be employed to label the states.  We find that 
doping occurs only for $2\mu > m$, at which point Cooper pair 
``fundamental fermions'' enter the system and effectively form a 
Luttinger liquid with a single gapless charge mode (with central 
charge $c=1$).  This phase (often denoted ``C1S0'')
still has a gap to spin excitations.
Previous work\cite{Balents96a,Fabrizio93,Schulz96a}\ 
has approached this phase via controlled
perturbative calculations in the interaction strength, 
at fixed doping $x$ {\it away} from half-filling.  
Here, we are considering a different order of limits, with fixed (albeit 
weak) interactions in the small doping limit, $x \rightarrow 0$.
In this limit, the Cooper-pairs being dilute behave
as hard-core bosons or free fermions.  Although the spin-gap is preserved 
in the doped state, it is {\sl discontinuous} as $x \rightarrow 
0^{+}$.  The discontinuity can be understood as the binding of an 
inserted spin-one magnon to a Cooper pair in the system to form a mass 
$\sqrt{3}m$ tensor particle, reduced by the binding energy 
$(2-\sqrt{3})m$ from its bare energy.  The spin-gap thus jumps from 
$\Delta_{s}(x=0) = m$ to $\Delta_{s}(x=0^{+}) = (\sqrt{3}-1)m$ upon 
doping.  Such binding of a pair to a magnon has been observed 
numerically in both Hubbard and t-J ladders by Scalapino and White  
.\cite{Scalapino97u1}\  Similarly, the energy to add an {\sl 
electron} (for the hole-doped system) jumps from $\Delta_{1-}(x=0) = 
3m/2$ to $\Delta_{1-}(x=0^{+}) = m/2$, {\sl the same} as the energy to 
add a single hole.  When many pairs are present, we have not succeeded 
in obtaining exact expressions for the spin and single-particle gaps, 
but argue that the spin gap should decrease with increasing doping, 
since the added magnon is attracted to an increasing density of Cooper 
pairs.  It seems likely, however, that integrability could be 
exploited even in this case to obtain exact results, and hope that 
some experts may explore this possibility in the future. 

Finally, we briefly address the behavior of the spin-spectal function
for the doped ladder at energies above the spin-gap.  In a recent
preprint SZH\cite{Scalapino97u}\ have argued that in this regime the
spin-spectral function for a model with exact SO(5) symmetry should
exhibit a sharp resonance at energy $2 \mu$ and momentum $(\pi,\pi)$,
the so-called $\pi-$resonance (introduced originally by Zhang to
explain the 42meV neutron scattering peak in the superconducting
Cuprates).  We show that a delta-function $\pi-$resonance requires, in
addition to SO(5) symmetry, the existence of a non-zero condensate
density in the superconducting phase.  Since condensation is not
possible in one-dimension, this precludes a delta-function
$\pi-$resonance.  Following a recent suggestion by
Zhang,\cite{Zhang:pc}\ we address briefly the possibility of a weaker
algebraic singularity in the spin spectral function.  Regardless of
the nature of the behavior in the vicinity of $\omega=2\mu$, we
expect spectral weight at energies below $2\mu$ but above the spin-gap 
$\Delta_s$ discussed above.

The remainder of the paper is organized as follows.  In 
Sec.~\ref{sec:model}\ we describe the model Hamiltonian for the 
interacting ladder, reduce it to the continuum limit, bosonize the 
nine distinct interaction channels, and apply the renormalization 
group (RG) transformation.  Sec.~\ref{sec:DMott}\ details the 
simplifications that occur upon RG scaling, presents the bosonized 
form of the Hamiltonian in the D-Mott phase, and for completeness 
demonstrates the short-range d-wave correlations found in 
Ref.~\onlinecite{Balents96a}\  The bulk of the field-theoretic analysis is 
contained in Sec.~\ref{sec:GN}.  By refermionizing the bosonized 
hamiltonian, we obtain the GN model
exposing the exact SO(8) symmetry,
and describe why this symmetry is hidden
in the original 
variables.  The triality symmetry is identified, and used to 
understand the degeneracy between the three mass $m$ octets.  To help 
in developing an intuition for the GN model, several approximate 
pictures are presented to understand the excitations: a mean field 
theory which is asymptotically exact for $N\rightarrow\infty$ in a generalized 
$SO(N)$ GN model, and a semi-classical theory based on the bosonized 
(sine-Gordon-like) form of the Hamiltonian.  We conclude 
Sec.~\ref{sec:GN}\ by proving the uniqueness of the ground-state in 
the D-Mott phase and determining the quantum numbers of the $24+28=52$ 
particles.  The latter task is complicated by the necessity of 
introducing Jordan-Wigner {\sl strings}, which are required to 
preserve gauge-invariance under an unphysical gauge symmetry 
introduced in bosonization.  The string operators modify the momenta 
of the certain excitations by a shift of $(\pi,\pi)$ from their naive 
values determined from the GN fermion operators.  With the 
field-theoretic analysis complete, we go on to discuss correlation 
functions in Sec.~\ref{sec:correlators}, giving detailed predictions 
for the single-particle spectral function, spin spectral function, 
optical conductivity, and various equal-time spatial correlators.  
Sec.~\ref{sec:SO5}\ describes the construction of general $SO(5)$ 
invariant models in weak-coupling, their phases, and the phase-diagram 
of the Scalapino-Zhang-Hanke model in weak-coupling.  Finally, 
Sec.~\ref{sec:doping}\ describes the behavior of the D-Mott phase upon 
doping, including the behavior of various gaps, and a discussion of 
the status of the $SO(5)$ ``$\pi$ resonance'' in one dimension.  
Various 
technical points and long equations are placed in the appendices.  
Appendix A gives the full set of nine RG equations at half-filling, 
Appendix B discusses gauge redundancy and the multiplicity of the 
ground state in different phases, Appendix C constructs spinor and 
vector representations of $SO(5)$, Appendix D relates $SO(5)$ and 
$SO(8)$ currents, and Appendix E gives the five RG equations in the 
reduced  $SO(5)$ subspace.

\section{Model}
\label{sec:model}
We consider electrons hopping on a two-leg ladder 
as shown in Fig. 1.
In the absence of interactions, the Hamiltonian consists
of the kinetic energy, which we assume contains only near-neighbor
hopping,
\begin{eqnarray} 
H_0 &=&\sum_{x,\alpha} \bigg\{ 
-t a^{\dag}_{1 \alpha} (x+1) a^{}_{1\alpha}(x) + (1 \rightarrow 2)
\nonumber\\
&&-t_{\perp}  a^{\dag}_{1 \alpha}(x) a^{}_{2 \alpha}(x) +
h.c.  \bigg\},
\label{kinetic}
\end{eqnarray} 
where $ a^{}_{\ell} ( a_{\ell}^{\dag} ) $ is an electron annihilation 
(creation) operator on leg $\ell$ of the ladder ($\ell=1,2$), $x$ is 
a 
discrete coordinate running along the ladder and $\alpha =\uparrow, 
\downarrow $ is a spin index.  The parameters $t$ and $t_{\perp}$ are 
hopping amplitudes along and between the leg's of the ladder.

Being interested in weak interactions, we first diagonalize the 
kinetic energy in terms of bonding and anti-bonding operators: 
$c_{i,\alpha}= (a_{1,\alpha} + (-1)^i a_{2,\alpha})/\sqrt{2}$, with 
$i=1,2$.  The Hamiltonian is then diagonalized in momentum space 
along 
the ladder, describing two decoupled (bonding and anti-bonding) 
bands.  
Focussing on the case at half-filling with one electron per site, 
both 
bands intersect the Fermi energy (at zero energy) provided $t_{\perp} 
< 2t$.  Moreover, due to a particle/hole symmetry present with near 
neighbor hopping only, the Fermi velocity $v_i$ in each band is the 
same, denoted hereafter as $v$.  It is convenient to linearize the 
spectrum around the Fermi points at 
$\pm k_{{\scriptscriptstyle F}i}$ (see Fig. 1), 
which at half-filling satisfy 
$k_{{\scriptscriptstyle F}1} + k_{{\scriptscriptstyle F}2} = \pi$.  
Upon expanding the 
electron operators as, 
\begin{equation}
  c^{}_{i\alpha} \sim c^{}_{{\scriptscriptstyle R} i \alpha}
 e^{ik_{{\scriptscriptstyle F}i}x} + 
  c^{}_{{\scriptscriptstyle L} i \alpha} 
e^{-ik_{{\scriptscriptstyle F}i}x},
  \label{decompose_electrons}
\end{equation}
the effective low energy expression for 
the kinetic energy takes the form, $H_0 = \int dx {\cal H}_0$, with 
Hamiltonian density,
\begin{equation}
{\cal H}_{0}= v \sum_{i,\alpha} [
c^{\dag}_{{\scriptscriptstyle R}i\alpha} i\partial_{x} 
c^{}_{{\scriptscriptstyle R}i\alpha}
- c^{\dag}_{{\scriptscriptstyle L}i\alpha}
i\partial_{x}c^{}_{{\scriptscriptstyle L}i\alpha} ] .
\label{kinetics_f}
\end{equation}

This Hamiltonian describes Dirac Fermions, with four flavors labelled 
by band and spin indices.  Since all flavors propagate both to the 
right and left with the {\it same} velocity, the model exhibits an 
enlarged symmetry.  Specifically, if the four right (and left) moving 
Dirac Fermions are decomposed into real and imaginary parts, 
$\psi^{}_{{\scriptscriptstyle P}i\alpha} = 
(\xi^{1}_{{\scriptscriptstyle P}i\alpha} + i 
\xi^{2}_{{\scriptscriptstyle P}i\alpha})/\sqrt{2}$ 
where $P=R/L$ and $\xi^{1}, \xi^{2}$ are Majorana fields, the eight 
right (and left) moving Majorana fields, denoted 
$\xi_{{\scriptscriptstyle PA}}$ with 
$A=1,2,...,8$ form an eight component vector.  The Hamiltonian 
density, when re-expressed in terms of these eight component vectors 
takes the simple form
\begin{equation}
{\cal H}_{0} = {v \over 2} \sum_{A=1}^8 [ 
\xi^{}_{{\scriptscriptstyle RA}} 
i \partial_x \xi^{}_{{\scriptscriptstyle RA}} 
- \xi^{}_{{\scriptscriptstyle LA}}
i \partial_x \xi^{}_{{\scriptscriptstyle LA}} ]   , 
\end{equation}
which is invariant under {\it independent} global
$SO(8)$ rotations among {\it either}
the right or left vector of Majorana fields.
This enlarged $O(8)_R \times O(8)_L$ symmetry is only present
at half-filling with particle/hole symmetry.

\begin{figure}[hbt]
\epsfxsize=8cm
\centerline{\epsfbox{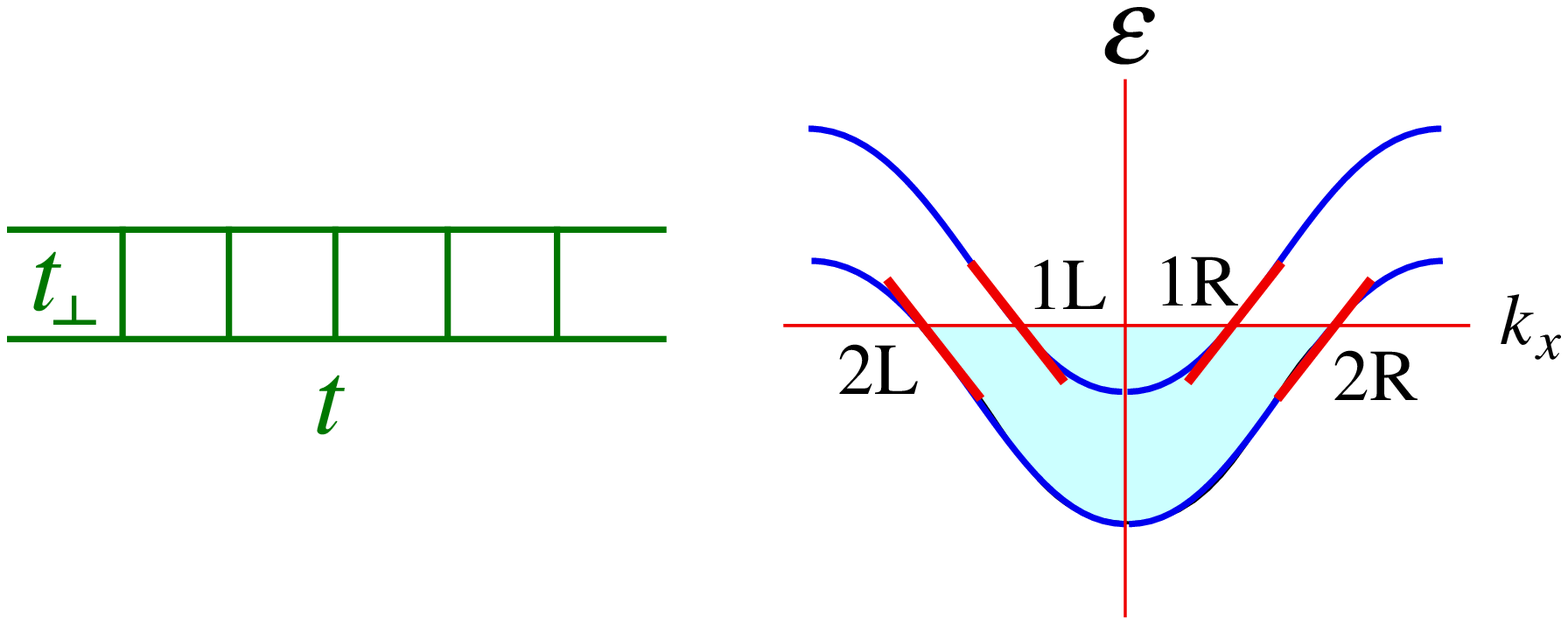}}
%\vspace{5pt}
{\noindent FIG. 1: A two-leg ladder and its band
structure. In the low-energy limit, the energy dispersion is
linearized near the Fermi points. The two resulting 
relativistic Dirac Fermions are distinguished by pseudospin indices
$i=1,2$ for the anti-bonding and bonding bands, respectively.}
\end{figure}

Electron-electron interactions scatter right-moving electrons into 
left-moving electrons and vice-versa, destroying this large 
symmetry.  For general 
spin-independent interactions the symmetry will be broken down to 
$U(1) \times SU(2)$, corresponding to total charge and spin 
conservation.  
In the following we consider general finite-ranged spin-independent 
interactions between the electrons hopping on the two-leg ladder.  We 
assume the typical interaction strength, $U$, is weak -- much smaller 
than the bandwidth.  We focus on the effects of the interactions to 
{\it leading} non-vanishing order in $U$.  In this limit it is 
legitimate to keep only those pieces of the interactions which 
scatter 
the low energy Dirac Fermions.  Of these, only those involving 
four-Fermions are marginal, the rest scaling rapidly to zero under 
renormalization.  Moreover, four-Fermion interactions which are 
chiral, say only scattering right movers, only renormalize Fermi 
velocities and can be neglected at leading order in small 
$U$.\cite{Balents96a,Lin97}\  All of 
the remaining four-Fermion interactions can be conveniently expressed 
in terms of currents, defined as
\begin{eqnarray}
 J_{ij} =& c^{\dag}_{i\alpha} c^{}_{j\alpha}, 
 \qquad
 \bbox{J}_{ij}=&\frac12 \; c^{\dag}_{i\alpha} 
 \bbox{\sigma}_{\alpha\beta} c^{}_{j\beta};
 \label{J_def}
 \\
  I_{ij} =& c_{i\alpha} \epsilon_{\alpha\beta} c_{j\beta},
 \qquad
 \bbox{I}_{ij}=&\frac12 \; c_{i\alpha} 
(\epsilon \bbox{\sigma})_{\alpha\beta} c_{j\beta},
\label{I_def}
\end{eqnarray}
where the $R,L$ subscript has been suppressed. 
Both $J$ and $I$ are invariant under global
$SU(2)$ spin rotations, whereas 
$\bbox{J}$ and $\bbox{I}$ rotate as $SU(2)$ vectors.
Due to Fermi statistics, some
of the currents are (anti-)symmetrical
\begin{equation}
I_{ij}=I_{ji} \qquad \bbox{I}_{ij} = -\bbox{I}_{ji},
\label{I_symm}
\end{equation}
so that 
$\bbox{I}_{ii}=0$ (no sum on $i$).

The full set of marginal momentum-conserving four-Fermion interactions
can be written
\begin{eqnarray} 
{\cal H}^{(1)}_{I} = && b^{\rho}_{ij} J_{{\scriptscriptstyle R}ij}
 J_{{\scriptscriptstyle L}ij} -
b^{\sigma}_{ij} \bbox{J}_{{\scriptscriptstyle R}ij} 
\cdot \bbox{J}_{{\scriptscriptstyle L}ij},
\nonumber\\
+&&f^{\rho}_{ij} J_{{\scriptscriptstyle R}ii} 
J_{{\scriptscriptstyle L}jj} -
f^{\sigma}_{ij} \bbox{J}_{{\scriptscriptstyle R}ii} 
\cdot \bbox{J}_{{\scriptscriptstyle L}jj}.
\label{int1}
\end{eqnarray} 
Here $f_{ij}$ and $b_{ij}$ denote the forward and
backward (Cooper) scattering amplitudes, respectively, between bands
$i$ and $j$.  Summation on $i, j=1,2$ is implied. To avoid double
counting, we set $f_{ii}=0$ (no sum on $i$). 
Hermiticity implies $b_{12}=b_{21}$ and parity symmetry
($R \leftrightarrow L$) gives $f_{12}=f_{21}$,
so that there
are generally eight independent couplings $b^{\rho,\sigma}_{11}$, 
$b^{\rho,\sigma}_{22}$, $b^{\rho,\sigma}_{12}$, 
and $f^{\rho,\sigma}_{12}$.
At half-filling with particle/hole symmetry
$b_{11} = b_{22}$.  Additional momentum non-conserving Umklapp
interactions of the form
\begin{equation} 
{\cal H}^{(2)}_{I} = 
u^{\rho}_{ij} I^{\dag}_{{\scriptscriptstyle R}ij} 
I_{{\scriptscriptstyle L}{\hat i}{\hat j}} -
u^{\sigma}_{ij} \bbox{I}^{\dag}_{{\scriptscriptstyle R}ij} \cdot 
\bbox{I}_{{\scriptscriptstyle L}{\hat i}{\hat j}}
+ {\rm h.c.}
\label{int2}
\end{equation} 
are also allowed, (here $\hat{1}=2, \hat{2}=1$). Because the currents
$(\bbox{I}_{ij}), I_{ij}$ are (anti-)symmetric, one can always choose
$u_{12} = u_{21}$ for convenience. We also take
$u^{\sigma}_{ii}=0$ since $\bbox{I}_{ii}=0$.
With particle/hole symmetry there are thus just three
independent Umklapp vertices, $u^{\rho}_{11}$, $u^{\rho}_{12}$, 
and $u^{\sigma}_{12}$.  Together with the six forward and backward
vertices, nine independent couplings are required to describe
the most general set of marginal non-chiral four-Fermion interactions
for a two-leg ladder with particle/hole symmetry at half-filling.

Since our analysis below makes heavy use
of abelian Bosonization,\cite{Shankar95,Emery79}\ 
it is convenient at this stage to
consider the Bosonized form of the general interacting theory.
To this end, the Dirac fermion fields are expressed in terms
of Boson fields as
\begin{equation}
c^{}_{{\scriptscriptstyle P}i\alpha}= 
\kappa^{}_{i\alpha} 
e^{i\phi^{}_{{\scriptscriptstyle P}i\alpha}},
\label{bosonization}
\end{equation}
where $P=R/L = \pm$.  To ensure that the Fermionic operators
anti-commute the Boson fields are taken to satisfy
\begin{eqnarray}
[\phi_{{\scriptscriptstyle P}i\alpha}(x),
\phi_{{\scriptscriptstyle P}j\beta}(x^\prime)]  =&  
iP \pi \delta_{ij}\delta_{\alpha\beta} &{\rm sgn}(x-x^\prime),
\label{cm1}
\\
\left[\phi_{{\scriptscriptstyle R}i\alpha}(x),
\phi_{{\scriptscriptstyle L}j\beta}(x^\prime)\right] =&
 i\pi \delta_{ij} \delta_{\alpha \beta}&.
\label{cm2}
\end{eqnarray}
Klein factors, satisfying
\begin{equation}
\{ \kappa^{}_{i\alpha}, \kappa^{}_{j\beta} 
\}=2\delta_{ij}\delta_{\alpha\beta},
\end{equation}
have been introduced so that the Fermionic operators 
in different bands or 
with different spins anticommute with one another.

It will also be convenient to define
a pair of conjugate non-chiral Boson fields for each flavor,
\begin{eqnarray}
\varphi_{i\alpha} &\equiv& \phi_{{\scriptscriptstyle R}i\alpha}
+\phi_{{\scriptscriptstyle L}i\alpha},
\\
\theta_{i\alpha} &\equiv& \phi_{{\scriptscriptstyle R}i\alpha}
-\phi_{{\scriptscriptstyle L}i\alpha}, 
\label{eq:thetadef}
\end{eqnarray}
which satisfy 
\begin{equation}
[\varphi(x),\theta(x^\prime)]=-i4\pi \Theta(x^\prime-x)  .
\label{thetaphicom}
\end{equation}
Here, and in the remainder of the paper, we denote by $\Theta(x)$ the 
heavyside step function to avoid confusion with the $\theta$ fields 
defined in Eq.~\ref{eq:thetadef}\ above.
The field 
$\theta_{i\alpha}$ is a displacement (or phonon) field and 
$\varphi_{i\alpha}$ is a phase field.

The Bosonized form for the kinetic energy 
Eq.~\ref{kinetics_f} is
\begin{equation}
{\cal H}_{0} = \frac{v}{8\pi}\sum_{i,\alpha}  [
 (\partial_{x}\theta_{i\alpha})^{2}
+ (\partial_{x}\varphi_{i\alpha})^{2} ],
\end{equation}
which describes density waves propagating 
in band $i$ and with spin $\alpha$.

This expression can be conveniently separated into
charge and spin modes, by defining
\begin{eqnarray}
\theta_{i \rho} &=& (\theta_{i \uparrow} 
+ \theta_{i \downarrow} )/\sqrt{2}  
\label{bosondef1}\\
\theta_{i \sigma } &=& (\theta_{i \uparrow} 
- \theta_{i \downarrow} )/\sqrt{2} ,
\end{eqnarray}
and similarly for $\varphi$.  The $\sqrt{2}$ ensures that these new 
fields
satisfy the same commutators, Eq. (\ref{thetaphicom}).
It is also convenient to combine the fields in the two bands
into a $\pm$ combination, by defining
\begin{equation}
\theta_{\mu \pm} = ( \theta_{1 \mu} \pm \theta_{2 \mu} ) / \sqrt{2}  ,
\end{equation}
where $\mu = \rho,\sigma$, and similarly for $\varphi$.
It will sometimes be convenient to employ charge/spin and flavor
decoupled {\it chiral} fields, defined as
\begin{equation}
\phi_{{\scriptscriptstyle P}\mu\pm} = 
( \varphi_{\mu\pm} + P \theta_{\mu\pm} )/2 ,
\label{bosondef2}
\end{equation}
with $P=R/L=\pm$.

The Hamiltonian density ${\cal H}_0$ can now be re-expressed in a 
charge/spin and flavor decoupled form,
\begin{equation}
{\cal H}_{0} = \frac{v}{8\pi}\sum_{\mu, \pm} [
 (\partial_{x}\theta_{\mu \pm})^{2}
+ (\partial_{x}\varphi_{\mu \pm})^{2} ].
\end{equation}
The fields $\theta_{\rho+}$ and $\varphi_{\rho +}$ describe the total 
charge and current fluctuations, since under Bosonization, 
$c^\dagger_{{\scriptscriptstyle P}i\alpha} 
c^{}_{{\scriptscriptstyle P}i\alpha} = \partial_x 
\theta_{\rho+}/\pi$ 
and $vP c^\dagger_{{\scriptscriptstyle P}i\alpha} 
c^{}_{{\scriptscriptstyle P}i\alpha} = 
\partial_x 
\varphi_{\rho+}/\pi$.

The interaction Hamiltonians can also be readily expressed in terms 
of the
Boson fields.  The momentum conserving terms in Eq.~\ref{int1}
can be decomposed
into two contributions,
${\cal H}_I^{(1)} = {\cal H}_I^{(1a)} + {\cal H}_I^{(1b)}$, the first 
2
involving gradients of the Boson fields,
\begin{equation}
{\cal H}^{(1a)}_{I}= \frac{1}{16\pi^{2}}\sum_{\mu \pm} A_{\mu \pm} [ 
(\partial_{x}\theta_{\mu \pm})^{2}
-(\partial_{x}\varphi_{\mu \pm})^{2} ],
\end{equation}
with coefficient $A_{\rho\pm}=2(c^{\rho}_{11} \pm f^{\rho}_{12})$ and
$A_{\sigma\pm}=-(c^{\sigma}_{11} \pm f^{\sigma}_{12})/2$,
whereas the second contribution involves cosines of the Boson fields:
\begin{eqnarray}
{\cal H}^{(1b)}_{I}&=&
-2\Gamma b^{\sigma}_{12}\cos\varphi_{\rho-}\cos\theta_{\sigma+}
\nonumber\\
&+&\cos\theta_{\sigma+}(2b^{\sigma}_{11}\cos\theta_{\sigma-}
+2 \Gamma f^{\sigma}_{12} \cos\varphi_{\sigma-})
\nonumber\\
&-&\cos\varphi_{\rho-}(\Gamma b^{+}_{12}\cos\theta_{\sigma-}
+b^{-}_{12}\cos\varphi_{\sigma-}),
\label{boson1}
\end{eqnarray}
with $b^{\pm}_{12} = b^{\sigma}_{12}\pm 4b^{\rho}_{12}$.
Similarly, the Umklapp interactions can be Bosonized as,
\begin{eqnarray}
{\cal H}_{I}^{(2)} &=&
-16 \Gamma u^{\rho}_{11}\cos\theta_{\rho+}\cos\varphi_{\rho-}
-4u^{\sigma}_{12}\cos\theta_{\rho+} \cos\theta_{\sigma+}
\nonumber\\
&-&\cos\theta_{\rho+}(2u^{+}_{12}\cos\theta_{\sigma-}
+2\Gamma u^{-}_{12} \cos\varphi_{\sigma-}),
\label{boson2}
\end{eqnarray}
with $u^{\pm} = u^{\sigma}_{12}\pm 4u^{\rho}_{12}$.  Here
$\Gamma = \kappa_{1 \uparrow} \kappa_{1 \downarrow} \kappa_{2 
\uparrow} \kappa_{2 \downarrow}$ is a product of Klein factors.  
Since 
$\Gamma^2 = 1$, we can take $\Gamma = \pm 1$.  Hereafter, we will put 
$\Gamma = 1$.

In the absence of electron-electron interactions, the Hamiltonian is
invariant under spatially constant shifts of any of the eight
non-chiral Boson fields, $\theta_{\mu\pm}$ and $\varphi_{\mu \pm}$.
With interactions {\it five} of the eight Boson fields enter as
arguments of cosines, but for the remaining three -- $\varphi_{\rho+},
\varphi_{\sigma+}$ and $\theta_{\rho-}$ -- this continuous shift
symmetry is still present.  For the first two fields, the conservation
law responsible for this symmetry is readily apparent.  Specifically,
the operators $\exp(iaQ)$ and $\exp(iaS_z)$, with $Q$ the total
electric charge and $S_z$ the total z-component of spin, generate
``translations" proportional to $a$ in the two fields
$\varphi_{\rho+}$ and $\varphi_{\sigma+}$.  To see this, we note that
$Q = \int dx \rho(x)$ with $\rho(x) = \partial_x \theta_{\rho+}/\pi$
the momentum conjugate to $\varphi_{\rho+}$, whereas $S_z$ can be
expressed as an integral of the momentum conjugate to
$\varphi_{\sigma+}$.  Since the total charge is conserved, $[Q, H]=0$,
the full Hamiltonian must therefore be invariant under
$\varphi_{\rho+} \rightarrow \varphi_{\rho+} + a$ for arbitrary
constant $a$, precluding a cosine term for this field.  Similarly,
conservation of $S_z$ implies invariance under $\varphi_{\sigma+}
\rightarrow \varphi_{\sigma+} + a$.  The conservation law responsible
for the symmetry under shifts of the third field, $\theta_{\sigma-}$,
is present only in the weak coupling limit.  To see this, consider the
operator, ${\cal P} = k_{{\scriptscriptstyle F}1} J_1 +
k_{{\scriptscriptstyle F}2} J_2$, with $J_i = \sum_\alpha
(N_{{\scriptscriptstyle R}i\alpha} - N_{{\scriptscriptstyle
    L}i\alpha})$, where $N_{{\scriptscriptstyle P}i\alpha}$ is the
total number of electrons in band $i$ with spin $\alpha$ and chirality
$P$.  At weak coupling with Fermi fields restricted to the vicinity of
$k_{{\scriptscriptstyle F}i}$, this operator is essentially the total
momentum.  Since the total momentum is conserved up to multiples of
$2\pi$, one has $\Delta {\cal P} = \pm 2\pi n = \pm
2n(k_{{\scriptscriptstyle F}1} + k_{{\scriptscriptstyle F}2})$ for
integer $n$.  Moreover, since the Fermi momenta
$k_{{\scriptscriptstyle F}i}$ are in general unequal and
incommensurate, this implies that $\Delta J_1 = \Delta J_2 = \pm 2n$,
or equivalently that $J_1 - J_2$ is conserved at weak coupling.  Since
$J_1 - J_2 = \int dx j(x)$ with $j(x) = \partial_x
\varphi_{\rho-}/\pi$ the momentum conjugate to $\theta_{\rho -}$, this
conservation law implies invariance under $\theta_{\rho -} \rightarrow
\theta_{\rho -} +a$.

The remaining 5 Boson fields, entering as arguments of various 
cosine terms, will tend to be pinned at the minima of these 
potentials.  Two of these 5 fields, $\theta_{\sigma-}$ and 
$\varphi_{\sigma -}$, are dual to one another so that the uncertainty 
principle precludes pinning both fields.  Since there are various 
competing terms in the potential seen by these 5 fields, minimization 
for a given set of bare interaction strengths is generally 
complicated.  For this reason we employ the weak coupling 
perturbative 
renormalization group transformation, derived in 
earlier work.\cite{Balents96a,Lin97}\  Upon 
systematically integrating out high-energy modes away from the Fermi 
points and then rescaling the spatial coordinate and Fermi fields, a 
set of renormalization group (RG) transformations can be derived for 
the interaction strengths.  Denoting the nine interaction strengths 
as 
$g_i$, the leading order RG flow equations take the general form, 
$\partial_\ell g_i = A_{ijk} g_j g_k$, valid up to order $g^3$.  For 
completeness the RG flow equations are given explicitly in Appendix 
A.  
Our approach is to integrate the RG flow equations, numerically if 
necessary, to determine which of the nine coupling constants are 
growing large.

Under a numerical integration of these nine flow equations it is 
found 
that some of the couplings remain small, while others tend to 
increase, 
sometimes after a sign change, and then eventually diverge.  Quite 
surprisingly, though, the ratios of the growing couplings tend to 
approach fixed constants, which are {\it indepependent} of the 
initial 
coupling strengths, at least over a wide range in the nine 
dimensional 
parameter space.  These constants can be determined by inserting the 
Ansatz,
\begin{equation}
g_i(\ell) = {g_{i0} \over {(\ell_d - \ell)}}  ,
\label{power_law}
\end{equation}
into the RG flow equations, to obtain nine {\it algebraic} equations 
quadratic in the constants $g_{i0}$.  There are various distinct 
solutions of these algebraic equations, or rays in the nine-dimensional 
space, which correspond to different possible phases.  But for 
generic 
{\it repulsive} interactions between the electrons on the two-leg 
ladder, a numerical integration reveals that the flows are 
essentially 
always attracted to one particular ray.  In the next sections we 
shall consider the properties of this phase, which, for reasons which 
will become apparent, we denote by {\sl D-Mott}.

\section{D-Mott phase}
\label{sec:DMott}

In the phase of interest, two of the nine coupling
constants, $b^{\rho}_{11}$ and 
$f^{\sigma}_{12}$, remain small,
while the other seven grow large with fixed ratios:
\begin{eqnarray}
b^{\rho}_{12} = \frac14 b^{\sigma}_{12} = f^{\rho}_{12} 
= -\frac14 b^{\sigma}_{11} = 
\\
2u^{\rho}_{11} = 2u^{\rho}_{12} = \frac12 u^{\sigma}_{12} = g >0  .
\label{ratio}
\end{eqnarray}
Once the ratio's are fixed, there is a single remaining coupling 
contant, denoted $g$, which measures the distance from the origin 
along a very special direction (or ``ray") in the nine dimensional 
space of couplings.  The RG equations reveal that as the flows scale 
towards strong coupling, they are {\it attracted} to this special 
direction.  If the initial bare interaction parameters are 
sufficiently weak, the RG flows have sufficient ``time" to 
renormalize 
onto this special ``ray", before scaling out of the regime of 
perturbative validity.  In this case, the low energy physics, on the 
scale of energy gaps which open in the spectrum, is {\it universal}, 
depending only on the properties of the physics along this special 
ray, and independent of the precise values of the bare interaction 
strengths.

To expose this universal weak coupling physics, we use Eq.~\ref{ratio}
to replace the nine independent coupling constants in the most 
general Hamiltonian with the {\it single} parameter $g$, measuring 
the distance along the special ray.  Doing so reveals a remarkable 
symmetry, which is most readily exposed in terms of a new set of 
Boson 
fields, defined by,
\begin{eqnarray}
(\theta, \varphi)_{1} =&(\theta, \varphi)_{\rho+},
\qquad 
(\theta, \varphi)_{2}=&(\theta, \varphi)_{\sigma+},
\nonumber\\
(\theta, \varphi)_{3}=&(\theta, \varphi)_{\sigma-}, 
\qquad
(\theta, \varphi)_{4}=&(\varphi, \theta)_{\rho-}.
\label{mott_field}
\end{eqnarray}
The first three are simply the charge/spin and flavor fields
defined earlier.  However, in the fourth pair of fields,
$\theta$ and $\varphi$ have been interchanged.
It will also be useful to consider {\it chiral} boson fields for this
new set, defined in the usual way,
\begin{equation}
\phi_{{\scriptscriptstyle P}a} = (\varphi_a + P \theta_a )/2  ,
\end{equation}
with $a=1,..,4$, and $P=R/L = \pm$ as before.  The first three of 
these
chiral fields satisfy the commutators Eq. (\ref{cm1}) and (\ref{cm2}).
But for the fourth field, since 
$\phi_{{\scriptscriptstyle P}4}= 
P \phi_{{\scriptscriptstyle P}\rho-}$,
the second commutator is modified to  $[\phi_{{\scriptscriptstyle R}4},
 \phi_{{\scriptscriptstyle L}4}]=-i 
\pi$.  

In terms of these new fields, the full interacting Hamiltonian
density along the special ray takes an exceedingly simple form:
${\cal H} = {\cal H}_0 + {\cal H}_I$, with
\begin{equation}
{\cal H}_0 = { v \over {8\pi}} \sum_a  [(\partial_x \theta_a )^2
+ ( \partial_x \varphi_a)^2 ]  ,
\label{free_boson}
\end{equation}
\begin{eqnarray}
{\cal H}_{I} &=& -\frac{g}{2\pi^{2}} \sum_{a}
\partial_{x}\phi_{{\scriptscriptstyle R}a}
\partial_{x}\phi_{{\scriptscriptstyle L}a}
\nonumber\\
&&-4g \sum_{a \neq b} \cos \theta_{a} \cos \theta_{b}.
\label{int_boson}
\end{eqnarray}
We now briefly discuss some of the general physical properties which 
follow from this Hamiltonian.  In the next sections we will explore 
in 
detail the symmetries present in the model, and the resulting 
implications.

Ground state properties of the above Hamiltonian can be inferred by 
employing semi-classical considerations.  Since the fields 
$\varphi_a$ 
enter quadratically, they can be integrated out, leaving an effective 
action in terms of the four fields $\theta_a$.  Since the single 
coupling constant $g$ is marginally relevant and flowing off to 
strong 
coupling, these fields will be pinned in the minima of the cosine 
potentials.  Specifically, there are two sets of semiclassical 
ground states with all $\theta_{a} =2n_{a}\pi$ or all 
$\theta_{a} =(2n_{a}+1)\pi$, where $n_{a}$ are integers.
Excitations will be 
separated from the ground state by a finite energy gap, since the 
fields are harmonically confined, and instanton excitations 
connecting 
different minima are also costly in energy.

Since both $\theta_{\sigma \pm}$ fields are pinned, so are the 
spin-fields in each band, $\theta_{i \sigma}$ ($i=1,2$).  Since 
$\partial_x \theta_{i\sigma}$ is proportional to the z-component of 
spin in band $i$, a pinning of these fields implies that the spin in 
each band vanishes, and excitations with non-zero spin are expected 
to 
cost finite energy: the spin gap. This can equivalently be 
interpreted as singlet pairing of electron pairs in each band. It is 
instructive to consider the pair field operator in band $i$:
\begin{equation}
\Delta_i = c^{}_{{\scriptscriptstyle R}i \uparrow} 
c^{}_{{\scriptscriptstyle L}i \downarrow} = 
\kappa_{i \uparrow} \kappa_{i \downarrow} 
e^{{i \over \sqrt{2}} (\varphi_{i\rho} + \theta_{i\sigma})}  .
\end{equation}
With $\theta_{i \sigma} \approx 0$, $\varphi_{i\rho}$ can be 
interpreted as the phase of the pair field in band $i$.  The relative 
phase of the pair field in the two bands follows by considering the 
product
\begin{equation}
\Delta^{\vphantom\dagger}_1 \Delta^\dagger_2 = -\Gamma 
e^{i \theta_{\sigma -}} e^{i\varphi_{\rho -}}  ,
\label{pairsign}
\end{equation}
with $\Gamma = \kappa_{1 \uparrow} \kappa_{1 \downarrow}\kappa_{2 
\uparrow} \kappa_{2 \downarrow} =1$.  Since $\theta_4 = \varphi_{\rho 
-}$ the relative phase is also pinned by the cosine potential, with a 
sign change in the relative pair field, $\Delta^{\vphantom\dagger}_1 
\Delta^\dagger_2 < 0$, corresponding to a D-wave symmetry.  Being at 
half-filling, the overall charge mode, $\theta_{\rho +}$ is also 
pinned -- there is a charge gap -- and the two-point pair field  
correlation function falls off exponentially with separation.  We 
refer to this phase as a ``D-Mott" phase, having D-wave pairing 
correlations coincident with a charge gap.  Upon doping the D-Mott 
phase away from half-filling, gapless charge fluctuations are 
expected 
in the $(\rho +)$ sector, and power-law D-wave pairing correlations 
develop.

It is worth noting that the fully gapped D-Mott phase has a very 
simple interpretation in the strong coupling limit.  Two electrons 
across each of the rungs of the two-legged ladder form singlets, of 
the usual form \mbox{$|\uparrow,\downarrow \rangle - 
|\downarrow,\uparrow\rangle$}, where the two states refer to electrons 
on leg 1 or 2, respectively.  This two-electron state can be 
re-written in the bonding anti-bonding basis, and takes the form, 
\mbox{$|\uparrow \downarrow , -\rangle - |-, \uparrow \downarrow\rangle$}, 
where the two states now refer to bonding and anti-bonding orbitals.  
This resembles a local Cooper pair, with a relative sign change 
between bonding and anti-bonding pairs: an approximate D-wave 
symmetry.

\section{$SO(8)$ Gross-Neveu model}
\label{sec:GN}

As shown above, the bosonized effective Hamiltonian
on energy scales of order the gap is exceptionally simple in the
D-Mott phase.  In this section, we show that this simplicity is
indicative of a higher symmetry, and explore its ramifications upon
the spectrum.

\subsection{Gross-Neveu Model}

An obvious symmetry of the bosonic action, Eqs.~\ref{free_boson}- 
\ref{int_boson}, is
permutation of the fields $\theta_a \rightarrow P_{ab}\theta_b$, where
$P_{ab}$ is a permutation matrix.  In fact, this is only a small
subset of the true invariances of the model.  As is often the case,
abelian bosonization masks the full symmetry group.  It can be brought
out, however, by a refermionization procedure.  We define 
``fundamental'' (Dirac) fermion operators $\psi_{{\scriptscriptstyle 
P} a}$
with $a=1,2,3,4$ via
\begin{eqnarray}
\psi_{{\scriptscriptstyle P} a} & = & \kappa^{}_a e^{i 
\phi_{{\scriptscriptstyle P} a}},   \qquad a= 1\ldots 3 \nonumber \\
\psi_{{\scriptscriptstyle P} 4} & = & P \kappa^{}_4 e^{i 
\phi_{{\scriptscriptstyle P} 4}},   
\label{ffs}
\end{eqnarray}
and $P= R,L = \pm 1$, as before.   The Klein factors are given by
\begin{eqnarray}
\kappa_1 = \kappa_{2\uparrow} & \qquad & 
\kappa_2 = \kappa_{1\uparrow}, \\
\kappa_3 = \kappa_{1\downarrow} & \qquad & 
\kappa_4 = \kappa_{2\downarrow}.
  \label{GN_Klein}
\end{eqnarray}
In the re-Fermionization of the fourth field we have chosen to include
a minus sign for the left mover.  This is convenient, due to the 
modified
commutators between the left and right fields: 
$[\phi_{{\scriptscriptstyle R}4},\phi_{{\scriptscriptstyle L}4}] 
=-i\pi$,
in contrast to the ``standard" form in Eq.~\ref{cm2}.

In these variables, the effective Hamiltonian density becomes
\begin{equation}
  {\cal H} = \psi_a^\dagger i \tau^z \partial_x \psi_a - g
  \left( \psi_a^\dagger \tau^y \psi^{\vphantom\dagger}_a\right)^2 ,
\label{GNDirac}
\end{equation}
where $\psi_a = (\psi_{{\scriptscriptstyle R}a},
\psi_{{\scriptscriptstyle L}a})$, and $\bbox{\tau}$ is a vector of
Pauli matrices acting in the $R,L$ space.
Here, summation over repeated indices, $a=1,2,..,4$ is implicit.  It 
is remarkable that the
Hamiltonian can be written locally in the ``fundamental'' fermion
variables, which are themselves highly {\sl non}-locally related to
the ``bare'' electron operators.

A further simplification arises upon changing to Majorana fields,
\begin{equation}
\psi_{{\scriptscriptstyle P}a} = {1 \over \sqrt{2}}\left(
    \eta_{{\scriptscriptstyle R} 2a} + i \eta_{{\scriptscriptstyle R}
      2a-1} \right).
\end{equation}  
The Hamiltonian density then takes the manifestly invariant form  
\begin{equation} 
  {\cal H} =   {1 \over 2} \eta_{{\scriptscriptstyle RA}}
  i\partial_x \eta_{{\scriptscriptstyle RA}} 
  - {1 \over 2} \eta_{{\scriptscriptstyle LA}} i\partial_x
  \eta_{{\scriptscriptstyle LA}} + g G_{\scriptscriptstyle
    R}^{{\scriptscriptstyle AB}} G_{\scriptscriptstyle
    L}^{{\scriptscriptstyle AB}} 
  ,   \label{O8GrossNeveu}
\end{equation}
where the currents are
\begin{equation}
  G_{\scriptscriptstyle P}^{{\scriptscriptstyle AB}} = i
  \eta_{{\scriptscriptstyle PA}} \eta_{{\scriptscriptstyle PB}},
  \qquad A \neq B, 
\end{equation}
and $A,B = 1\ldots 8$.

\subsection{$SO(8)$ Symmetry}

Eq.~\ref{O8GrossNeveu}\ is the standard form for the $SO(8)$
Gross-Neveu model, which has been intensively studied in the
literature.\cite{Gross74,Zamolodchikov79,Shankar78,Karowski81,Dashen75,Shankar80,Shankar81}\ 
We first discuss its manifest symmetry properties.

The 28 currents $G_{\scriptscriptstyle P}^{{\scriptscriptstyle AB}}$
generate chiral $SO(8)$ transformations.  For $g=0$,
Eq.~\ref{O8GrossNeveu}\ has two independent symmetries under separate
rotations of the left- and right-moving fields.  For $g\neq 0$,
however, only simultaneous rotations of both chiralities are allowed.
More precisely, the unitary operators
\begin{equation}
  U(\chi_{{\scriptscriptstyle AB}}) = e^{i \chi_{{\scriptscriptstyle
        AB}} \int\! dx\, \left( G_{\scriptscriptstyle
        R}^{{\scriptscriptstyle AB}} + 
      G_{\scriptscriptstyle L}^{{\scriptscriptstyle AB}} \right)}, 
\end{equation} 
generate global orthogonal transformations of the Majorana fields, 
\begin{equation} 
  U^\dagger(\chi) \eta_{{\scriptscriptstyle PA}} U(\chi) = {\cal
    O}_{{\scriptscriptstyle AB}}(\chi) \eta_{{\scriptscriptstyle PB}},
    \label{Utransf} 
\end{equation} 
where the orthogonal matrix ${\cal O}(\chi)$ is given by
\begin{equation}
  {\cal O}(\chi) = e^{i \chi_{{\scriptscriptstyle AB}}
    T_{{\scriptscriptstyle AB}}}. 
\end{equation}
Here the $T_{\scriptscriptstyle AB}$ ($A>B$) are the 28 generators of
$SO(8)$ in the fundamental representation, with matrix elements
$\left[T_{{\scriptscriptstyle AB}}\right]_{{\scriptscriptstyle CD}} =
i(\delta_{{\scriptscriptstyle AC}} \delta_{{\scriptscriptstyle BD}} -
\delta_{{\scriptscriptstyle AD}} \delta_{{\scriptscriptstyle BC}})/2$.
Eq.~\ref{Utransf}\ indicates that the $\eta_{{\scriptscriptstyle PA}}$
transform as $SO(8)$ vectors.  Similarly, the currents
$G_{\scriptscriptstyle P}^{{\scriptscriptstyle AB}}$ are rank 2 
$SO(8)$
tensors.

It is worth noting that despite the non-local relation between the
fundamental and bare fermion operators, the $SO(8)$ symmetry remains
local in the bare electron basis.  This follows from the fact that the
chiral $SO(8)$ currents in the two bases are actually linearly 
related,
i.e.
\begin{equation}
  G_{\scriptscriptstyle P}^{{\scriptscriptstyle AB}} =
  M_{\scriptscriptstyle P}^{{\scriptscriptstyle ABCD}}
  \tilde{G}_{\scriptscriptstyle P}^{{\scriptscriptstyle CD}}, 
\end{equation}
where $\tilde{G}_{\scriptscriptstyle P}^{{\scriptscriptstyle AB}} = i
\xi_{{\scriptscriptstyle PA}} \xi_{{\scriptscriptstyle
    PB}}$, and the bare Majorana operators are defined by
\begin{eqnarray}
  c_{{\scriptscriptstyle P}1\uparrow}&=& {1 \over \sqrt{2}}\left(
    \xi_{{\scriptscriptstyle P}2} + i 
    \xi_{{\scriptscriptstyle P}1}\right),
\\
  c_{{\scriptscriptstyle P}1\downarrow} &=& {1 \over 
    \sqrt{2}}\left( \xi_{{\scriptscriptstyle P}4} + i
    \xi_{{\scriptscriptstyle P}3}\right), 
\\ 
  c_{{\scriptscriptstyle P}2\uparrow} &=& {1 \over \sqrt{2}}\left(
    \xi_{{\scriptscriptstyle P}6} + i 
    \xi_{{\scriptscriptstyle P}5}\right),
\\
  c_{{\scriptscriptstyle P}2\downarrow} &=& {1 \over 
    \sqrt{2}}\left( \xi_{{\scriptscriptstyle P}8} + i
    \xi_{{\scriptscriptstyle P}7}\right).   
\end{eqnarray}
The precise forms of the tensors $M_{\scriptscriptstyle P}$ are
complicated and not particularly enlightening.  Nevertheless, the {\sl
  existence} of the linear relation between currents implies that the
unitary operator $U(\chi)$ also generates local rotations of the bare
electron fields.  In these variables, however, the $SO(8)$ symmetry is
{\sl hidden}, because $M_{\scriptscriptstyle R} \neq
M_{\scriptscriptstyle L}$, which implies different rotations must be
performed amongst right- and left-moving electron operators.

Finally, it is instructive to see how the conservation of total 
charge and spin,
corresponding to a global $U(1) \times SU(2)$ symmetry, is embedded in the
larger $SO(8)$ symmetry.  To this end, consider the total electron
charge operator,
$Q$, which in terms of the low energy fields can be written,
\begin{equation}
Q = 2 \int dx \sum_P \psi^\dagger_{{\scriptscriptstyle P}1} 
\psi^{}_{{\scriptscriptstyle P}1} = 
2 \int dx (G_{\scriptscriptstyle R}^{21} + G_{\scriptscriptstyle L}^{21}),
\end{equation}
where $\psi_{{\scriptscriptstyle P}1}$ 
is a fundamental Gross-Neveu fermion.
The $U(1)$ charge symmetry is thus seen to be equivalent to
the $SO(2)$ symmetry of rotations in the $1-2$ plane of the 
eight-dimensional vector space. Similarly, the total spin operator
\begin{equation}
  \bbox{S} = \int \! dx \, [ \bbox{J}_{\scriptscriptstyle R}(x) +
  \bbox{J}_{\scriptscriptstyle L}(x) ],
\label{totalspin}
\end{equation}
with $\bbox{J}_{\scriptscriptstyle P}(x) =
\bbox{J}_{{\scriptscriptstyle P}ii}(x)$, can be re-expressed
in terms of $SO(8)$ generators by using,
\begin{equation}
  J^a_{\scriptscriptstyle P}(x) = \epsilon^{abc} 
G_{\scriptscriptstyle P}^{bc}  ,
\end{equation}
with $a,b,c = 3,4,5 = x,y,z$.  Thus we see the equivalence
between the $SU(2)$ spin rotations and $SO(3)$ rotations in
the 3-dimensional sub-space $3-4-5$ of the eight dimensional
vector space.  Rotations in the five-dimensional subspace
$1-2-3-4-5$, correspond to global $SO(5)$ rotations which
unify the charge and spin degrees of freedom.

In the absence of interactions in the
Gross-Neveu model, all of the excitations including spin remain 
massless.
In this case there is an independent $SU(2)$ spin symmetry
in the right and left moving sectors.  The spin currents
$\bbox{J}_{\scriptscriptstyle P}$ can then be shown
to satisfy,
\begin{eqnarray}
[J^a_{\scriptscriptstyle P}(x), 
J^b_{\scriptscriptstyle P}(x^\prime)] =
\delta(x-x^\prime) i \epsilon^{abc} 
J^c_{\scriptscriptstyle P}(x) + \\
i {P \over 2\pi} k \delta_{ab}
\delta^\prime(x-x^\prime)  ,
\end{eqnarray}
with $a,b,c  = x,y,z$ and $k=2$.  This is
referred to as an $SU(2)$ current algebra at level ($k$) two.

We conclude this subsection by answering a question which may have
occurred to the alert reader: why is the symmetry of the model $SO(8)$
rather than $O(8)$?  Based on Eq.~\ref{O8GrossNeveu}, it would appear
that {\sl any} transformation of the form $\eta_{\scriptscriptstyle
  PA} \rightarrow {\cal O}_{\scriptscriptstyle AB}
\eta_{\scriptscriptstyle PB}$ would leave the Hamiltonian invariant,
including {\sl improper} rotations with ${\rm det} {\cal O} = -1$.
The presence of such improper rotations means $O(8) = SO(8) \times
{\cal Z}_2$, since any orthogonal matrix can be factored into a
product of matrix with determinant one and a {\sl particular}
(reflection) matrix, e.g. ${\cal O}^{\rm r}_{\scriptscriptstyle AB} =
\delta_{\scriptscriptstyle AB} - 2 \delta_{\scriptscriptstyle A1}
\delta_{\scriptscriptstyle B1}$.  We have already shown above that the
$SO(8)$ symmetry is physical -- i.e. the symmetry generators act
within the Hilbert space of the physical electrons.  It is
straightforward to show that the ${\cal Z}_2$ reflection is however,
{\sl unphysical}.  To see this, imagine performing the  ${\cal Z}_2$
reflection effected by ${\cal O}^{\rm r}$ above, which takes
$\eta_{\scriptscriptstyle P1}
\rightarrow - \eta_{\scriptscriptstyle P1}$.  Using the bosonization
rules, this corresponds to $\theta_1 \rightarrow - \theta_1$ and
$\varphi_1 \rightarrow - \varphi_1$.  Returning to the physical
fields, one finds that the bare electron operators transform much more 
non-trivially:
\begin{equation}
  c_{{\scriptscriptstyle P} i\alpha}^{\vphantom\dagger}
  \stackrel{{\cal Z}_2}{\longrightarrow} c_{{\scriptscriptstyle P}
    i\alpha}^{\vphantom\dagger} \psi_{\scriptscriptstyle P1}^\dagger.
\end{equation}
As we shall show in Sec.~IV.E.3, a single GN fermion operator, such as
$\psi_{\scriptscriptstyle P1}^\dagger$, is unphysical.  The ${\cal
  Z}_2$ reflection thus takes a physical electron operator into an
unphysical one, which implies that the symmetry cannot be effected by
a unitary operator within the Hilbert space of the electrons.  For
this reason, the true symmetry group of the ladder model is $SO(8)$.

\begin{figure}[h]
\epsfxsize=6cm
\centerline{\epsfbox{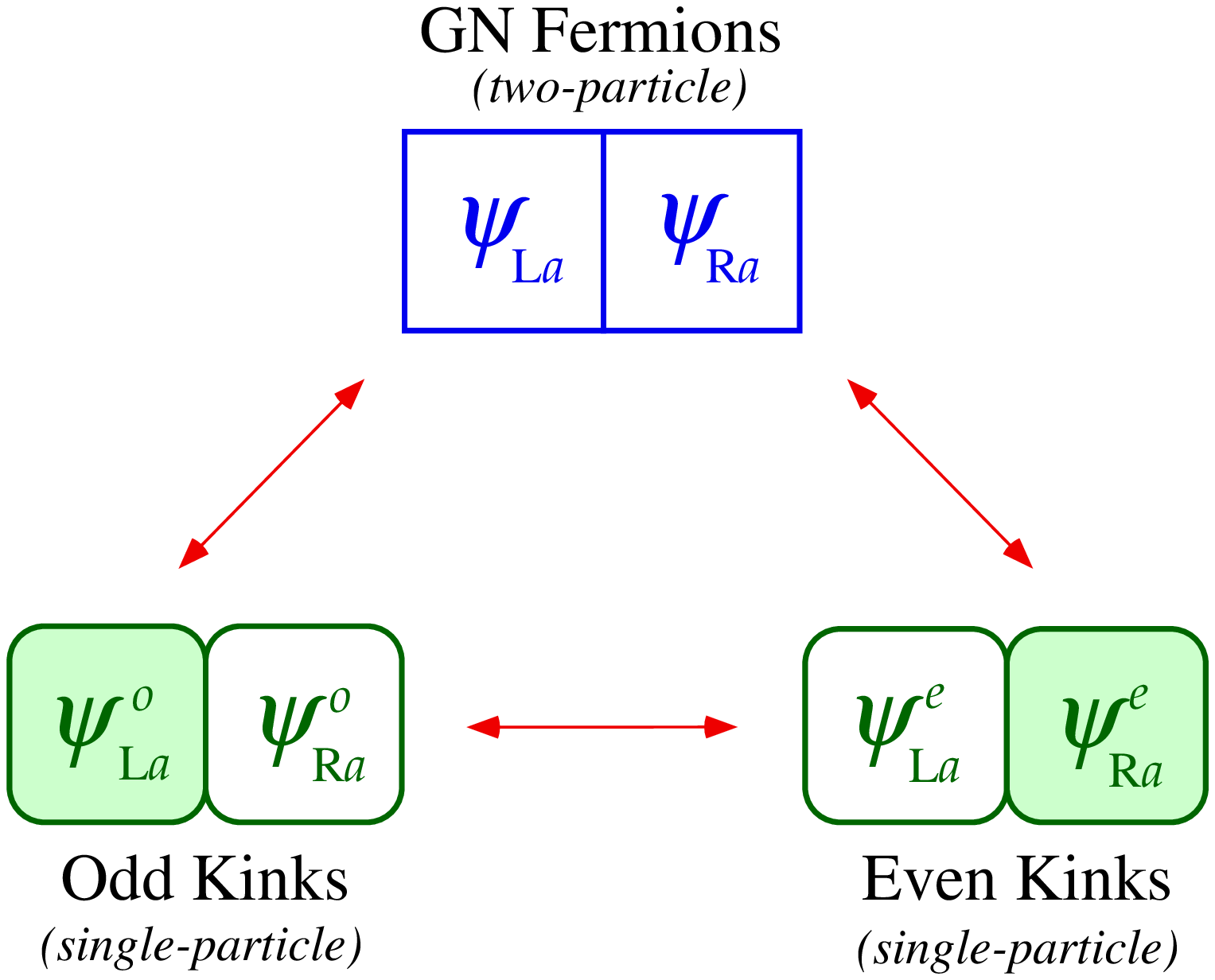}}
%\vspace{5pt}
{\noindent FIG.  2: Triality between GN fermions, even kinks and 
odd kinks. The $SO(8)$ GN Hamiltonian is identical in terms of these 
three sets of fermionic operators. Operators in the gray areas are
physical and gauge independent (see Sec.~IV.E), while the other fermion 
operators must be ``dressed'' by an appropriate Jordan-Wigner string
to remain in the physical Hilbert space.}
\end{figure}

\subsection{Triality}

Most of the above properties hold more generally for the $SO(N)$ GN
model, even for $N \neq 8$.  However, the case $N=8$ is extremely
special, and in fact possesses an additional {\sl triality} symmetry
not found for other $N$.  A useful reference is
Ref.~\onlinecite{Shankar80,Shankar81}.

To expose the additional symmetry, we return to the sine-Gordon
formulation.  Essentially, the triality operation trades the original
basis $\{ \theta_a \}$ in the four-dimensional space of boson fields
for either one of two other orthogonal bases.  Explicitly, the two
alternate choices are the even and odd fields $\theta^{e/o}_a$, where 
\begin{eqnarray}
  \theta^{e/o}_1 & = & (\theta_1 + \theta_2 + \theta_3 \pm
  \theta_4)/2, \\
  \theta^{e/o}_2 & = & (\theta_1 + \theta_2 - \theta_3 \mp 
\theta_4)/2, \\
  \theta^{e/o}_3 & = & (\theta_1 - \theta_2 + \theta_3 \mp 
\theta_4)/2, \\
  \theta^{e/o}_4 & = & (\theta_1 - \theta_2 - \theta_3 \pm
  \theta_4)/2.
  \label{evenoddfields}
\end{eqnarray}
Here the upper and lower signs apply to the even and odd fields,
respectively, and identical definitions hold for the dual
$\varphi^{e/o}_a$ and chiral $\phi_{{\scriptscriptstyle P}a}^{e/o}$ 
bosons.  
The bosonized
Hamiltonian in Eqs.~\ref{free_boson}, \ref{int_boson}
 is invariant under either change of
variables, i.e.
\begin{equation}
  H[\theta_a] = H[\theta^e_a] = H[\theta^o_a].
\end{equation}
For each of these bases, an inequivalent refermionization is possible,
analogous to the introduction of the fundamental fermions in
Eq.~\ref{ffs}.  In particular, the Hamiltonian is unchanged in form
when rewritten in terms of either the even or odd fermion operators,
\begin{equation}
  \psi^{e/o}_{{\scriptscriptstyle P} a} = \kappa_a^{e/o} e^{i
    \phi^{e/o}_{{\scriptscriptstyle P} a}}. 
\end{equation}
It should be noted that the set of even and odd fermion operators
contains all the bare electron fields.  In particular,
\begin{eqnarray}
  \psi_{{\scriptscriptstyle R}1}^e = c_{{\scriptscriptstyle
      R}1\uparrow}, & \qquad & \psi_{{\scriptscriptstyle L}1}^o = 
  c_{{\scriptscriptstyle L}1\uparrow},\label{kink1} \\
  \psi_{{\scriptscriptstyle R}2}^e = c_{{\scriptscriptstyle
      R}2\uparrow}, & \qquad & \psi_{{\scriptscriptstyle L}2}^o = 
  c_{{\scriptscriptstyle L}2\uparrow}, \\ 
  \psi_{{\scriptscriptstyle R}3}^e = c_{{\scriptscriptstyle
      R}2\downarrow}, & \qquad & \psi_{{\scriptscriptstyle L}3}^o = 
  c_{{\scriptscriptstyle L}2\downarrow}, \\ 
  \psi_{{\scriptscriptstyle R}4}^e = c_{{\scriptscriptstyle
      R}1\downarrow}, & \qquad & \psi_{{\scriptscriptstyle L}4}^o = 
  c_{{\scriptscriptstyle L}1\downarrow}.\label{kink4} 
\end{eqnarray} 
The other eight even and odd fields ($\psi_{{\scriptscriptstyle
    L}a}^e$ and $\psi_{{\scriptscriptstyle R}a}^o$) 
are not simply related, however, to the electron fields. 
 
\subsection{Conventional Gross-Neveu Excitation spectrum} 
 
The $SO(N)$ GN model is integrable, and the excitation spectrum is
known exactly.  To organize the presentation, we divide the discussion
of the excitation spectrum into two parts.  In this subsection, we
summarize known results for the conventional GN model.  The precise
nature of the excitations for the two-leg ladder model, however,
differs from those in the conventional GN model.  This difference
arises from the non-local relation between the electron and GN fields.
Excitations within the GN model must be slightly modified to satisfy
gauge invariance with respect to some unphyical degrees of freedom
introduced in the mapping.  These modifications and the resulting
spectrum in the D-Mott phase are described in the subsequent
subsection.
 
Within the GN model, the excitations are of course organized into 
$SO(N)$ multiplets, but are further constrained for the case of 
interest, $N=8$, by triality.  In this subsection, we discuss the 
lowest-lying states, their multiplet structures and quantum numbers, 
and give some useful physical pictures to aid in understanding their 
properties. 
 
\subsubsection{Results from integrability} 
 
The lowest-lying excitations are organized into three $SO(8)$ vector 
multiplets, which are degenerate due to triality, for a total of 24 
particles.  Four of the 28 global $SO(8)$ generators may be chosen diagonal (to 
form the Cartan subalgebra).  We will label the particles by the 
values of the four associated charges, denoted by the ordered quadruplet $(N_1,N_2,N_3,N_4)$, and defined by
\begin{equation} 
  N_a = \int \! dx \, \psi_a^\dagger \psi^{}_a, 
\end{equation}
(no sum on $a$). 
In this notation, one $SO(8)$ multiplet contains the states (known as 
{\sl fundamental fermions}) with only one of the four $N_a = \pm1$, 
and all others equal to zero.  The remaining 16 degenerate states have
$N_a = \pm 1/2$ for $a=1,2,3,4$, which are divided into those with an even number of
$N_a = + 1/2$ (the {\sl even kinks}) and the remainder with an odd
number of $N_a = + 1/2$ (the {\sl odd kinks}).  The reasons for this
terminology will become apparent later in this section.  
Each particle has a mass $m$
and disperses (due to Lorentz invariance) as
$\epsilon_1(q) = \sqrt{m^2 + q^2}$, with momentum $q$.  Since the electron
band operators
$c_{{\scriptscriptstyle P}i\alpha}$ are defined relative to their
Fermi momenta $k_{{\scriptscriptstyle F}i}$, 
the actual momenta of each particle
is offset from the GN model momentum, $q$, by some amount.  We will return
to these ``base'' momenta later in this subsection, as well as to the
other physical quantum numbers of the excitations.
 
At somewhat higher energies there is another multiplet of 28
``particles", which transform as an antisymmetric second rank $SO(8)$
tensor.  This multiplet can be viewed as two-particle bound states of
the fundamental Gross-Neveu fermions, or equivalently under triality
as bound even-even or odd-odd kinks.  Indeed, of these 28 states, 24
have two zero charges and two $N_a = \pm 1$.  The other four are bound
states of a fundamental fermion with it's anti-particle (an excition
in the semiconductor picture, below), so they do not carry any of the four
quantum numbers.  Each of the 28 ``particle" states has a mass $m_2 =
\sqrt{3} m$.  Finally, for energies $\epsilon > \epsilon_c(q) =
2\sqrt{m^2 + q^2/4}$, a two-particle continuum of (unbound) scattering
states exists.

\begin{figure}[hbt]
\epsfxsize=6cm
\centerline{\epsfbox{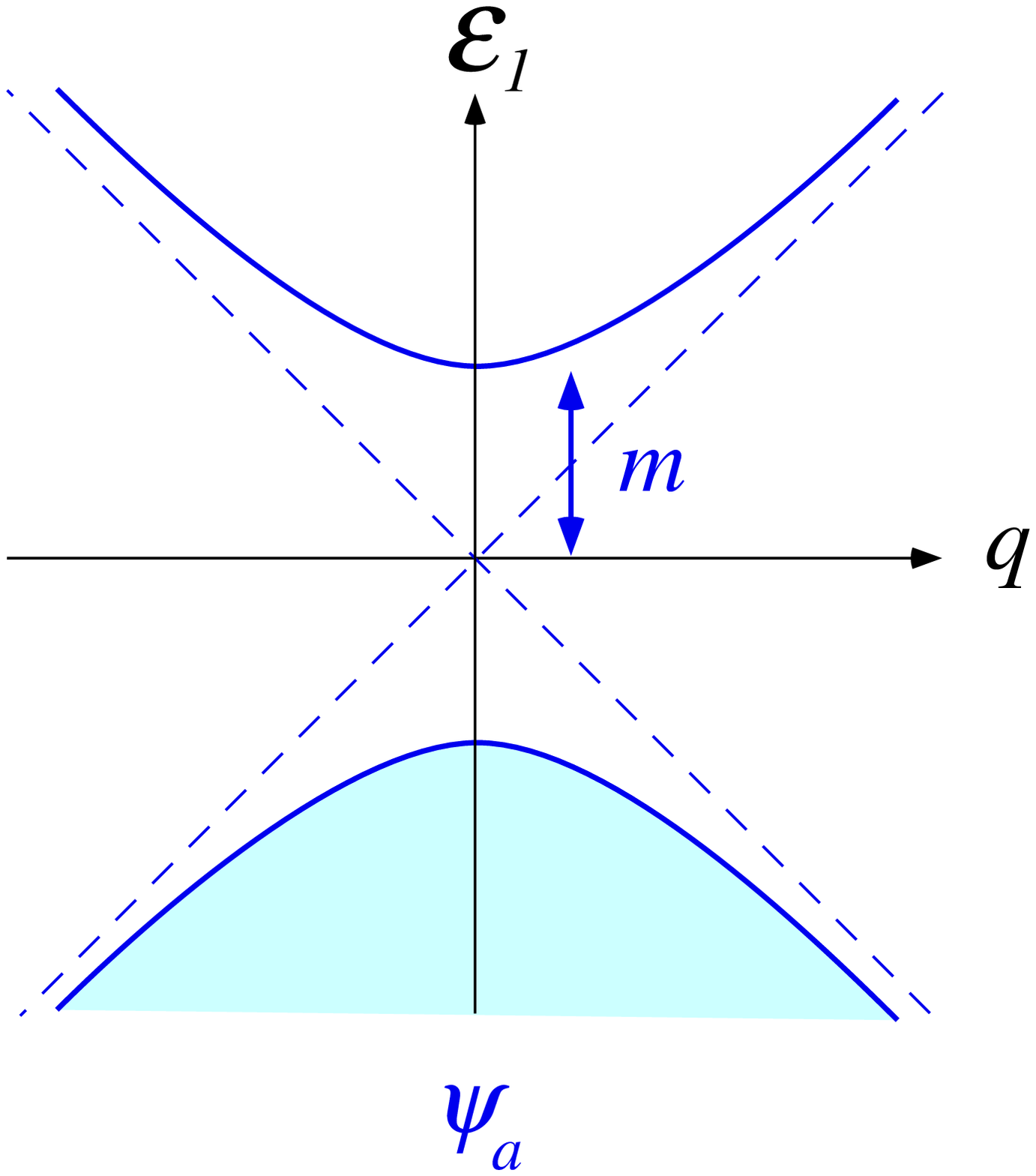}}
\vspace{5pt}
{\noindent FIG. 3: The mean field picture of the $SO(N)$ GN model. 
There are four flavors of relativistic massive fermions $\psi_{a}$, 
with dispersion $\epsilon_1(q) =\pm \sqrt{m^2 + q^2}$. The negative
energy bands are filled, while the positve energy bands are empty.
As in a semi-conductor, the positive and negative energy bands are 
separated by a finite gap $2m$.  
}
\end{figure}

\subsubsection{Mean-field picture} 
 
It is instructive to see how these excitations arise in a mean-field 
treatment of the GN interaction.  The mean-field treatment becomes 
exact for the $SO(N)$ generalization of the GN model for large even 
$N$.  To carry it out, we employ the Dirac fermion version of the 
Hamiltonian, Eq.~\ref{GNDirac}.  In the mean-field approximation, the 
bilinear $\psi_a^{\dagger}\tau^{y}\psi_{a}^{}$ 
acquires an expectation value, and the ``quasiparticle'' Hamiltonian 
density becomes 
\begin{equation} 
        {\cal H}_{\rm MF} =  \psi_{a}^{\dagger} 
        i\tau^{z}\partial_{x}\psi_{a}^{\vphantom\dagger} - 
        \Delta \psi_a^{\dagger}\tau^{y}\psi_{a}^{\vphantom\dagger} ,
        \label{HMF}
\end{equation}
where $\Delta = 2g\langle
\psi_a^{\dagger}\tau^{y}\psi_{a}^{\vphantom\dagger} \rangle$ is a
mean-field gap parameter.  The mean-field Hamiltonian is simply that
of four massive Dirac equations.  It is easily
diagonized in momentum space, using $\psi_a(q) = \exp (i \Omega(q)
\tau^x /2 )\tilde \psi_a$, where $\Omega(q) = \cot^{-1} (vq/\Delta)$,
which gives
\begin{equation}
  H_{\rm MF} = \int\! {{dq} \over {2\pi}} \, \epsilon_1(q)
  \tilde\psi_a^\dagger \tau^z \tilde\psi_a^{\vphantom\dagger},
\end{equation}
with $\epsilon_1(q) =\sqrt{m^2 + q^2}$ and 
the mass $m=\Delta$.  From the diagonalized form it is straightforward 
to determine the MF estimate,
\begin{equation}
  m_{\rm\scriptscriptstyle MF} 
  = 2\Lambda e^{-\pi/Ng},
\end{equation}
for the general $SO(N)$ case, where $\Lambda \sim t$ is a momentum
cut-off.  The exponential dependence on $g$ can be understood from the
marginality of the interactions under the RG scaling transformation.
The picture is that of a ``semi-conductor'', as indicated
schematically in Fig.~3.  These massive Dirac particles and their
antiparticles may be identified with the fundamental fermion $SO(8)$
vector multiplet.  The even and odd kinks likewise arise from applying
the same decoupling to the even and odd fermion representations of the
Hamiltonian.

While Eq.~\ref{HMF}\ is correct for $SO(\infty)$, it requires
corrections otherwise.  For finite $N$, the chiral ``order parameter''
$\Delta$ fluctuates around its vacuum value, and these fluctuations
generate {\sl attractive} interactions between the GN fermions.  The
attractive interactions lead to the formation of 
two-particle bound states, whose mass $m_{2} = 2m[1 - (\pi^2/2N^2) +O(N^{-4})]$
approaches twice the fermion
mass for $N \gg 1$, due to the weakness of the interactions in this
limit.  For $SO(8)$, however, the inter-fermion interactions are not
weak, and the bound states have the strongly reduced mass $m_{2} =
\sqrt{3} m$.  The 28 bound states of two fermions form the
abovementioned rank 2 tensor multiplet.  A priori, one might expect
three such multiplets, arising from bound states of the three sets of
fermions.  We will see, however, in the next section that this does
not lead to any new particle content.  Indeed, the particles in the
tensor multiplet can be equally well viewed as bound states of
fundamental, even, or odd fermions.

\subsubsection{Semi-classical picture}

These excitations can be readily understood in the semi-classical 
limit of the Bosonized Hamiltonian.  In this language, particles 
correspond to classical solitons, in which the phase fields 
$\theta_{a}$ connect different vacuum (classical minimum energy) 
values at $x=\pm\infty$.  The winding numbers of these solitons have a
direct connection to the $SO(8)$ charges, since by bosonization
\begin{eqnarray}
  \Delta \theta_a & = & \theta(\infty) - \theta(-\infty) =
  \int_{-\infty}^\infty 
  \!\! dx \, \partial_x  \theta_a \nonumber \\
  & = & 2\pi N_a.
\end{eqnarray}
Thus each of the GN particles labelled by the four quantum numbers
$(N_1,N_2,N_3,N_4)$ can be
readily transcribed into a semi-classical soliton.  The fundamental
fermions are those configurations in which {\sl one} of the four phase
fields $\theta_{a}$ changes by $\pm 2\pi$.  The second type of soliton
changes all four $\theta_{a}$ fields by $\pm \pi$, which changes
$\cos\theta_a \rightarrow -\cos\theta_a$, but leaves the vacuum energy
unchanged.  The $2^{4} = 16$ possible ``kinks'' form the
semi-classical analog of the even and odd kink $SO(8)$ octets.

\begin{figure}[hbt]
\epsfxsize=7cm
\centerline{\epsfbox{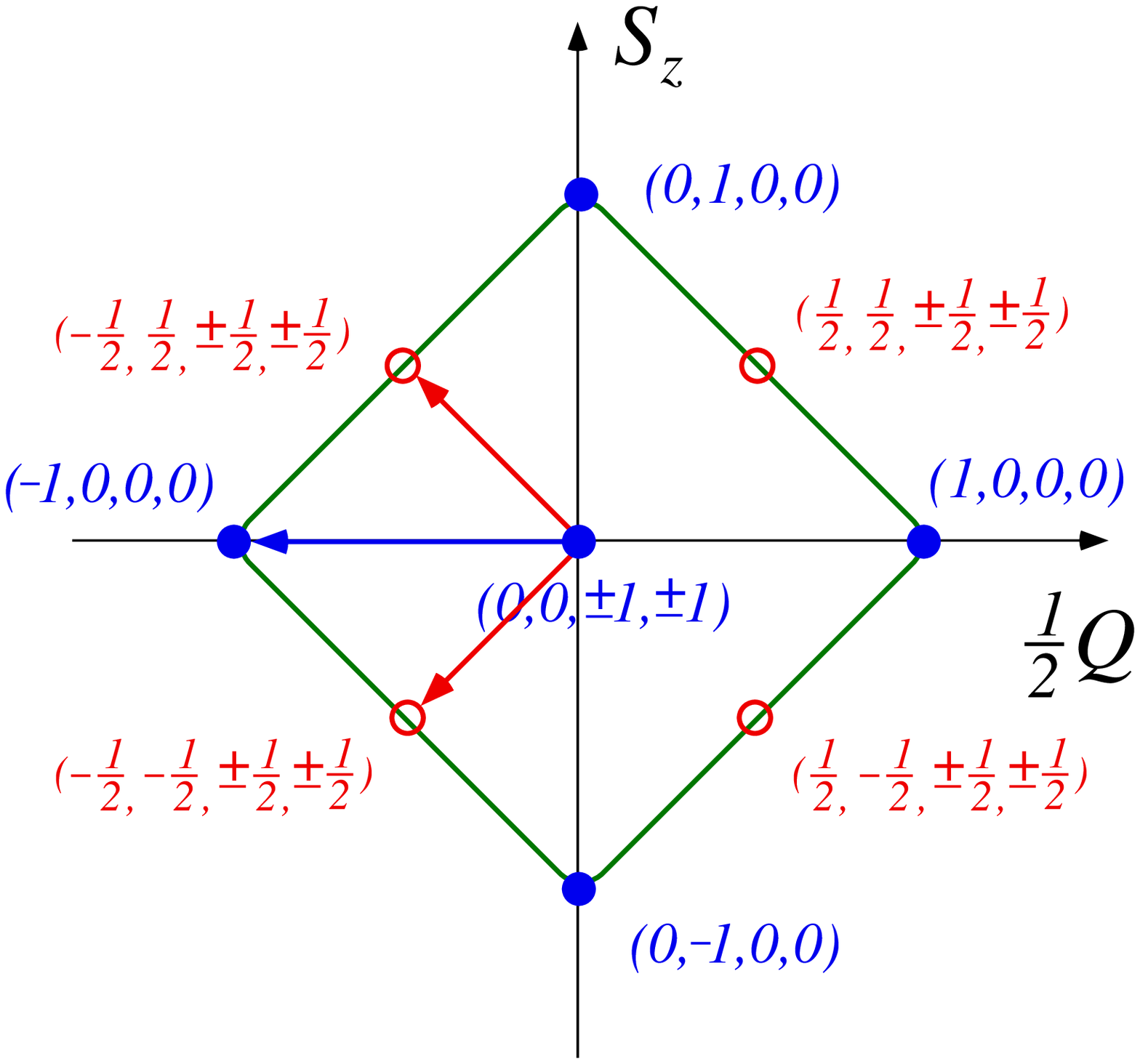}}
\vspace{5pt}
{\noindent FIG. 4: The 24 mass $m$ excitations of the $SO(8)$ GN
  model, projected into the $(N_{1}, N_{2})=(Q/2, S_{z})$ plane.  Full 
  and open circles indicate the ``fundamental'' fermions and kinks,
  respectively.  All 24 excitations lie on the unit sphere in the full 
  four-dimensional Cartan space.  The equivalence of a fundamental
  fermion and an even and odd kink can be seen graphically by simple
  vectorial addition, e.g. the odd kink $(-1,1,1,1)/2$ and the even kink 
$-(1,1,1,1)/2$ add to form the GN fermion $(-1,0,0,0)$.}
\end{figure}

While the even and odd kinks exist for general $N$, for the special
case of $SO(8)$, triality implies that the kinks and fundamental
solitons are on an equal footing.  This is most easily seen using a
simple graphical construction.  Construct an $N/2$-dimensional space
(for $N$ even) with axes $\theta_a$.  In this space, draw a lattice
consisting of a point for each vacuum configuration of the fields.
All possible solitons can be represented on this lattice as lines
connecting different points to the origin (see Fig. 4).  The
fundamental fermions are then the line segments to the neighboring
points along the axes.  For $N=8$, however, there are another 16
points equidistant to the origin, which represent the even and odd
kinks (for $N>8$, these are further from the origin, while for $N<8$
they are in fact closer).  In this case, the even or odd kink segments
form equally good orthonormal axes in this space, and viewed in this
basis, the three sets of particles cyclically exchange their roles.

One can also compose these particles by vectorial addition.  For
instance, an even and an odd kink can be added to form a fundamental
fermion.  The 2-particle bound states may also be visualized in this
way, by adding e.g. two different fundamental fermion vectors.  From
this construction it is easy to see that any such 2-particle state can
be equally well composed from two even or two odd kinks.  There is
thus only a single 28-fold tensor multiplet of 2-particle bound 
states.

\subsection{Consequences for the D-Mott phase}

We are now in a position to discuss the nature of the ground state and
excitation spectrum in the D-Mott phase, using the technology of the
GN model.  

\subsubsection{Gauge redundancy} 

To proceed, we must first describe the boundary conditions and
gauge-redundancy needed to fully specify the model.  Since the phase
fields were originally introduced to bosonize the (physical)
electron operators, the
chiral electron phases are defined only moduli $2\pi$,
\begin{equation}
  \phi_{{\scriptscriptstyle P}i\alpha}(x,\tau) \leftrightarrow
  \phi_{{\scriptscriptstyle P}i\alpha}(x,\tau)  + 2\pi {\cal
    A}_{{\scriptscriptstyle P}i\alpha}(x,\tau),
  \label{gauge_invariance}
\end{equation}
where the ${\cal A}_{{\scriptscriptstyle P}i\alpha}$ are integers.  
These
integers describe a sort of gauge redundancy in the description:
semiclassical phase configurations which differ only by a different
choice of ${\cal A}_{{\scriptscriptstyle P}i\alpha}$ are to be 
treated as
identical quantum states.  Furthermore, as for any gauge theory,
physical operators must be gauge-invariant, i.e. unchanged under the
operation in Eq.~\ref{gauge_invariance}, which can be performed {\sl
  locally}.

\subsubsection{Uniqueness of the ground state}

From both the standpoint of the fermionic GN Hamiltonian and its
bosonized sine-Gordon form, the system appears to exhibit broken
symmetry.  The conventional GN model has a spontaneously broken
``chiral'' symmetry: the hamiltonian is invariant under the chiral
transformation $\psi_a \rightarrow \tau^z \psi_a$, however, the chiral
order parameter $\Delta = 2g \langle \psi_a^\dagger \tau^y
\psi_a^{\vphantom\dagger}\rangle \neq 0$ and changes sign under this
transformation.  In the bosonization language, this transformation
corresponds to $\theta_a \rightarrow \theta_a + \pi$ (for all $a$),
which takes $\cos\theta_a \rightarrow - \cos\theta_a$.  The bosonic
model appears to have even more broken symmetries, i.e. there is a
countably infinite set of semiclassical vacua, related by the
additional transformations $\theta_a \rightarrow \theta_a + 2\pi n_a$,
for integer $n_a$.

On physical grounds, however, we expect that the D-Mott phase has no 
broken symmetry, and consequently a unique ground state.  To reconcile 
this apparent discrepancy, we must account for the fact that the 
phases $\theta_a$ are not gauge invariant.  Indeed apparently 
different vacua may represent the same physical state in a different 
gauge.  To establish the physical
equivalence between different vacua
is a rather tedious and 
technical exercise, although
straightforward.
In Appendix \ref{app:gauge} we carry through this exercise and demonstrate
that {\it all} of the semi-calssical vacua do indeed
correspond to the same physical state.  Thus, as expected,
there are no broken symmetries in the 
D-Mott phase and the ground state is unique.

\subsubsection{Quantum numbers}

To connect the GN results with the physical two-leg ladder system, we now
consider the quantum numbers of the various excitations.  Each quantum
number corresponds to some conserved quantity in the system.  The most
physically interesting are the charge, spin, momentum along the $x$
direction, and parity (or equivalently, momentum in the $y$
direction).  The charge and spin are conserved quantities
corresponding to continuous global symmetries, so we can work directly
with the Hermitian generators
\begin{eqnarray}
  Q & = & \int \! dx\, c_{{\scriptscriptstyle P}j\alpha}^\dagger
    c_{{\scriptscriptstyle P}j\alpha}^{\vphantom\dagger}, \\
  \bbox{S} & = & \int \! dx\, c_{{\scriptscriptstyle 
P}j\alpha}^\dagger
  {\bbox{\sigma}_{\alpha\beta} \over 2}
  c_{{\scriptscriptstyle P}j\beta}^{\vphantom\dagger}.
\end{eqnarray}
Since the translational and leg-interchange symmetries are discrete,
we should really speak of the unitary operators themselves.  Since
right- and left-moving particles in band $j$ carry quasi-momentum $\pm
k_{{\scriptscriptstyle F}j}$, respectively, the translation operator is simply
\begin{equation}
  \hat{T}_x = e^{i \sum_j k_{{\scriptscriptstyle F}j} 
\left( N_{{\scriptscriptstyle R}j} - 
N_{{\scriptscriptstyle L}j} \right)},
\end{equation} 
where $N_{{\scriptscriptstyle P}j} = \int dx \sum_\alpha
c_{{\scriptscriptstyle P}j\alpha}^\dagger 
c_{{\scriptscriptstyle P}j\alpha}^{\vphantom\dagger}$ is the total 
number of
electrons in band 
$j$ with chirality $P$.  The quasi-momentum $P_x$ is defined by 
$\hat{T}_x = \exp (i P_x)$.  Because the anti-bonding (band $1$) 
operators have $k_y = \pi$, the parity or translation operator in the
$y$ direction is
\begin{equation}
  \hat{T}_y = e^{i\pi N_1},
\end{equation}
where $N_1 = N_{{\scriptscriptstyle R}1} + N_{{\scriptscriptstyle L}1}$.  

In the weak-coupling limit, $U \ll t, t_{\perp}$, there
are two additional conserved quantities, the band spin difference
$S_{12}^z = S_1^z - S_2^z$ and the relative band chirality $P_{12} =
(N_{{\scriptscriptstyle R}1} - N_{{\scriptscriptstyle L}1} -
N_{{\scriptscriptstyle R}2} + N_{{\scriptscriptstyle L}2})/2$.   
 
It is useful to rewrite these expressions in terms of the bosonized phase variables.  Because the symmetry generators involve
spatial integrals of fermionic bilinears, they can be expressed in
terms of the winding numbers $\Delta\theta_a$ and their conjugates
\begin{equation}
  \Delta\varphi_{a} = \int_{-\infty}^{\infty} \! dx \, 
\partial_x \varphi_a(x) =
  \varphi_a(\infty) -\varphi_a(-\infty).
\end{equation}
Using Eqs.~\ref{bosondef1}-\ref{bosondef2}, we find
\begin{eqnarray}
  Q & = & {{\Delta\theta_1} \over \pi}, \label{eq:PairWinding}\\
  S^z & = & {{\Delta\theta_2} \over {2\pi}}, \\
  S_{12}^z & = & {{\Delta\theta_3} \over {2\pi}}, \\
  P_{12} & = & {{\Delta\theta_4} \over {2\pi}}, \\
  \hat{T}_x & = & \exp\left[\frac{i}{2} \left(\Delta\varphi_1
  +{{(k_{{\scriptscriptstyle F}1} - k_{{\scriptscriptstyle F}2})} 
\over \pi} \Delta\theta_4\right)
  \right], \label{Tx}\\
  \hat{T}_y & = & \exp\left[ \frac{i}{2} \left(\Delta\theta_1 +
    \Delta\varphi_4\right)\right].  \label{Ty}
\end{eqnarray}

As discussed in the previous section, the winding numbers of the
$\theta_a$ are just the $SO(8)$ conserved charges.  Thus the first 
four
conserved quantities can be directly transcribed for all the GN
excitations, i.e. $(Q,S^z,S_{12}^z,P_{12}) = (2N_1,N_2,N_3,N_4)$.
These are tabulated in the first three columns in Table 1.  

The momentum and parity of the particles are more complicated,
however, because Eqs.~\ref{Tx}--\ref{Ty}\ contain the conjugate fields
$\Delta\varphi_1$, $\Delta\varphi_4$.  As such, $P_x$ and $P_y$ are not
simply determined from the $SO(8)$ charges $N_a$.  The additional
physics required is the operator content of the original electron
problem.  

To see how this comes in, let us imagine a local operator ${\cal
  O}^\dagger (\{ N_a \}; x)$, which creates the particle with charges 
$\{
N_a \}$ when acting on the ground state, i.e.
\begin{equation}
  {\cal O}^\dagger (\{ N_a \};x) |0\rangle = |\{ N_a \}; x\rangle,
\end{equation}
where $|\{N_a \}\rangle$ is the quantum state with one excited $\{ N_a
\}$ particle localized at $x$.  Now consider the exponential of a
phase field $\varphi_a$.  It can be rewritten as the line integral of 
the
momentum conjugate to $\theta_a$, i.e.
\begin{equation}
  e^{-i N_a \varphi_a(x)/2} = e^{i 2\pi N_a\int_x^{{\scriptscriptstyle
        \infty}} \! dx\, \Pi_a(x)}, 
\end{equation}
where $\Pi_a(x) = \partial_x \varphi_a/(4\pi)$ is the momentum 
conjugate
to $\theta_a$, i.e. $[\theta_a(x), \Pi_b(x')] = i
\delta_{ab}\delta(x-x')$.  Since the momentum $\Pi_a$ generates
translations of the phase $\theta_a$, the exponential operator creates
a soliton of size $2\pi N_a$ located at the point $x$.  In order to
have the correct winding numbers, the desired quantum operator must
thus have the form
\begin{equation}
  {\cal O}^\dagger (\{ N_a \};x) = \Lambda[\{ \theta_a \}; \{ N_a \} ]
  e^{-i N_a \varphi_a /2 }.
\end{equation}
Here we have included an arbitrary function $\Lambda$ of the
$\theta_a$ fields, which does not wind the phase and thus does not
affect the $SO(8)$ charges.

To determine $\Lambda$, we next impose gauge invariance.  Consider
first the operators which create the fundamental fermions, with only
one non-zero $N_a = \pm 1$.  The creation operator takes the form
\begin{equation}
  {\cal O}^\pm_a = \tilde\Lambda_a[\{\theta_a \}] e^{ \mp i
    (\varphi_a + \theta_a)/2},
  \label{GN_create}
\end{equation}
where we have removed a factor $e^{\mp i \theta_a/2}$ from $\Lambda$
to define $\tilde\Lambda_a^\pm$.  This is desirable because the last
factor (up to a Klein factor $\kappa_a$) is simply the GN fermion operator
$\psi_{{\scriptscriptstyle R}a}^\dagger$ ($\psi_{{\scriptscriptstyle
    R}a}^{\vphantom\dagger}$ for the lower sign).  Now ${\cal
  O}^\pm_a$ must be invariant under all possible gauge
transformations, Eq.~\ref{gauge_invariance}.  It is a straightforward
exercise to show that the most general form for $\tilde\Lambda_a^\pm$
is the same for all the fundamental fermions, and is given by
\begin{equation}
  \tilde\Lambda = {\cal O}_s 
  \left.\sum_{\{ k_a \}}\right.^\prime \lambda_{\{ k_a \}} e^{i \sum_a
    k_a \theta_a 
    },
  \label{generalstring}
\end{equation}
where $\left.\sum\right.^\prime$ indicates a sum over all quadruplets
of integers with $\sum_a k_a$ even, and
\begin{equation}
  {\cal O}_s = e^{{i \over 2}(\theta_1+\theta_2+\theta_3-\theta_4)}.
\end{equation}
Note that $\tilde\Lambda$ {\sl does not} include a term proportional
to unity, which implies that a single GN fermion is by itself not
gauge-invariant and hence unphysical.  Instead, physical particles
have an attached operator ${\cal O}_s$ (or its counterparts
with extra factors from the $\left.\sum\right.^\prime$ term).  ${\cal 
O}_s$
represents a Jordan-Wigner ``string'', and can be rewritten
only non-locally in terms of the fermion fields.  It modifies the
momentum and statistics of the fundamental fermions to those of the
physical excitations.

It is now straightforward to determine the quasi-momentum and parity
of the fundamental fermions using the translation operators in
Eqs.~\ref{Tx}--\ref{Ty}.  In particular, we must have
\begin{eqnarray}
  \hat{T}_x {\cal O}^\pm_a \hat{T}_x^{-1} & = & e^{i P_x(a)} {\cal
    O}^\pm_a, \label{xtrans} \\
  \hat{T}_y {\cal O}^\pm_a \hat{T}_y^{-1} & = & e^{i P_y(a)} {\cal
    O}^\pm_a. \label{ytrans}
\end{eqnarray}  
The left-hand sides of Eqs.~\ref{xtrans}--\ref{ytrans}\ 
can be evaluated by employing the commutators of the Bose fields
to obtain:
\begin{eqnarray}
\hat{T}_x \theta_a \hat{T}_x^{-1} & = & \theta_a + 2\pi \delta_{a1} ,
\\
\hat{T}_x \varphi_a \hat{T}_x^{-1} & = & \varphi_a 
+2(k_{{\scriptscriptstyle F}1}-k_{{\scriptscriptstyle F}2})\delta_{a4} ,
\end{eqnarray}
and
\begin{eqnarray}
\hat{T}_y \theta_a \hat{T}_y^{-1} & = & \theta_a + 2\pi \delta_{a4} ,
\\
\hat{T}_y \varphi_a \hat{T}_y^{-1} & = & \varphi_a +2\pi \delta_{a1} .
\end{eqnarray}
Eq.~\ref{xtrans}\ can be written as a product of three terms:
\begin{eqnarray}
  \hat{T}_x {\cal O}^\pm_a \hat{T}_x^{-1} & = & \hat{T}_x \left(
    \left.\sum\right.^\prime \right)
  \hat{T}_x^{-1} \times \hat{T}_x {\cal O}_s \hat{T}_x^{-1} \nonumber
  \\
  & & \times \hat{T}_x e^{ \mp i (\varphi_a + 
\theta_a)/2}\hat{T}_x^{-1}.
  \label{translation_decomposition}
\end{eqnarray}
Consider the first term.  Using the above commutators one can readily
show that the sum in Eq.~\ref{generalstring}
is invariant under $x-$translations.
The string, however,
carries momentum $P_x = \pi$:
\begin{equation}
  \hat{T}_x {\cal O}_s \hat{T}_x^{-1} = - {\cal O}_s = e^{i\pi} {\cal
    O}_s. 
\end{equation}
This momentum must be added to the ``bare'' momentum of the GN
fermion, given by the last term in
Eq.~\ref{translation_decomposition}.  A similar calculation for
$\hat{T}_y$ shows that $\sum^\prime$ is again invariant, but ${\cal
  O}_s$ carries transverse momentum $\pi$.  The resulting net momenta
of the fundamental solitons are summarized in the last two columns of
Table. 1.

Similar manipulations hold for the kink excitations.  In particular,
the even kink creation operators must obey
\begin{equation}
  {\cal O}^{e\pm}_a =  \left.\sum_{\{ k_a \}}\right.^\prime
  \lambda_{\{ k_a \}} 
  e^{i \sum_a k_a \theta_a } \times e^{\mp i(\varphi_a^e +
    \theta_a^e)/2},
  \label{even_create}
\end{equation}
and similarly for the odd kinks,
\begin{equation}
  {\cal O}^{o\pm}_a =  \left.\sum_{\{ k_a \}}\right.^\prime
  \lambda_{\{ k_a \}} 
  e^{i \sum_a k_a \theta_a } \times e^{\mp i(\varphi_a^o -
    \theta_a^o)/2}.
  \label{odd_create}
\end{equation}
Note that the choice to factor out the right-moving even and GN
fermions in Eqs.~\ref{GN_create},\ref{even_create}\ and left-moving
odd fermions in Eq.~\ref{odd_create}\ is arbitrary.  A right-mover can
always be converted to a left-mover and vice-cersa by absorption of an
$e^{\pm i \theta}$ factor into a redefinition of the ``string'' part
of the soliton creation operator.  For the even and odd kinks, the
above choice is particularly convenient, since the right-moving even
fields and left-moving odd fields are {\sl exactly} bare electron
operators, and hence manifestly physical.  We see from
Eqs.~\ref{even_create}--\ref{odd_create}\ that the kinks have the
quantum numbers of bare electrons.  A remarkable consequence of this
result is that the number of single-electron degrees of freedom has
effectively {\sl doubled} relative to the free fermi gas, since each
of the 16 kinks can be created with arbitrary momentum (relative to
its base momentum of $\pm k_{{\scriptscriptstyle F}i}$), 
including particles {\sl below}
the former fermi sea and holes {\sl above} it.  The momenta calculated
from Eqs.~\ref{even_create}--\ref{odd_create}\ complete the last two
columns of Table 1.

\vskip 0.1in
  \begin{tabular*}{3.2in}{@{\extracolsep{\fill}}c|cccc}
    \hline\hline 
    $(N_1,N_2,N_3,N_4)$ & $Q$ & $S^z$ & $P_x$ & $P_y$ \\
    \tableline
    $(1, 0,0,0)$ & $2$ & $0$ & $0$ & $0$ \\
    $(0,1,0,0)$ & $0$ & $1$ & $\pi$ & $\pi$ \\
    $(0,0,1,0)$ & $0$ & $0$ & $\pi$ & $\pi$ \\
    $(0,0,0,1)$ & $0$ & $0$ & $2 k_{{\scriptscriptstyle F}1}$ & $0$ \\\hline
    $(1,1,1,1)/2$ & $1$ & ${1 \over 2}$ & $k_{{\scriptscriptstyle F}1}$ & $\pi$
    \\
    $(1,-1,-1,1)/2$ & $1$ & $-{1 \over 2}$ & $k_{{\scriptscriptstyle F}1}$ & $\pi$
    \\ 
    $(1,1,-1,-1)/2$ & $1$ & ${1 \over 2}$ & $k_{{\scriptscriptstyle F}2}$ & $0$
    \\ 
    $(1,-1,1,-1)/2$ & $1$ & $-{1 \over 2}$ & $k_{{\scriptscriptstyle F}2}$ & $0$
    \\\hline 
    $(1,1,1,-1)/2$ & $1$ & ${1 \over 2}$ & $-k_{{\scriptscriptstyle F}1}$ & $\pi$
    \\
    $(1,-1,-1,-1)/2$ & $1$ & $-{1 \over 2}$ & $-k_{{\scriptscriptstyle F}1}$ & 
$\pi$
    \\ 
    $(1,1,-1,1)/2$ & $1$ & ${1 \over 2}$ & $-k_{{\scriptscriptstyle F}2}$ & $0$
    \\ 
    $(1,-1,1,1)/2$ & $1$ & $-{1 \over 2}$ & $-k_{{\scriptscriptstyle F}2}$ & $0$ 
\\
    \hline\hline
  \end{tabular*}
  \\ \vskip 0.1in {\noindent Table 1: Physical quantum numbers of the
    mass $m$ particles labelled by their four $U(1)$ charges.  The
    antiparticles are obtained by changing the sign of all the quantum
    numbers.}
  \label{tab:ffqns}

\subsection{$SU(2)$ invariance and spin multiplets}

We conclude this section with a remark on $SU(2)$ invariance and the
excitations with spin.  In Table 1, we have classified the mass $m$
excitations in the D-Mott phase by the four $U(1)$ charges ($SO(8)$
Cartan generators) $N_a$.  Another almost equivalent choice is to
label the particles by their charge, momentum, {\sl total spin} $S^2 =
s(s+1)$ and spin projection $S^z$.  It is fairly trivial to relabel
the kinks in this way, reflecting their natural correspondence with
single-particle spin-$1/2$ excitations.  They may be grouped into
doublets with $s=1/2$ and $S^z = \pm 1/2$, e.g. the $(1,1,1,1)/2$ and
$(1,-1,-1,1)/2$ kinks form a spin-$1/2$ doublet with $Q=1 e$ and
$\bbox{P} = (k_{{\scriptscriptstyle F}1},\pi)$.  For the GN fermions, 
this is less trivial.
The $(\pm 1,0,0,0)$ solitons are spin zero, and correspond to charge
$\pm 2e$ Cooper pairs with zero momentum.  Similarly, the $(0,0,0,\pm
1)$ fermions carry neither charge nor spin, and may be regarded as
dressed particle-hole pairs carrying only momentum.  The remaining
four solitons are more nontrivial, however.  Their spin content may be
brought out by 
re-fermionizing the total spin operator, as in Eq.~\ref{totalspin},
\begin{equation}
  \bbox{S} = \int \! dx \, [ \bbox{J}_{\scriptscriptstyle R}(x) +
  \bbox{J}_{\scriptscriptstyle L}(x) ],
\end{equation}
with chiral currents,
\begin{equation}
  J^A_{\scriptscriptstyle P}(x)  = i \epsilon^{ABC}
  \eta_{\scriptscriptstyle P B}\eta_{\scriptscriptstyle P C} ,
\end{equation}
with $A,B,C=3,4,5$.
Thus the three Majorana fields $\eta_{\scriptscriptstyle PA}$, with $A
= 3,4,5$, transform as a triplet of spin $s=1$ operators, i.e.
\begin{equation}
  [S^A, \eta_{\scriptscriptstyle PB}] = -
  i\epsilon^{\scriptscriptstyle ABC} \eta_{\scriptscriptstyle P C} . 
\end{equation}
The other $5$ GN Majorana fields
commute with $\bbox{S}$, and hence represent spin-singlet operators.  

As was shown in the previous subsection, the physical GN excitations
consist not of GN fermions but rather required an attached string
${\cal O}_s$.  Fortunately, the string does not carry any spin, i.e.
\begin{equation}
 {\cal O}_s \bbox{S} {\cal O}_s^\dagger = \bbox{S}.
\end{equation}
Thus the true soliton excitations (GN fermions+strings) satisfy the
same transformation rules with respect to spin as the bare Majorana
fermions.  The four remaining solitons $(0,\pm 1,0,0)$ and $(0,0,\pm
1,0)$, which involve the four Majorana Fermions $\eta_{\scriptscriptstyle A}$
with $A=3,4,5,6$, can 
therefore be decomposed into an $s=1$ triplet and a spin-zero
singlet (both with $Q=0$ and $\bbox{P}=(\pi,\pi)$).  The triplet can
be regarded as a minimum energy magnon, while the singlet can be
grouped with the $(0,0,0,\pm 1$) solitons as another particle-hole
excitation. 

With the $SU(2)$ invariance realized, we can tabulate the particles in
the GN model in a slightly different way, classifying them by their
$SU(2)$ multiplet (i.e. $s=0$, $1/2$ or $1$), charge, and momentum.
To label the particles classified in this way, the abelian
$(N_1,N_2,N_3,N_4)$ notation is no longer convenient, since it does
not respect the $SU(2)$ invariance.  Instead, we can schematically
indicate the 8 particles in the vector multiplet by 
$\eta_{\scriptscriptstyle A}$ and the 28
in the tensor multiplet by $\eta_{\scriptscriptstyle A} 
\eta_{\scriptscriptstyle B}$ (remembering that 
$\eta_{\scriptscriptstyle B}\eta_{\scriptscriptstyle A} =
 -\eta_{\scriptscriptstyle A} \eta_{\scriptscriptstyle B}$).
For convenience, we list the 8 GN fermions
and the 28 mass $\sqrt{3}m$ bound states in this way in Table 2 (we
do not list the remaining  16 particles, since they have the quantum
numbers of electrons and are easily remembered in this way). 

\vskip 0.1in
  \begin{tabular*}{3.2in}{@{\extracolsep{\fill}}l|cccc}
    \hline\hline 
    Label & $Q$ & $s$ & $P_x$ & $P_y$ \\
    \tableline
    $\eta_{\scriptscriptstyle 1}$,$\eta_{\scriptscriptstyle 2}$ & 
$\pm 2$ & $0$ & $0$ & $0$ \\
    $\eta_{\scriptscriptstyle 3}$,$\eta_{\scriptscriptstyle
      4}$,$\eta_{\scriptscriptstyle 5}$ & $0$ & $1$ & $\pi$ & $\pi$ 
\\ 
    $\eta_{\scriptscriptstyle 6}$ & $0$ & $0$ & $\pi$ & $\pi$ \\
    $\eta_{\scriptscriptstyle 7}$,$\eta_{\scriptscriptstyle 8}$ & $0$ 
& $0$ & $\pm 2k_{{\scriptscriptstyle F}1}$ & $0$ \\
    \tableline
    $\eta_{\scriptscriptstyle 1}\eta_{\scriptscriptstyle 2}$ & $0$ & 
$0$ & $0$ & $0$ \\
    $\eta_{\scriptscriptstyle 7}\eta_{\scriptscriptstyle 8}$ & $0$ & 
$0$ & $0$ & $0$ \\
    $\eta_{\scriptscriptstyle A}\eta_{\scriptscriptstyle B}$, $A=1,2$;
    $B=7,8$ & $\pm 
    2$ & $0$ & $\pm 2k_{{\scriptscriptstyle F}1}$ & $0$ \\ 
    $\eta_{\scriptscriptstyle A} \eta_{\scriptscriptstyle 6}$, 
$A=1,2$ & $\pm 2$ & $0$ &
    $\pi$ & $\pi$ \\ 
    $\eta_{\scriptscriptstyle 6} \eta_{\scriptscriptstyle A}$, 
$A=7,8$ & $0$ & $0$ &
    $\pm(k_{{\scriptscriptstyle F}1}-k_{{\scriptscriptstyle F}2})$ 
    & $\pi$ \\ 
    $\eta_{\scriptscriptstyle A} \eta_{\scriptscriptstyle B}$, $A \neq
    B = 3,4,5$ & $0$ & $1$ & $0$ 
    & $0$ \\
    $\eta_{\scriptscriptstyle A} \eta_{\scriptscriptstyle 6}$, 
$A=3,4,5$ & $0$ & $1$ & $0$
    & $0$ \\
    $\eta_{\scriptscriptstyle A} \eta_{\scriptscriptstyle B}$,
    $A=1,2$; $B=3,4,5$ & $\pm 2$ & $1$ & $\pi$ & 
    $\pi$ \\
    $\eta_{\scriptscriptstyle A} \eta_{\scriptscriptstyle B}$,
    $A=3,4,5$; $B=7,8$ & $0$ & $1$ & 
$\pm(k_{{\scriptscriptstyle F}1}-k_{{\scriptscriptstyle F}2})$ & 
    $\pi$ \\
    \hline\hline
  \end{tabular*}
  \\ \vskip 0.1in {\noindent Table 2: Physical quantum numbers of the
    mass $m$ (above horizontal line) and mass $\sqrt{3}m$ (below
    horizontal line) particles.  }
  \label{tab:su2qns}
  \vskip 0.1in

\section{Correlation functions and physical properties}
\label{sec:correlators}

We have seen that in the weak-coupling limit, the two-leg ladder
possesses an enhanced symmetry.  The effective theory in this limit is
the $SO(8)$ GN model, which is both exactly integrable and exhibits a
remarkable ``triality''.  In this section we will discuss a variety of the
resulting physical consequences.

The most remarkable consequence of triality is the equality of the
single-particle and two-particle gaps.\cite{Shankar80,Shankar81}\
The 16 kinks have the same quantum
numbers as the bare electrons at the former Fermi surface.  The single-particle
gap, defined as the minimum energy needed to add an electron or hole
to the system, is thus simply $\Delta_1 = m$.  The 8 GN fermions,
however, have charge $Q=\pm 2$ or $Q=0$, and thus represent
excitations corresponding to an even number of electrons and/or holes.
For instance, electron or hole pairs can be added with zero net
momentum via the $(\pm 1,0,0,0)$ solitons, while spin-1 excitations
may be added with momentum $(\pi,\pi)$ via the $(0,\pm 1,0,0)$ and
$(0,0,\pm 1,0)$ solitons (more precisely the $\eta_5$ state).
The gap for all 8 minimal energy 2-particle excitations is also
$\Delta_2 = m$.

The equality of the single-particle and two-particle gaps is in marked
contrast to the behavior of other more conventional insulators.  In a
band insulator (such as the two-leg ladder at half-filling
with $t_\perp \gg t$), the
single-particle gap is just the band gap, while the two-particle gaps
are twice as large: $\Delta_2 = 2 \Delta_1$.  Another familiar case is
the strong-interaction limit $U \gg t$.  In this case, the single
particle gap is huge $\Delta_1 \sim U$, while the lowest two-particle
(e.g. spin) gaps are much smaller $\Delta_2 \sim t^2/U \ll \Delta_1$
or indeed vanishing ($\Delta_2 = 0$) for ordered or quasi-long-range
ordered antiferromagnets (e.g. $d=2$ or single-chain Hubbard models).
The detailed mathematical mapping between the GN, odd, and even
fermion fields allows us to extend the relationship between the
single-particle and two-particle properties beyond the values of the
gaps, as we detail below.

First, we will discuss several correlation functions which
characterize the spin and charge dynamics of the system.  The most
interesting of these are the single-particle spectral function,
measurable by electron photo-emission, and the dynamic spin structure
factor, which is probed by inelastic neutron scattering.  Other interesting
correlators include the current-current correlation function, which
determines the conductivity, the density-density correlation function
and pairing correlation function, which can be measured in
numerical simulations.

\subsection{Single-particle spectral function}

First consider the single-particle Greens function,
\begin{equation}
  G_1(\bbox{k},\tau) = \sum_{\ell x} e^{-i k_y \ell -ik_x x} T_\tau
  \langle 0| 
  a_{1+\ell \alpha}^{\vphantom\dagger}(x,\tau) a_{1\alpha}^\dagger 
(0,0)
  |0\rangle, 
\end{equation}
where $T_\tau$ is the (Euclidean) time-ordering symbol and $\bbox{k} =
(k_x,k_y)$.  The right-hand side is independent of $\alpha$ by
spin-rotational invariance; however, we choose to define the spectral
function for fixed $\alpha$, i.e. no sum is implied above.  In
general, the single-particle spectral functions can be extracted from
$G_1$ by Fourier transformation.  Defining $G_1(\bbox{k},i\omega) =
\int d\tau\, G_1(\bbox{k},\tau) \exp( i\omega\tau)$, one finds
\begin{equation}
  {1 \over \pi} {\rm Im} G_1(\bbox{k},i\omega \rightarrow 
\omega+i\delta) \rightarrow
  A_{1p}(\bbox{k},\omega) + A_{1h}(-\bbox{k},-\omega),
\end{equation}
where the particle and hole spectral functions are
\begin{eqnarray}
  A_{1p}(\bbox{k},\omega) & = & \sum_n \left| \langle n |
    a_{1\alpha}^\dagger 
    |0\rangle\right|^2  \delta(\bbox{k}-\bbox{k}_n) 
  \delta(\omega-E_n), \label{psf}\\
  A_{1h}(\bbox{k},\omega) & = & \sum_n \left| \langle n |
    a_{1\alpha}^{\vphantom\dagger}
    |0\rangle\right|^2 \delta(\bbox{k}-\bbox{k}_n) 
   \delta(\omega-E_n). \label{hsf}
\end{eqnarray}
Here we have abbreviated $\delta(\bbox{k}) \equiv
2\pi\delta(k_x) \delta_{k_y,0}$.  The task is then to
evaluate $G_1(\bbox{k},\tau)$.  
In the weak-coupling limit studied here, this
is obtained in two stages.  We first integrate out the electron fields
$c_{i\alpha}^\dagger(k),c_{i\alpha}^{\vphantom\dagger}(k)$, for
$|k-k_{{\scriptscriptstyle F}i}| > \Lambda$, 
which can be accomplished perturbatively in
the interactions, since the energy denominators are finite away from
the Fermi momenta.  The perturbative corrections to the free-electron
$G_1(k,\tau)$ are therefore small in these regions.  Within the
cut-off region of width $2\Lambda$, we must employ the full RG
treatment.  The RG scales the problem onto the GN model, which thus
applies at the lowest energies.  

For the electron spectral function, the non-interacting result 
$A_{1p/h}^0$ is
trivial, since single electron states are exact eigenexcitations.  The
result is 
\begin{eqnarray}
  A_{1p}^0(\bbox{k},\omega) & = & \delta(\omega- \varepsilon_1({\bf 
k}))
  \theta(\omega), \label{nintp}\\ 
  A_{1h}^0(\bbox{k},\omega) & = & \delta(\omega+\varepsilon_1({\bf 
k}))
  \theta(\omega), \label{ninth}
\end{eqnarray}
where $\varepsilon_1({\bf k}) = -t\cos k_x - {t_\perp \over 2} \cos
k_y$.  Interactions of course modify this form somewhat, leading to
some spectral weight away from $\omega = \varepsilon_1({\bf k})$, and
a broadening of the delta-function peak in $A_{1p/h}$ for some
momenta.  In weak-coupling, away from small $\omega$, however, both
effects are small. We will return to them after we consider the
behavior of the spectral function for small frequencies.

The low-frequency limit of $A_{1p}$ is dominated by momentum near the
Fermi points.  Transforming to the slowly-varying Luttinger
fields, we have
\begin{equation}
  G_1(Pk_{{\scriptscriptstyle F}i} + q, k_{yi}; \tau) \approx \int \! dx \,
  e^{-i q x} \langle c_{{\scriptscriptstyle P}i\alpha}^{}(x,\tau)
  c_{{\scriptscriptstyle P}i\alpha}^\dagger(0,0) \rangle,
  \label{G1GNform}
\end{equation}
for $q \ll 1$, where $k_{yi} = (2-i)\pi$.  
Unfortunately,
integrability does {\sl not} give exact forms for the time-dependent
correlation functions in Eq.~\ref{G1GNform}.  A considerable amount
can be learned, however, from the exact excitation spectrum and from
approximate methods.   

The spectrum determines the {\sl support} of $A_{1p/h}({\bf
  k},\omega)$.  From Eq.~\ref{psf}, it is clear that $A_{1p/h}({\bf
  k},\omega)$ is non-zero only when there exists an excitation (or
more than one) with momentum ${\bf k}$, spin $s=1/2$, charge $\pm e$
and energy $\omega$.  From Table 1, we see that the sixteen kinks have
exactly the appropriate quantum numbers for all possible momenta near
the four (i.e. two pairs) fermi points with either $S^z = \pm 1/2$ and
charge $\pm e$.  We therefore expect that $A_{1p/h}({\bf k},\omega)$
first becomes non-zero for $\omega = \sqrt{(k_x-k_{{\scriptscriptstyle
      F}i})^2 + (k_y - k_{yi})^2 + m^2}$.  Since the kinks are
isolated particles with a fixed energy-momentum relation, these
excitations give a sharp delta-function peak in $A_{p/h}$.  It is
natural to identify this peak as the continuation of the
non-interacting delta-function in Eqs.~\ref{nintp}-\ref{ninth}\ to the
region near the Fermi points.  At higher energies other states should
contribute to the spectral weight.  A quick consideration of the
quantum numbers is sufficient to conclude that none of the mass
$\sqrt{3}m$ bound states have the appropriate quantum numbers (e.g.
all have charge zero or $\pm 2e$).  The next lowest-lying excitations
with the quantum numbers of individual electrons are in fact
scattering (unbound) states of a kink and a GN fermion.  For instance,
a $(1,1,1,-1)/2$ kink and a $(0,0,0,1)$ fermion can form a scattering
state with the quantum numbers of a spin up electron with momentum
$(k_{{\scriptscriptstyle F}1}+q,\pi)$, with $q \ll 1$.  Similarly,
other combinations of kinks with the $(0,0,0,\pm 1)$ GN fermions
contribute to the single-particle spectral weight at $(\pm
3k_{{\scriptscriptstyle F}1} + q,\pi)$, $(\pm k_{{\scriptscriptstyle
    F}2}+q,0)$, and $(\pm 3k_{{\scriptscriptstyle F}2}+q,0)$.  All
these form continua with $\omega > \epsilon_c(q) = 2\sqrt{m^2 +
  (q/2)^2}$ at each momenta, since the energy at a particular momenta
can always be increased continuously by shifting the kink and the
bound-state momenta in opposite directions.  Further excitation of
more than one $(0,0,0,\pm 1)$ quanta leads to spectral weight at all
momenta separated by an even multiple of $2k_{{\scriptscriptstyle
    F}1}$, i.e.  ${\bf k} = ((2n+1)k_{{\scriptscriptstyle F}1},\pi)$.
The excitation gap for such a point increases, however, by an
additional factor of $m$ as each GN fermion (i.e. factor of
$(2k_{{\scriptscriptstyle F}1},0)$ away from the Fermi points) is
added.  Furthermore, the higher harmonic contributions to the spectral
function are expected to have small amplitudes, as they involve
multiple scatterings of the original injected electron (see below).

\begin{figure}[hbt]
\epsfxsize=7cm
\centerline{\epsfbox{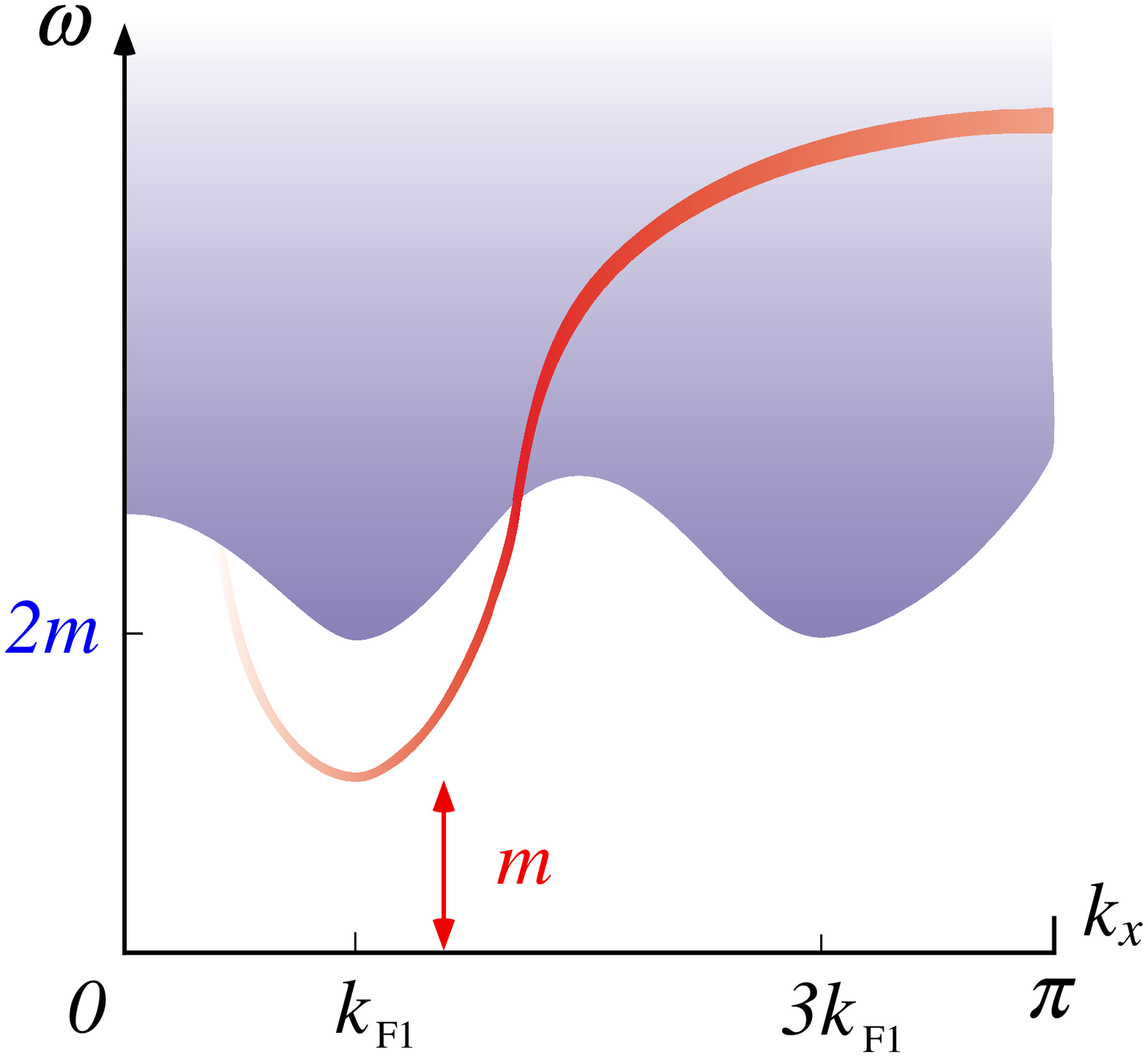}} \vspace{5pt} {\noindent FIG.
  5: Schematic plot of the single-particle electron spectral function
  $A_{1p}(k_x, k_y=\pi,\omega)$. The curve below the continuum
  indicates a sharp resonance, i.e. a delta-function peak in $A_{1p}$,
  which accquires a finite width once it passes inside the continuum
  due to the onset of decay processes. The continuum above energy $2m$
  coincides with the creation of (unbound) scattering states of a
  single kink and a GN fermion.}
\end{figure}

To understand the magnitude of $A_{1p/h}({\bf k},\omega)$ in the
allowed regions requires a knowledge of the matrix elements in
Eq.~\ref{psf}, or of the full Green's function in Eq.~\ref{G1GNform}.
Since exact results are unavailable for these quantities, we consider
instead the mean-field approximation.  Without loss of generality, let
us consider momenta near the particular Fermi point ${\bf k} \approx
(k_{{\scriptscriptstyle F}1},\pi)$ and spin $S^z = +1/2$.  
Using Eq.~\ref{G1GNform}\ and
Eqs.~\ref{kink1}-\ref{kink4}, 
the bare electron operators can be rewritten
exactly in terms of even Fermion fields,
\begin{equation}
  \langle c_{{\scriptscriptstyle R}1\alpha}^{}(x,\tau)
  c_{{\scriptscriptstyle R}1\alpha}^\dagger(0,0) \rangle =
  \langle \psi_{{\scriptscriptstyle R}1}^e(x,\tau)
  \psi_{{\scriptscriptstyle R}1}^{e\dagger}(0,0)\rangle.
\end{equation}
In the mean-field approximation, the exact eigenstates are created by
the rotated operators for particles 
$\tilde\psi_{{\scriptscriptstyle R}a}^{e\dagger}(k)$
and holes 
$\tilde\psi_{{\scriptscriptstyle L}a}^{e}(k)$ (see Sec.~IV.D.2), so that the
above expectation value can be computed by a simple rotation.  One
finds 
\begin{eqnarray}
  A_{1p}^{\rm MF}&&(k_{{\scriptscriptstyle F}1}+q,\pi,\omega)  = 
\nonumber \\
  & & {1 \over 2} \left( 1 + {q
      \over \sqrt{q^2+m^2}} \right) \delta(\omega - \sqrt{q^2+m^2}).
  \label{MFsf}
\end{eqnarray}
The $q$--dependent factor out front arises from the rotation to the
$\tilde\psi$ spinor, and is analogous to a ``coherence factor'' in
superconductivity.  Eq.~\ref{MFsf}\ captures a simple and appealing
physical effect.  Although single-kink states exists for all momenta,
their contribution to the spectral weight is suppressed for $q \ll -m$
by the ``coherence factor'' above.  Such a negative momentum
corresponds to the addition of an electron at a momentum {\sl inside}
the Fermi sea.  Interactions deplete the Fermi sea slightly near the
Fermi points, allowing electrons to be added, but with a weight that
vanishes as $q\rightarrow -\infty$, i.e.  deep within the sea.
Similarly, the hole spectral function has weight {\sl outside} the
Fermi sea, since some excited particles exist as part of the ground
state.  Unfortunately, the mean-field approximation is not
sophisticated enough to capture the continuum at $\omega >
\epsilon_c(q)$, since the $\tilde\psi$ fields create exact eigenstates
in this limit.  Thus the continuum is not present in Eq.~\ref{MFsf}.
On physical grounds, however, we expect that it will be similarly
suppressed for momenta inside the former Fermi sea.

Having determined the behavior of $A_{1p/h}$ near the Fermi points and
momenta separated from them by even multiples of 
$(2k_{{\scriptscriptstyle F}1},0)$, we
return to the question of the spectral weight away from these points,
at energies away from the non-interacting single-particle energy
$\varepsilon_1({\bf k})$.  In the non-interacting system, states with
one added electron {\sl plus} additional neutral electron-hole pairs
in fact form a continuum away from the single-particle energy.
Consider, for instance, $k_y = \pi$.  The lowest energy continuum of
states with a single added particle-hole pair (plus one electron)
consists of those states in which both electrons are infinitesimally
above the Fermi surface ($k_1 = k_2 = k_{{\scriptscriptstyle F}1}$)
 and the hole makes up
the missing momentum ($k_3 = 2k_{{\scriptscriptstyle F}1}-k$). 
This begins at
$\varepsilon_2(k_x,\pi) = 
\cos(2k_{{\scriptscriptstyle F}1}-k_x) - t_\perp/2$, which is
{\sl below} the single-particle energy (i.e. $\varepsilon_2 <
\varepsilon_1$) for $k_{{\scriptscriptstyle F}1} 
< k_x <3k_{{\scriptscriptstyle F}1}$, crossing zero again at
$k_x=3k_{{\scriptscriptstyle F}1}$.  It does not contribute to 
the single-particle spectral
function, however, due to orthogonality.  In an interacting Fermi
liquid, we would expect that an added electron can scatter into these
states (i.e. emit a low energy particle-hole pair), and some weight
would appear in $A_{1p/h}$ associated with the continuum states.

In the weakly-interacting ladder, an added electron away from the
Fermi points can also scatter to create neutral excitations, and some
weight will appear in the regions near the non-interacting continuum
for $\omega \gtrsim \varepsilon_2({\bf k})$.  This continuum away from
the Fermi points should merge smoothly into the continuum above
$\omega = 2m$ at the Fermi points (and the higher harmonics, e.g.
$(\pm 3k_{{\scriptscriptstyle F}1},\pi)$).  
Clearly, since the single-particle energy begins
{\sl below } the continuum near the Fermi momenta and it is above the
continuum far away, it must cross into the continuum at some point.
Where it is below the continuum, we expect that the spectral function
retains a sharp delta-function peak.  Once it passes above, however,
the single-particle mode can decay into the continuum states, and
should acquire a small width.  Putting this behavior together with the
spectral function near the Fermi points and higher harmonics, we
arrive at the schematic single-particle spectral function illustrated
in Fig.~5.  The most dramatic feature is the sharp delta-function
peak near the Fermi points which crosses into the continuum and
acquires a width at higher energies.

\subsection{Spin structure factor}

The spin spectral function can be defined in a similar way to the
single-particle one.  Consider the structure function
\begin{eqnarray}
  {\cal S}^{ij}({\bf k},i\omega) & = & {1 \over 2} \sum_{\ell\ell'
    x}\int\!d\tau\, e^{-ik_x 
    x-ik_y (\ell-\ell') + i\omega\tau} \nonumber \\
  & & \times T_\tau \langle 0| S^i_\ell(x,\tau)
  S^j_{\ell'}(0,0)|0\rangle,
\end{eqnarray}
where the lattice spin operator is ${\bf S}_\ell(x) =
a_{\ell\alpha}^\dagger {\bbox{\sigma}_{\alpha\beta} \over 2}
a_{\ell\beta}^{\vphantom\dagger}$.  The spin spectral function $A_s$ 
is obtained from this in the usual way
\begin{equation}
  {1 \over \pi}{\rm Im} \, {\cal S}^{ij}(\bbox{k},i\omega \rightarrow
  \omega+i\delta) 
  \rightarrow 
  A_s(\bbox{k},\omega)\delta^{ij}, \qquad {\rm for}\;\; \omega>0.
\end{equation}
The spectral decomposition of $A_s$ is
\begin{equation}
  A_s({\bf k},\omega) = {1 \over 3} \sum_n \left| \langle 0| S^i
  |n\rangle \right|^{2} \delta({\bf k} - {\bf k}_n)
  \delta(\omega - E_n).
\end{equation}

As for the single-particle case, we expect that the spin spectral
function will be approximately equal to its non-interacting value for
$\omega \gg m$.  A straightforward calculation for the non-interacting problem gives
\begin{eqnarray}
  A_s^0&&({\bf k},\omega)  = {1 \over {8\pi}} \sum_{a,b = \pm 1}
  {\Theta(1-|r|) \over {\sin {k_x \over 2} \sqrt{1-r^2}}} \nonumber \\
  && \hspace{-0.8cm} \times  \Bigg\{\Theta\left[b\cos {k_x \over 2}
    \sqrt{1-r^2} + r 
    \sin {k_x \over 2} + 
    a t_\perp/2 \right] \nonumber \\
  && \hspace{-0.8cm} -\Theta\left[b\cos {k_x \over 2} \sqrt{1-r^2}
    - r \sin {k_x \over 2} + 
    a {t_\perp \over 2}\cos k_y  \right]\bigg\},
  \label{nonintS}
\end{eqnarray}
where $r = (\omega + a{t_\perp \over 2}(\cos k_y -1) )/(2\sin {k_x
  \over 2})$.  Clearly, the non-interacting spin spectral function is
considerably more complex than its single-particle counterpart.  This
is because the neutral spin-one excitations in the non-interacting
system are particle-hole pairs, and thus comprise a continuous
spectrum.  The result in Eq.~\ref{nonintS}\ is plotted in
Fig. 6.  

\begin{figure}[hbt]
\epsfxsize=7cm
\centerline{\epsfbox{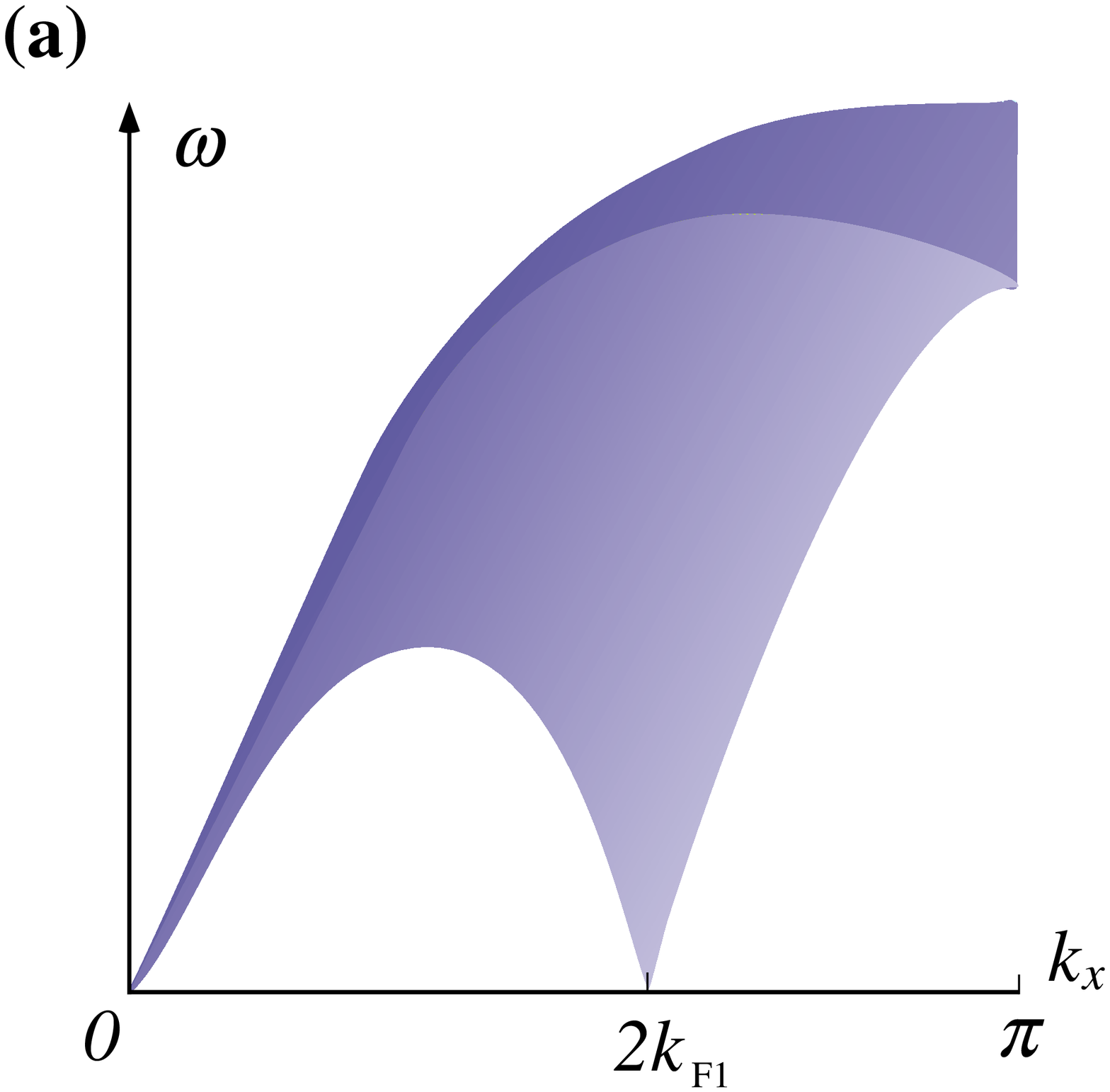}}
\epsfxsize=7cm
\centerline{\epsfbox{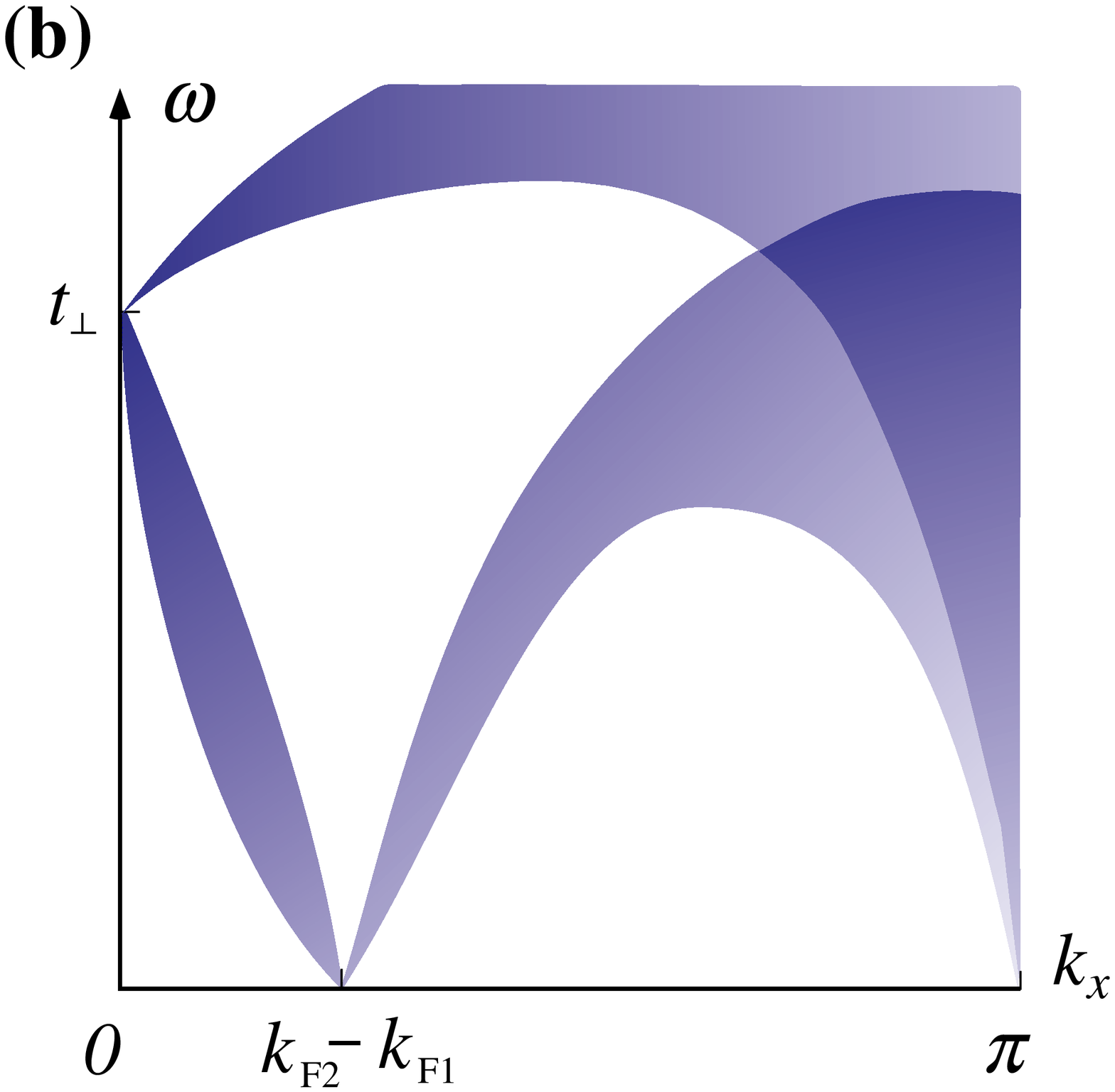}}
\vspace{5pt}
{\noindent FIG. 6: Intensity plot of the non-interacting spin
spectral function at (a) $k_y=0$, (b) $k_y=\pi$.  The darkness
is proportional to the spectral weight, white indicating
regions of phase space in which no particle-hole pairs exist. }
\end{figure}

For $k_y=0$, the particle
and hole must come from the same band.  In this case low energy
excitations exist near $k_x=0$, when both are taken near the same
Fermi point, and near $k_x = \pm 2k_{{\scriptscriptstyle F}1}$ 
($k_x = \pm 2k_{{\scriptscriptstyle F}2}$ are
equivalent to these values modulo $2\pi$), when the particle and hole
are taken from opposite Fermi points (still in the same band). Two
branches of continua (from the bonding and anti-bonding bands) form as
the momenta of the particles and holes are varied.  For $k_y=\pi$, the
particle and hole must come from opposite bands.  In this case, the
$k_x=0$ states have an energy of exactly $t_\perp$, since this
requires a vertical transition.  Low energy excitations exist near
$k_x = \pm (k_{{\scriptscriptstyle F}2}-k_{{\scriptscriptstyle F}1})$, 
due to particle-hole pairs taken from two
right-moving or two left-moving Fermi points.  They also exist near
$k_x = \pi$, when the particle and hole are taken from opposite sides
of the Fermi surface.  Extending these two branches of excitations
gives the form shown in Fig. 6(b).

\begin{figure}[hbt]
\epsfxsize=7cm
\centerline{\epsfbox{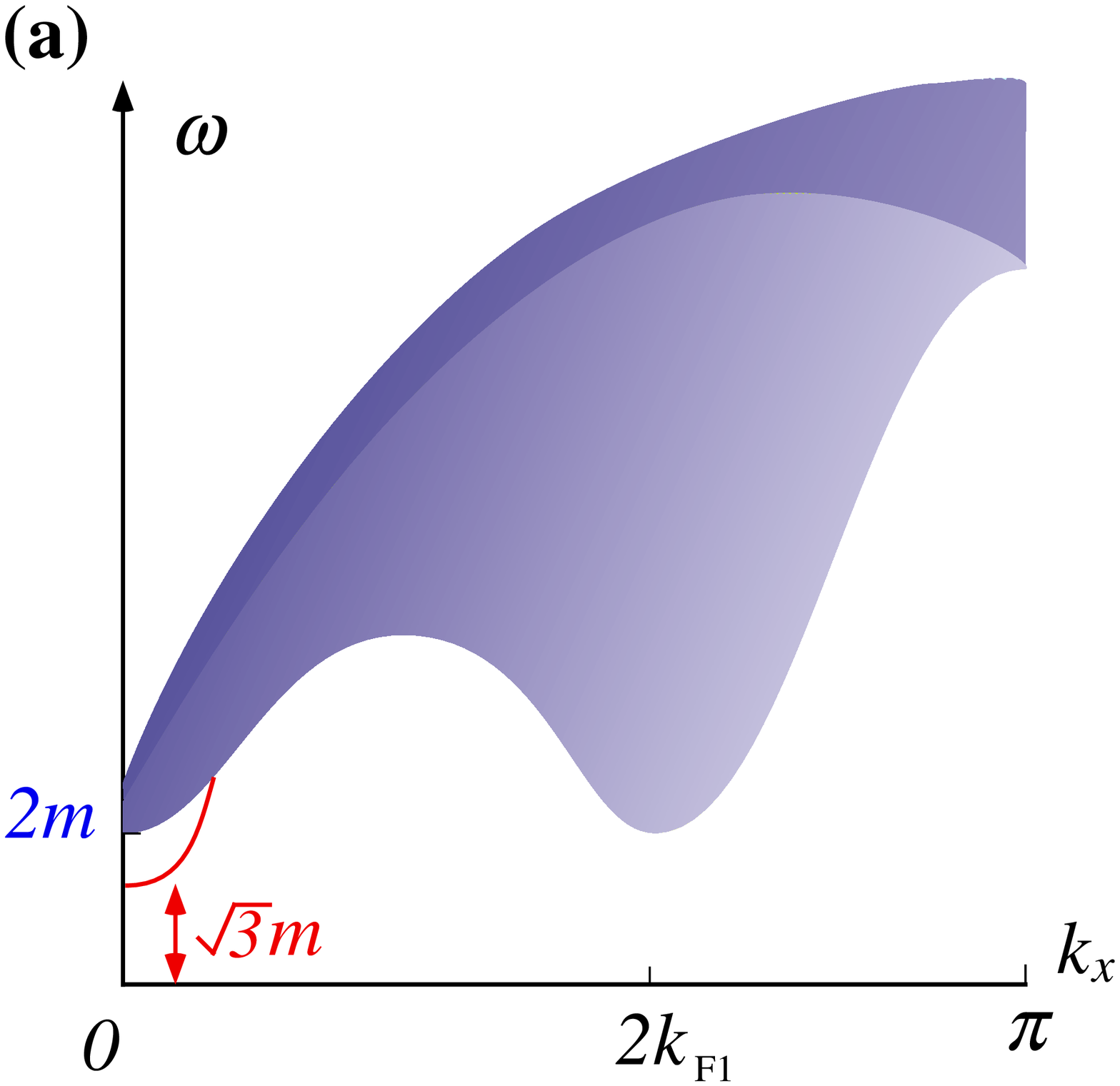}}
\epsfxsize=7cm
\centerline{\epsfbox{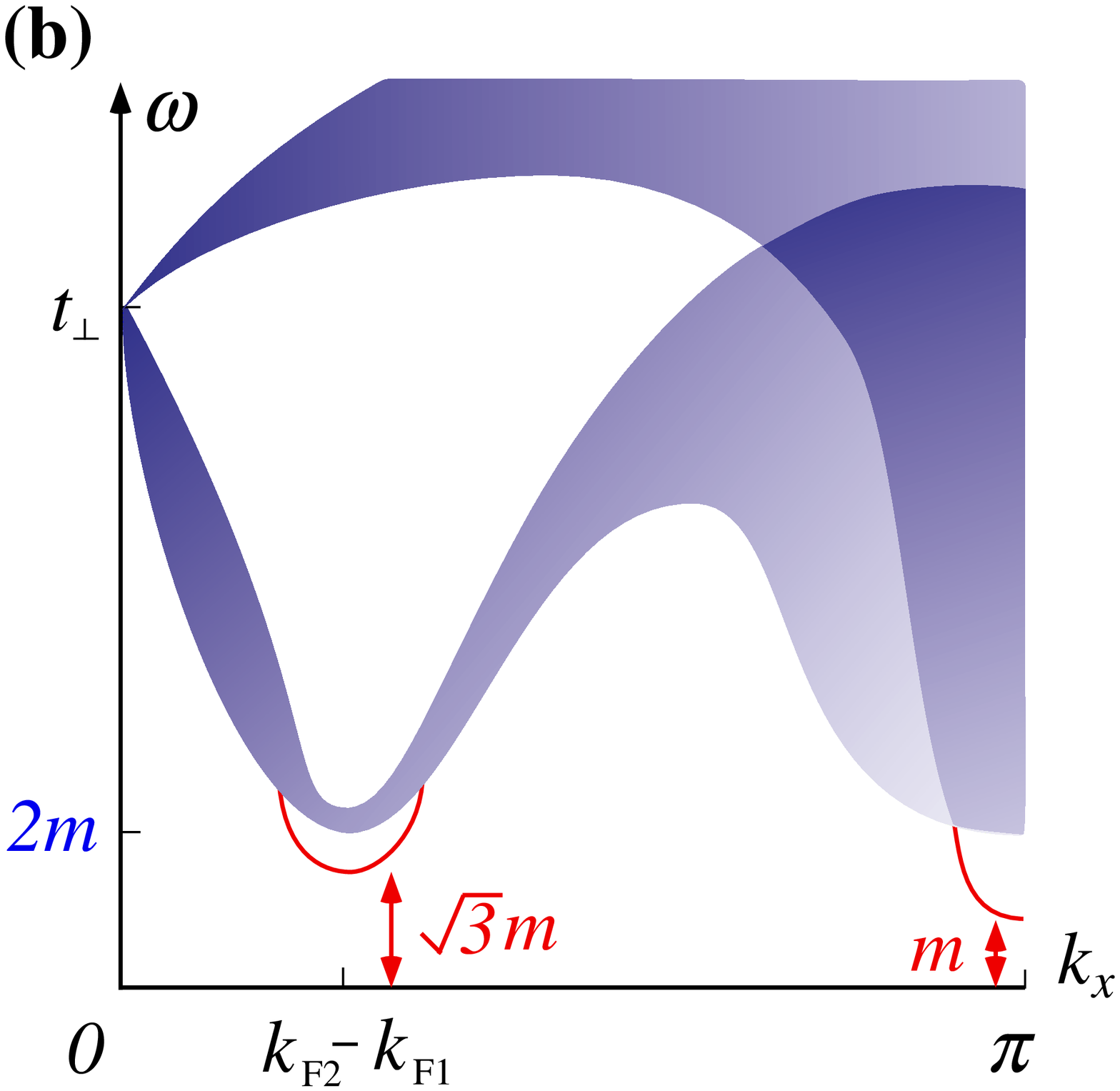}}
\vspace{5pt}
{\noindent FIG.  7: Intensity plot of the interacting spin
spectral function at (a) $k_y=0$, (b) $k_y=\pi$ in the presence of 
interactions. In the low-energy portion, various gaps develop and 
excitations with sharp delta-function peaks are present (see
Sec.~V.B).  Note that the minimum energy spin excitations occur at
${\bf k} = (\pi,\pi)$.}
\end{figure}

As for the single-particle spectral function, introducing interactions
allows for additional structure, and the low-lying excitations are
raised up to energies of order $m$.  In particular, from
Table. 2, we see that the lowest lying neutral triplet states
are the mass $m$ GN fermions $\eta_{\scriptscriptstyle 3,4,5}$ at 
${\bf k} =(\pi,\pi)$.  The next highest energy neutral triplets are the mass
$\sqrt{3}m$ bound states.  The $\eta_{\scriptscriptstyle
  3,4,5}\eta_{\scriptscriptstyle 7,8}$ have momenta $(\pm
(k_{{\scriptscriptstyle F}1}-k_{{\scriptscriptstyle F}2}),\pi)$, 
while the $\eta_{\scriptscriptstyle 3,4,5}
\eta_{\scriptscriptstyle 4,5,3}$ and $\eta_{\scriptscriptstyle
  3,4,5}\eta_{\scriptscriptstyle 6}$ have momentum $(0,0)$.  
Above these exist 
continua dispersing like $\epsilon_c(q) = 2\sqrt{m^2 + (q/2)^2}$
away from all the aforementioned points {\sl and} 
$(\pm 2k_{{\scriptscriptstyle F}1},0)$
(the excitations at these last points arise from certain pairs of
unbound kinks). Since there are no sharp resonances (delta-function
peaks) in the non-interacting limit, we expect that the mass $m$ and
$\sqrt{3}m$ peaks must broaden at higher energies to merge into the
continua found there.  A schematic form is shown in Fig. 7.

\subsection{Optical conductivity}

Another quantity of considerable experimental relevance is the optical
conductivity.  We are interested in the real part of the conductivity,
defined by
\begin{equation}
  {\rm Re} \,\sigma(\omega) = {\rm Im} \left[ {\Pi(i\omega \rightarrow
    \omega + i \delta)} \over \omega \right],
\label{optical}
\end{equation}
where the (${\bf k}=\bbox{0}$) current-current correlator is 
\begin{equation}
  \Pi(i\omega) = {1 \over 2}\sum_{\ell\ell'}\int \! dx d\tau
  e^{i\omega\tau} \langle T_\tau J_\ell(x,\tau) J_{\ell'}(0,0)\rangle.
  \label{BigPi}
\end{equation}

The electrical current operator is
\begin{equation}
  J_\ell(x) = {e \over {2i}} \left(
    a_\ell^\dagger(x)a_\ell^{\vphantom\dagger}(x+1) - 
    a_\ell^\dagger(x+1)a_\ell^{\vphantom\dagger}(x) \right).
\end{equation}
To evaluate Eq.~\ref{BigPi}, only the slowly varying (${\bf k = 0}$)
component of the current is needed.  Decomposing the lattice fields
into their continuum components using Eq.~\ref{decompose_electrons}\
and then applying the bosonization and refermionization rules, one
finds the long-wavelength form
\begin{eqnarray}
  J_\ell & \sim & \sin k_{{\scriptscriptstyle F}1} \partial_x \phi_1 \\
  & = & {\sin k_{{\scriptscriptstyle F}1} \over {2\pi}} 
\psi^\dagger_1 \tau^z
  \psi^{\vphantom\dagger}_1 \\ 
  & = & {\sin k_{{\scriptscriptstyle F}1} \over {2\pi}}
\left( G^{\scriptscriptstyle 12}_{\scriptscriptstyle R} -
    G^{\scriptscriptstyle 12}_{\scriptscriptstyle L} \right).
\end{eqnarray}

From this form, the current operator clearly excites the mass 
$\sqrt{3}m$
$\eta_1\eta_2$ bound states, as well as higher energy continuum
scattering states with energies above $2m$.  Since Eq.~\ref{optical}\
is nothing but $1/\omega$ times the spectral function of $J$,  the
zero temperature optical conductivity is thus zero for $\omega <
\sqrt{3}m$, has a sharp (delta-function) peak at $\omega = \sqrt{3}m$,
and a threshold with continuous weight for $\omega > 2m$.  Based on
the mean-field picture, we expect the spectral weight in the
two-particle continuum to have a square root singularity (due to the
van Hove singularity at the bottom of the band -- see,
e.g. Ref.~\onlinecite{Giamarchi97}), i.e. 
\begin{equation}
  \sigma(\omega) \approx A \delta(\omega-\sqrt{3}m) + B \sqrt{m \over
    {\omega-2m}} \Theta(\omega-2m).
\end{equation}

At non-zero temperatures it is difficult to determine $\sigma(\omega)$
from the Kubo formula, Eq.~\ref{optical}.  Instead, the general
features can be argued on more conventional transport grounds,
borrowing heavily from recent results of Damle and
Sachdev\cite{Damle97u}\  for {\sl
  spin} dynamics of gapped 2-leg ladders.  The important physical
effect for $T > 0$ is the presence of a non-zero equilibrium
concentration ($\propto e^{-m/T}$) of activated excitations.  In the
semi-conductor analogy, these are activated particles and holes.  In
principle, all 24 mass $m$ states have identical equilbrium
concentrations; for $T \ll m$ we expect that we can neglect the much
smaller activated densities ($O(e^{-\sqrt{3}m/T})$) of bound states.

In this case the low-frequency conductivity can be estimated using a
simple Drude argument.  We focus on the charged species of mass $m$,
i.e. the $\eta_{1,2}$ fundamental fermions and the kinks.  Each of
these contributes in parallel to the 
conductivity a term of the Drude form, 
\begin{equation}
  \sigma(\omega) = {\sigma_0 \over {1+ i\omega\tau}},
\end{equation}
with $\sigma_0 = n \tau/m$, $\tau$ a scattering time, and $n$ 
the density of thermally excited carriers.  Using the Boltzmann
distribution, we have $n \sim \sqrt{mT} e^{-m/T}$ for $T \ll m$.  The
average separation between particles $l(T) = 1/n$ is thus much larger
than their typical wavelength $\lambda(T) \sim 2\pi/p \sim
1/\sqrt{mT}$, obtained by equipartition.  The particles thus behave
essentially classically except during a collision, when they 
scatter strongly (as known, e.g. from the exact S-matrices for the
$SO(N)$ GN model), and their scattering time is determined simply by
the time between collisions: $\tau \sim l(T)m /p \sim T^{-1}e^{m/T}$.
The exponential dependences in the dc conductivity thus cancel, and
$\sigma_0 \sim 1/\sqrt{mT}$ (the same result is obtained from the
Einstein relation $\sigma_0 = {\partial n \over \partial\mu} D$).  In
principle, the dimensionless numerical prefactor in this relation
could be obtained using the methods of Ref.~\onlinecite{Damle97u}, but
we content ourselves here simply with the scaling form.  Note that
although the height of the Drude peak diverges as $T \rightarrow 0$,
its width shrinks much more rapidly (exponentially), and the weight at
$\omega=0$ is negligible at low temperatures.  

Turning to the
higher-frequency features (for $\omega \approx \sqrt{3}m$ and $\omega
\gtrsim 2m$), we expect that scattering between the {\sl injected} bound
states or particle-hole pairs and the thermally excited carriers
will occur on the same characteristic timescale, $\tau$.  These peaks
therefore also acquire exponentially small widths ($O(1/\tau)$) for $T
\ll \Delta$.  Since the overall spectral weight in ${\rm
  Re}\,\sigma(\omega)$ must be conserved, we expect the heights of these
features to diverge much more strongly than the Drude peak, i.e.
$\sigma(\omega=\sqrt{3}m,T) \sim \tau \sim e^{m/T}$ and $\sigma(\omega
= 2m,T) \sim \sqrt{\tau} \sim e^{m/2T}$.  Indeed, the optical
conductivity presumably satisfies universal scaling forms near these
points, i.e.
\begin{equation}
  {\rm Re}\, \sigma(\omega) \sim \cases{ \tau
    \Sigma_{\scriptscriptstyle 1}
    [(\omega-\sqrt{3}m) \tau], & $|\omega-\sqrt{3}m| \ll m$ \cr
    \sqrt{\tau} \Sigma_{\scriptscriptstyle 2}[(\omega-2m)\tau] &
    $|\omega - 2m| \ll m$ \cr},
\end{equation}
for $T \ll m$, where $\Sigma_{\scriptscriptstyle 1}$ and
$\Sigma_{\scriptscriptstyle 2}$ are universal scaling functions.  A
schematic illustration of the optical conductivity at finite
temperature is given in Fig. 8.

\begin{figure}[hbt]
\epsfxsize=7cm
\centerline{\epsfbox{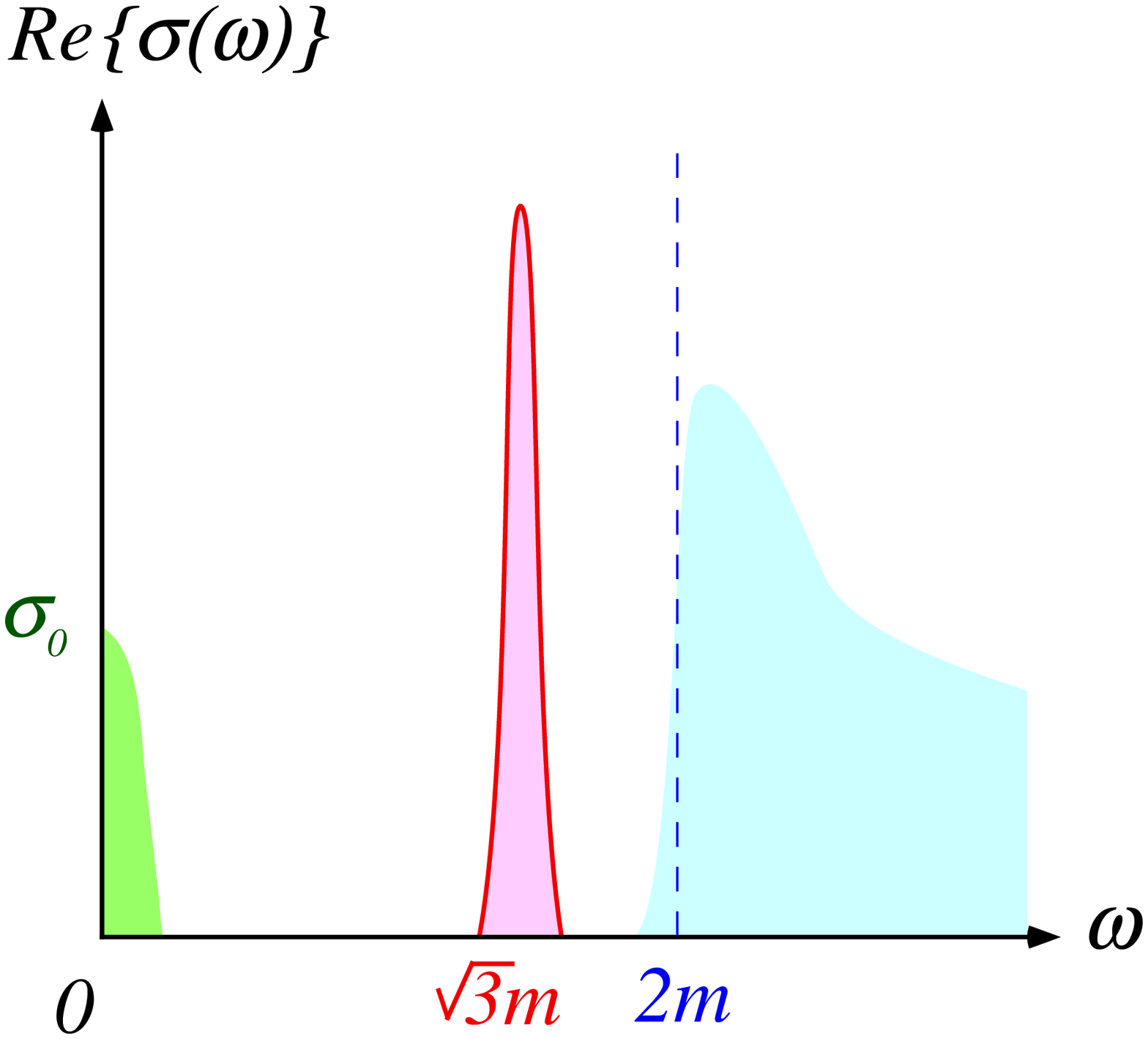}}
\vspace{5pt}
{\noindent FIG. 8: Optical conductivity at finite temperature. At low
  temperatures, all the features become exponentially sharp, with a
  width $\delta\omega \sim e^{-m/T}$.  In this limit, the ``exciton''
  peak at $\omega = \sqrt{3}m$ retains a non-zero weight, sharpening into 
  a delta-function, and the peak near $\omega = 2m$ is also
  exponentially high.  By contrast, the Drude peak at $\omega=0$ has
  vanishing weight at low temperatures, its height diverging only as
  $\sigma_0 \sim 1/\sqrt{mT}$.}
\end{figure}

\subsection{Equal-time spatial correlators}
\label{subsec:correlator}

Numerous other correlators can be measured at equal times in numerical
simulations, and sometimes experimentally (e.g. static structure
factors).  The properties of essentially any such correlator can be
deduced from the GN spectrum, as summarized in Table~2.
Due to the Lorentz invariance of the GN model, intermediate states
with a finite energy $\epsilon$ give rise to exponentially decaying
{\sl spatial} correlation functions with the corresponding length
$\xi_\epsilon = v/\epsilon$.

For completeness, we quote two examples here.  The pair-field
correlator, defined by $\Delta(x) =
a_{1\uparrow}(x)a_{2\downarrow}(x)$ has the correlation function
\begin{eqnarray}
  \langle \Delta^{\vphantom\dagger}(x) \Delta^\dagger(0) \rangle &
  \sim & A_1
     e^{-m|x|/v} + \Bigg[A_2 (-1)^x \nonumber \\
  & & \hspace{-1.0in} + A_3 e^{i2k_{{\scriptscriptstyle F}1}x}
    + A_4 e^{-i2k_{{\scriptscriptstyle F}1}x} \Bigg]
 e^{-\sqrt{3}m|x|/v} + \cdots,
\label{pairpair}
\end{eqnarray}
for $|x| \gg 1$. Here the first term comes from the mass $m$ ``Cooper
pairs'' and the second from the corresponding bound states.
In the prefactors to the exponentials, $A_j$, we have
neglected sub-dominant spatial dependences, which generally will
have power law forms.  For example, due to the one-dimensional
van-Hove singularity for adding a pair above the threshhold energy
$m$, one expects $A_1(x) \sim x^{-1/2}$ for large $x$.
Similarly, the real-space density-density correlation function is
\begin{eqnarray}
  \langle n_\ell(x) n_1(0) \rangle  & \sim & [ B_1 (-1)^{x +
    \ell} + B_2 \cos(2k_{F1}x) ] e^{-m|x|/v}  +  \nonumber\\ 
  &  & \hspace{-1.0in} \Bigg[ B_3 + B_4 (-1)^\ell \cos
    (k_{{\scriptscriptstyle F}1}-k_{{\scriptscriptstyle F}2})x \Bigg] 
  e^{-\sqrt{3}m|x|/v} + \cdots
\end{eqnarray}
Here $n_\ell(x) = a_{\ell\alpha}^\dagger(x)
a_{\ell\alpha}^{\vphantom\dagger}(x) - 1$.  The real-space spin-spin
correlation function has an identical form, except with $B_2=0$.

\section{Generic interactions and $SO(5)$ symmetry}
\label{sec:SO5}

In the previous sections, we focused on the properties of the D-Mott
phase which occurs with generic predominantly repulsive interactions
in the two-leg ladder.  In weak coupling this phase exhibits a
remarkable $SO(8)$ symmetry with dramatic physical consequences for
both two-particle and single-particle properties.  As remarked in the
introduction, there exists an $SO(5)$ subalgebra of the full $SO(8)$
group whose vector representation ``unifies'' superconductivity and
antiferromagnetism.  Thus all the consequences of this $SO(5)$
symmetry, proposed by Zhang as a phenomenological model for the
cuprates, are shared by the D-Mott phase.  A number of authors have
proposed {\sl exactly} $SO(5)$-invariant lattice models including, in
a recent paper by Scalapino, Zhang and Hanke (SZH), a two-leg ladder
model.\cite{Scalapino97u}\  
SZH derived a complex phase diagram for this model in the
strong coupling limit in a space including both repulsive and
attractive interactions.  In this section, we will develop the
necessary technology and study  in weak-coupling both the SZH model
and other generic $SO(5)$-invariant 
two-leg ladder systems.\cite{Shelton97u,Arrigoni97u}\  

In fact, the focus on $SO(5)$-invariant models is less restrictive
than might be naively expected.  Indeed, in our numerical studies of
the full RG equations at half-filling, 
{\sl all} weak coupling two-leg ladder models we
have studied ({\sl including those with attractive interactions })
scale under the RG onto the $SO(5)$-invariant manifold.  {\it Within} this
manifold, we have observed five attractors, including the D-Mott phase
and a Luttinger liquid (C2S2) phase continuously connected to the
non-interacting Fermi liquid, in which all the elementary excitations
remain gapless.  The remaining three attractors are all massive
phases, and
comprise an {\sl S-Mott} phase similar to the D-Mott phase but with
approximate ``s-wave'' pair correlations, a {\sl charge-density-wave}
(CDW) phase with a density-wave at ${\bf k} = (\pi,\pi)$, and a {\sl
spin-Peierls} (SP) phase without a density wave but with kinetic energy
modulated at the same wavevector.  We group the D-Mott with the latter
three to form four {\sl dominant phases}.  We will see that (in weak
coupling) while all of these dominant phases share Zhang's $SO(5)$
symmetry, each one possesses a {\sl distinct} higher $SO(8)$
symmetry.  The different $SO(8)$ symmetries are related in rather
simple ways which have ramifications for the critical points between
the different dominant phases.

\subsection{$SO(5)$ Symmetry}

We begin by reviewing some basic properties of the $SO(5)$ symmetry,
demonstrating in the process the relation to the $SO(8)$ symmetry
already discussed.  The $SO(5)$ algebra was originally designed to
rotate the five-component {\sl vector} containing the real and
imaginary parts of the D-wave pair field and the three components of
the staggered magnetization.  A set of operators which performs this
function was introduced by Zhang\cite{Zhang97}
 -- these are the 10 generators of
$SO(5)$, which are conveniently grouped into the antisymmetric matrix
\begin{equation}
  K_{\scriptscriptstyle AB}= \left[
    \begin{array}{ccccc}
      0& & & & \\
      Q_{p}&0&&&\\
      {\rm Re}\Pi_{x}&-{\rm Im}\Pi_{x}&0& &\\
      {\rm Re}\Pi_{y}&-{\rm Im}\Pi_{y}&S_{z}&0 & \\
      {\rm Re}\Pi_{z}&-{\rm Im}\Pi_{z}&-S_{y}&S_{x} &0 \\
    \end{array}
  \right],
  \label{eq:SO5gens}
\end{equation}
where $A,B =1\ldots 5$ spans the matrix of generators and 
$K_{\scriptscriptstyle AB}= - K_{\scriptscriptstyle BA}$.
The various components are defined as bi-linears in electron
operators,
\begin{eqnarray}
  Q_p &=& \frac12 \sum_{\bbox{k}}  \left( a_\alpha^\dagger(\bbox{k})
    a_\alpha^{}(\bbox{k}) - 1 \right),
  \\
  \bbox{S} &= &\frac12 \sum_{\bbox{k}}a^{\dag}_{\alpha}(\bbox{k})
  \bbox{\sigma}_{\alpha \beta} a^{}_{\beta}(\bbox{k}),
  \\
  \bbox{\Pi} &= &\frac12 \sum_{\bbox{k}} \phi_{\bbox{k}}
  a_{\alpha}(-\bbox{k}+\bbox{N}) 
  (\sigma_{y}\bbox{\sigma})_{\alpha \beta}
  a^{}_{\beta}(\bbox{k}),
\label{pi-op}
\end{eqnarray}
where $\bbox{N} = (\pi,\pi)$ is the ``nesting'' vector and $Q_p$ is
the charge measured in the number of {\sl pairs} of electrons relative
to half-filling.  Here $a_\alpha({\bf k})$ is the Fourier transform of
the lattice electron operator and a spin sum is implicit.  To see that
the $SO(5)$ symmetry is just the first subgroup of $SO(8)$, we need
simply take the continuum limit of Eq.~\ref{eq:SO5gens}.
Straightforward but lengthy decomposition of the lattice electron
fields into their slowly varying components (as in Section II) and
consequent bosonization and refermionization (as in Section IV) gives
the exceedingly simple result
\begin{equation}
  K_{\scriptscriptstyle AB} = \int \! dx\, \left(
    G_{\scriptscriptstyle R}^{\scriptscriptstyle AB} +
    G_{\scriptscriptstyle L}^{\scriptscriptstyle AB} \right),
\label{5-8}
\end{equation}
where the $G_{\scriptscriptstyle P}^{\scriptscriptstyle AB}$
are precisely the $SO(8)$ generators introduced in Section IV.  
Since only the five-dimensional upper-left block of the full matrix
of SO(8) generators enter in Eq.~\ref{5-8}, the
$SO(5)$ symmetry rotates the first
five Majorana fermions $\eta_{\scriptscriptstyle A}, A=1\ldots 5$.
As discussed in Section IV, the first five components
of this vector representation contain both the
pair field ($\sim \eta_{1,2}$) and the staggered 
magnetization
($\sim \eta_{3,4,5}$).  

\subsection{Microscopically $SO(5)$-invariant models and $SO(5)$ 
spinors}

We now turn to a discussion of {\sl microscopically}
$SO(5)$-invariant ladder Hamiltonians.  A particular example is the
SZH model, which is the most general $SO(5)$-symmetric two-leg ladder
Hamiltonian with nearest-neighbor hopping and only intra-rung two-body
interactions.  The interaction terms on each rung of the ladder
take the form
\begin{eqnarray}
  H_{int}&=& U \sum_{\ell}\{(n_{\ell\uparrow}-\frac12)
  (n_{\ell\downarrow}-\frac12)\}
  \nonumber\\
  &&+V(n_{1}-1)(n_{2}-1)
  +J \bbox{S}_{1} \cdot \bbox{S}_{2},
  \label{rung-model}
\end{eqnarray}
where $\ell=1,2$ refer to the two legs.
$SO(5)$ symmetry requires a single constraint on the
three couplings: $J=4(U+V)$.  For $U,V
\gg t,t_\perp$ the hopping $t$ can be treated perturbatively, and SZH
have determined the (quite complex) phase diagram in the $U-V$ plane.
With the weak-coupling RG, we can attempt to complete the phase
diagram by exploring the opposite limit, $U,V \ll
t,t_\perp$. (One can hope to
determine the behavior at intermediate coupling
$U,V \sim t, t_\perp$
by interpolation.)  Further, we can explore the generic
behavior of other weakly-interacting $SO(5)$-invariant two-leg
ladder systems which contain, for example,
inter-rung interactions.

\begin{figure}[hbt]
\epsfxsize=7cm
\centerline{\epsfbox{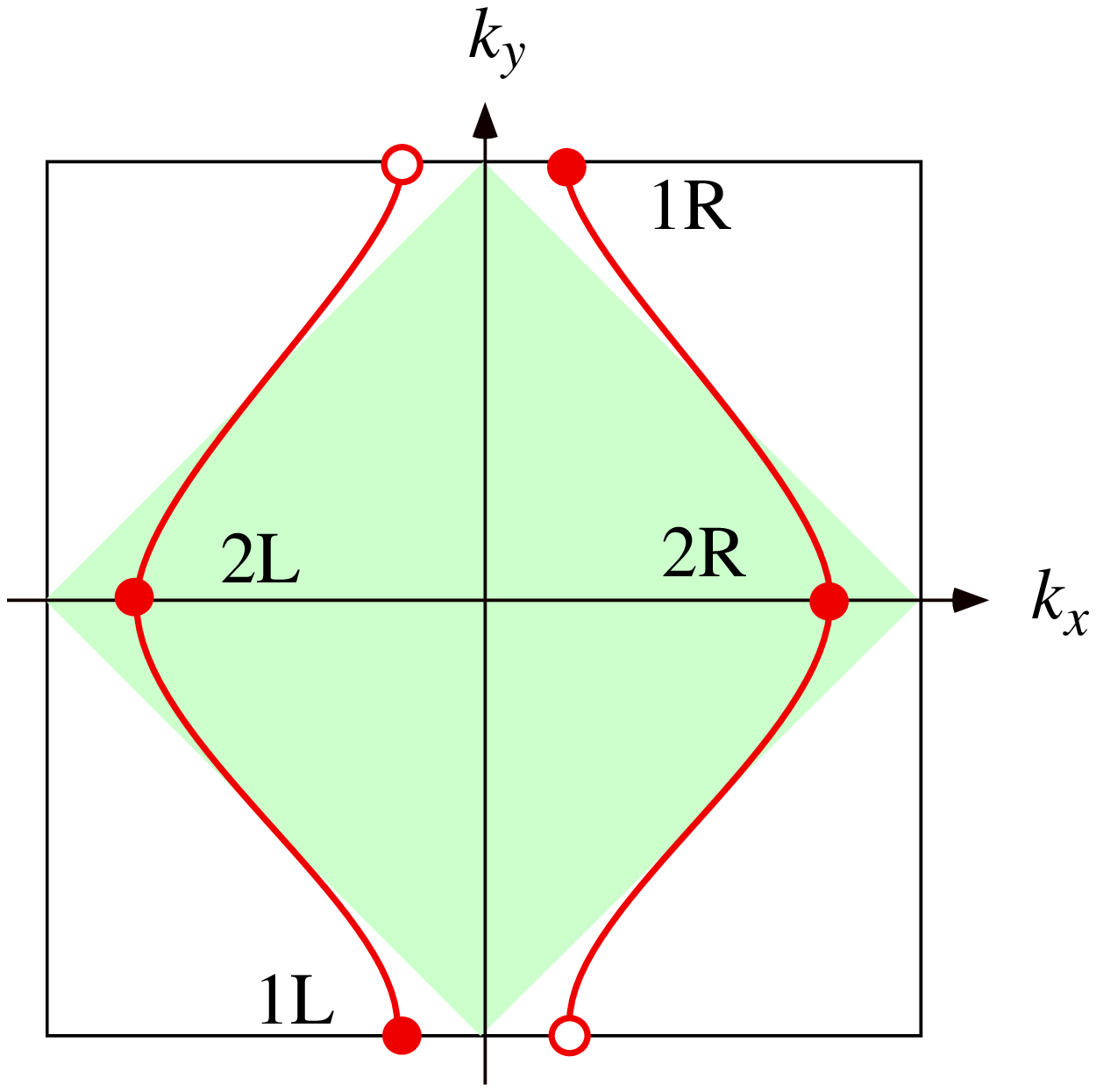}} \vspace{5pt} {\noindent
  FIG.  9: The folded Brillouin zone for the $SO(5)$ spinor.  The
  allowed momenta are chosen to be in the grey area. For a two-leg
  model, the transverse momentum $k_{y}$ takes two values $0, \pi$. In
  our convention, only $k_{y}=0$ excitations are inside the folded
  Brillouin zone.}
\end{figure}

To do so, we need a means of {\sl constructing} $SO(5)$-invariant
models in weak coupling.  For lattice models, such constructions have
been discussed by Henley\cite{Henley97u}\ and Rabello et.
al.,\cite{Rabello97u}\ and applied by SZH to the two-leg ladder.  Here we
generalize these methods to the {\sl chiral} fermions operators which
appear in the linearized continuum model obtained in the
weakly-interacting limit.  Since the Hamiltonian is built from
electron operators, we need to introduce {\sl spinor} representations
of $SO(5)$.  We begin with the lattice construction of Rabello et.
al.,\cite{Rabello97u}\ defining the four-component spinor as
\begin{equation}
  \Psi(\bbox{k}) = \left(
    \begin{array}{c}
      a_{\alpha}(\bbox{k})\\
      \phi_{\bbox{k}} a^{\dag}_{\alpha}(-\bbox{k}+\bbox{N})
    \end{array}
  \right),
\label{Rabello_spinor}
\end{equation}
where $\bbox{N}=(\pi, \pi)$ is the nesting vector of the Fermi
surface.  To avoid double-counting, the allowed momentum $\bbox{k}$ in
the spinor only runs in the ``folded'' Brillouin zone, whose size is
half of the original one, as shown in Fig. 9.  For the 
two-leg ladder model, in which the only
transverse momenta are $k_y=0,\pi$, it is possible to specify the
folded Brillouin zone by simply setting $k_y=0$ in the above spinor.
The factor
$\phi_{\bbox{k}}$ (which in the general two-dimensional case is a
non-trivial function with absolute value one) can be taken to be unity
with this convention.  Re-expressing the spinor in terms of the 
band electron operators and Fourier transforming gives
\begin{equation}
  \Psi(x) = \int \!dk_{x}\, \Psi(k_{x},0) e^{ik_{x}x} = \left(
    \begin{array}{c}
      c_{2\alpha}(x)\\
      (-1)^{x}c^{\dag}_{1\alpha}(x)
    \end{array}
  \right).
  \label{spinor:lattice}
\end{equation}
In the continuum limit valid for weak coupling at low energies,
a chiral decomposition
is possible:
\begin{equation}
  \Psi(x) \approx \Psi_{{\scriptscriptstyle R}} e^{ik_{{\scriptscriptstyle F}2}x} 
  +\Psi_{{\scriptscriptstyle L}} e^{-ik_{{\scriptscriptstyle F}2}x},
\end{equation}
with chiral spinors defined by  
\begin{equation}
  \Psi_{{\scriptscriptstyle P}}(x) = \left(
    \begin{array}{c}
      c^{}_{{\scriptscriptstyle P}2\alpha}(x)\\
      c^{\dag}_{{\scriptscriptstyle P}1\alpha}(x)
    \end{array}
  \right).
  \label{spinor-comp}
\end{equation}
To obtain Eq.~\ref{spinor-comp}, the $(-1)^{x}$ factor in
Eq.~\ref{spinor:lattice}\ was cancelled using the relation
$k_{{\scriptscriptstyle F}1}+k_{{\scriptscriptstyle F}2}=\pi$.  

The advantage of the spinor basis over the electron band operators
$c_{{\scriptscriptstyle P}i\alpha}$ is that they transform 
simply under $SO(5)$.  In
particular, under a unitary transformation generated by the operator
\begin{equation}
  U(\theta_{\scriptscriptstyle AB}) = \exp (i
  \theta_{\scriptscriptstyle AB} K_{\scriptscriptstyle AB}), 
\label{element}
\end{equation}
where $A,B=1\ldots 5$, the 
spinors $\Psi_{{\scriptscriptstyle P}}$ transform 
according to 
\begin{equation}
\Psi'_{{\scriptscriptstyle P}a} = 
U^{\dag}(\theta) \Psi_{{\scriptscriptstyle P}a} U(\theta) 
= [T(\theta)]_{ab } \Psi_{{\scriptscriptstyle P}b},
\end{equation} 
where the spinor indices $a,b=1\ldots 4$, and $T(\theta) = \exp
(i\theta_{\scriptscriptstyle AB} \Gamma^{\scriptscriptstyle AB}) $ is
the rotational matrix for a spinor.  Here $\Gamma^{\scriptscriptstyle
  AB} = i[\Gamma^{\scriptscriptstyle A},\Gamma^{\scriptscriptstyle
  B}]/4$ where the $\Gamma^{\scriptscriptstyle A}$ are five
generalized (4 by 4) Dirac matrices, discussed in detail in
Appendix~\ref{app:spinor}.  They satisfy the usual Clifford algebra
\begin{equation}
  \{ \Gamma^{\scriptscriptstyle A}, \Gamma^{\scriptscriptstyle B} \} =
  2 \delta_{\scriptscriptstyle AB}.
\label{gamma}
\end{equation}

Using the spinors, we can break down all fermion bilinears into
irreducible representations of $SO(5)$, i.e. generalized currents.
Three ``normal'' sets, which involve one $\Psi^\dagger$ and one $\Psi$
spinor, carry net momentum zero or $(\pi,\pi)$:
\begin{eqnarray}
{\cal J}_{{\scriptscriptstyle P}} 
&\equiv& \Psi^{\dag}_{{\scriptscriptstyle P}a} 
\Psi^{}_{{\scriptscriptstyle P}a},
\label{eq:Jscalar}
\\
{\cal J}^{\scriptscriptstyle A}_{{\scriptscriptstyle P}} &\equiv& 
\Psi^{\dag}_{{\scriptscriptstyle P}a}
(\Gamma^{\scriptscriptstyle A})_{ab}\Psi^{}_{{\scriptscriptstyle P}b},
\label{eq:Jvector}
\\
{\cal J}^{\scriptscriptstyle AB}_{{\scriptscriptstyle P}} &\equiv& 
\Psi^{\dag}_{{\scriptscriptstyle P}a} 
(\Gamma^{\scriptscriptstyle AB})_{ab}
\Psi^{}_{{\scriptscriptstyle P}b}.
\label{eq:Jtensor}
\end{eqnarray}
The three currents in Eqs.~\ref{eq:Jscalar}-\ref{eq:Jtensor}\ 
transform as an $SO(5)$ scalar, vector, and rank 2 antisymmetric
tensor, respectively.  A second set of currents (and their hermitian
conjugates) appear ``anomalous'', and carry net momentum $(\pm
2k_{{\scriptscriptstyle F}2},0)$ or $(\pi \pm
2k_{{\scriptscriptstyle F}2},\pi)$ : 
\begin{eqnarray}
{\cal I}_{{\scriptscriptstyle P}} &\equiv& 
\Psi^{}_{{\scriptscriptstyle P}a} R_{ab}
\Psi^{}_{{\scriptscriptstyle P}b}, 
\label{eq:Iscalar}
\\
{\cal I}^{\scriptscriptstyle A}_{{\scriptscriptstyle P}} &\equiv& 
\Psi^{}_{{\scriptscriptstyle P}a} 
(R\Gamma^{\scriptscriptstyle A})_{ab}\Psi^{}_{{\scriptscriptstyle P}b}.
\label{eq:Ivector} 
\end{eqnarray}
These two currents, which transform as a scalar and a vector
under $SO(5)$, require the introduction of the matrix
\begin{equation}
  R=\left( 
    \begin{array}{cc}
      0 & {\bf 1}\\
      -{\bf 1}&0
    \end{array} 
  \right),
\end{equation}
where ${\bf 1}$ is the two by two identity matrix.  Note that it is
straightforward to show that the matrices $R\Gamma^{\scriptscriptstyle
  AB}$ are symmetric, so that a non-vanishing anomalous tensor current
cannot be defined.  A simple counting verifies that the above set of
currents completely spans the space of electron bilinears.  There are
$1+5+10 = 16$ currents in Eqs.~\ref{eq:Jscalar}-\ref{eq:Jtensor}, and
an additional $2\cdot(1+5) = 12$ currents in
Eq.~\ref{eq:Iscalar}-\ref{eq:Ivector}\ and their complex conjugates,
for a total of $28=8\cdot7/2$ independent bilinears.

In weak coupling, we must generically consider all Hermitian products
of two bilinears which are (1) invariant under $SO(5)$ and (2)
conserve quasi-momentum.  Neglecting purely chiral terms (which, as in
Sec.~\ref{sec:model}, only renormalize velocities), there are then
five allowed couplings.  The interaction Hamiltonian density takes the
form
\begin{eqnarray}
  {\cal H}_{int} & = & g_{s} {\cal J}_{{\scriptscriptstyle R}} 
{\cal J}_{{\scriptscriptstyle L}} 
+g_{v} {\cal J}^{{\scriptscriptstyle A}}_{{\scriptscriptstyle R}}
{\cal J}^{{\scriptscriptstyle A}}_{{\scriptscriptstyle L}}
+g_{t} {\cal J}^{{\scriptscriptstyle AB}}_{{\scriptscriptstyle R}} 
{\cal J}^{{\scriptscriptstyle AB}}_{{\scriptscriptstyle L}} 
\nonumber \\
& + & h_{s} \bigg\{ {\cal I}_{{\scriptscriptstyle R}} 
{\cal I}_{{\scriptscriptstyle L}} + {\rm h.c.} \bigg\} 
+h_{v} \bigg\{ {\cal I}^{{\scriptscriptstyle A}}_{{\scriptscriptstyle R}} 
{\cal I}^{{\scriptscriptstyle A}}_{{\scriptscriptstyle L}} + {\rm h.c.} 
\bigg\}.
\label{SO5-gen}
\end{eqnarray}
Note that momentum conservation forbids forming a quartic interaction
from one normal and one anomalous current.

The above Hamiltonian represents the most general $SO(5)$ invariant
ladder theory with weak
interactions.  The five coupling contants ($g_s, g_v, g_t, h_s, h_v$) specify
a five-dimensional manifold within the
more general nine-dimensional space of $U(1)\times SU(2)$ symmetric
theories.  This manifold is determined
explicitly by a set of linear equations, given 
in App.~\ref{app:so5currents}, which relate the five $SO(5)$
invariant couplings to the $9$ $U(1)\times SU(2)$ couplings
which were introduced in Sec. II.
Because the $SO(5)$ manifold posesses higher symmetry,
it closes under an RG transformation.  
The five RG equations describing the flows {\it within} the
$SO(5)$ manifold are given explicitly in 
App.~\ref{app:so5rg}.

\subsection{SZH Model and Four Dominant Phases}

The weak coupling phase diagram for the SZH
model can now be obtained, by numerical integration
of the $SO(5)$ invariant RG flow equations (Eqs.~\ref{eq:so5rg1}-\ref{eq:so5rg5}).
The initial (bare) values of the $SO(5)$
coupling constants are obtained by taking the continuum limit of the SZH model.
For {\it each} initial set of bare parameters,
the phase is determined
by bosonizing those couplings which grow large under
the RG transformation, as described in Sec. II.  
The resulting weak coupling phase diagram is shown in Fig. 10.

Four new phases appear in addition to the D-Mott phase which occurs
for predominantly repulsive interactions.  In the ``C2S2'' region in
Fig. 10, all five couplings scale to zero under the RG.  This {\sl
  Luttinger liquid} phase thus retains all the gapless modes (2 charge
and 2 spin, hence C2S2) of the original non-interacting electron
system, and thereby has (an approximate) chiral $SO(8)$ symmetry.

\begin{figure}[hbt]
\epsfxsize=8cm
\centerline{\epsfbox{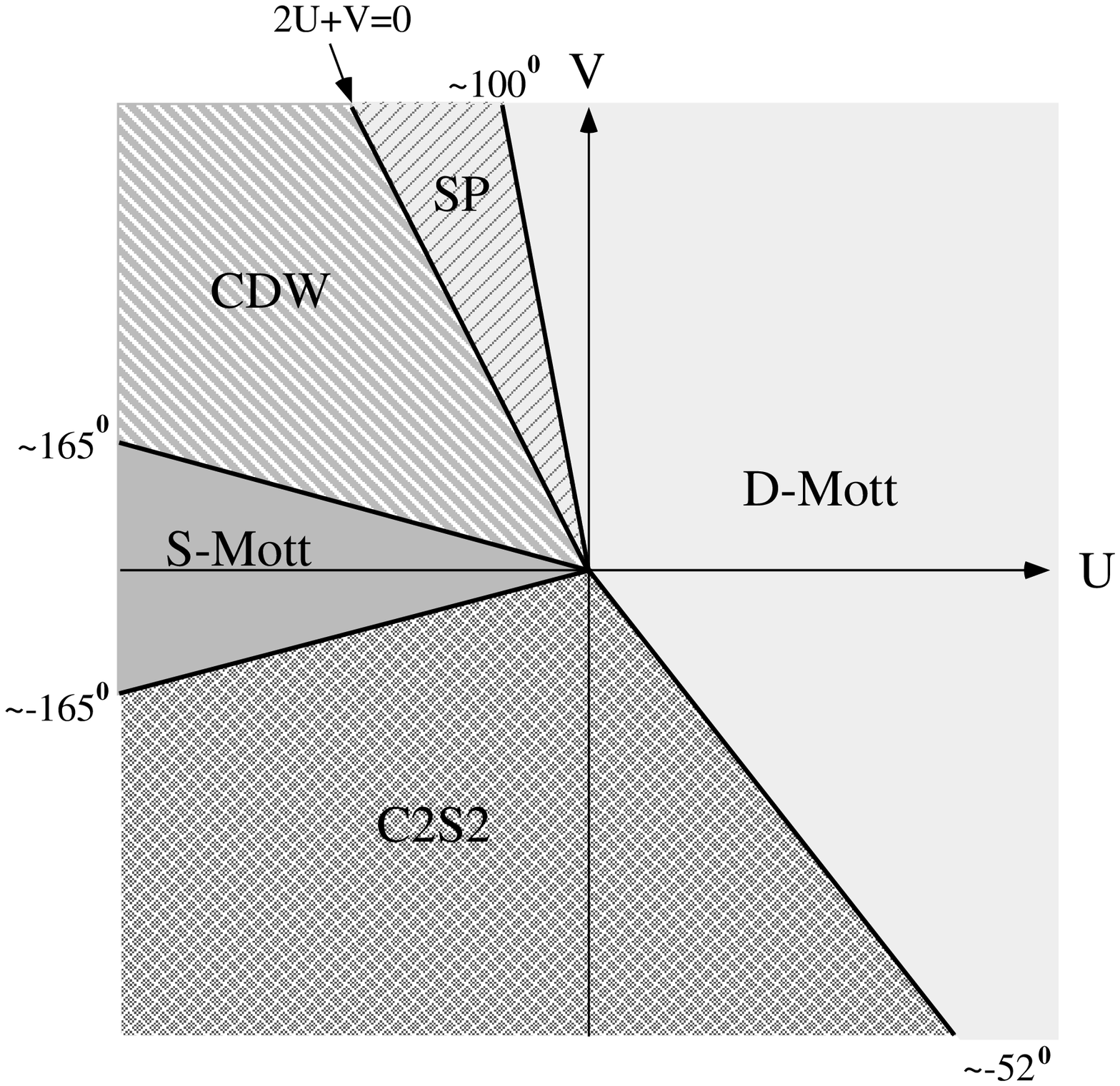}} \vspace{5pt} {\noindent FIG.
  10: Phase diagram of the $SO(5)$ symmetric SZH model plotted in the
  $U$--$V$ plane with $J=4(U+V)$ and $U,V \ll t=t_\perp$.}
\end{figure}

We group the other three states together with the D-Mott as four {\sl
dominant phases}.  In the S-Mott phase, the interactions diverge in
the same way as in the D-Mott case, given by Eqs.~\ref{ratio}, with
the modification $b^{\rho}_{12}, b^{\sigma}_{12}, u^{\rho}_{11} \to
-b^{\rho}_{12}, -b^{\sigma}_{12}, -u^{\rho}_{11}$.  In the
$SO(5)$-invariant notation, this corresponds to changing the sign of
$h_s$ and $h_v$.   Semiclassically, the only change
in the behavior is that $\langle \varphi_{\rho-} \rangle_{\rm S-Mott} 
=
\langle \varphi_{\rho-}\rangle_{\rm D-Mott} + \pi = \pi$.  The
$\theta_{\sigma\pm}$ and $\theta_{\rho+}$ fields are unaffected, so
that the S-Mott phase still has short-range pairing.  It is, however,
of approximate {\sl s-wave} symmetry, with $\Delta_1 \Delta_2^\dagger
> 0$ due to the shift in $\varphi_{\rho-}$, as can be seen from
Eq.~\ref{pairsign}.  It is interesting that the strong-coupling
``s-wave'' paired state on a rung, $|\uparrow\downarrow,-\rangle +
|-,\uparrow\downarrow\rangle$ is identical in the ladder leg and band
bases, and corresponds to an {\it on-site}
pairing or singlet state.  In contrast,
pairing in the strong-coupling D-Mott state
is across the rung of the ladder,
as depicted schematically
in Fig. 11.

In the SP and CDW phases, the ratios of diverging couplings are
somewhat different.  In particular, $b^{\sigma}_{11}, b^{\rho}_{11}$
are irrelevant and
\begin{eqnarray}
f^{\rho}_{12} = -\frac14 f^{\sigma}_{12} = (\mp) b^{\rho}_{12} 
= (\pm) \frac14 b^{\sigma}_{12} = 
\\
\frac12 u^{\sigma}_{12}=-2u^{\rho}_{12}=
(\pm) 2u^{\rho}_{11} = g >0,
\end{eqnarray}
where the upper and lower signs hold in the SP and CDW phases,
respectively.  These modifications imply a fairly dramatic change in
the behavior relative to the D-Mott and S-Mott states.  In fact, the SP
and CDW are {\sl dual} to the D-Mott and S-Mott, respectively, in the
following sense: each is obtained from its dual counterpart by
interchanging $\varphi_{\sigma-}$ and $\theta_{\sigma-}$.  Because of
this interchange, the pair fields fluctuate wildly even locally, and
$\langle \Delta_1^{\vphantom\dagger} \Delta_2^\dagger \rangle_{\rm SP}
= \langle \Delta_1^{\vphantom\dagger} \Delta_2^\dagger \rangle_{\rm
  CDW} = 0$.  Instead, these two phases break discrete ${\cal Z}_2$
symmetries. 

To explore this in detail, consider the order parameters
\begin{eqnarray}
  B_\ell(x) &=& \frac12 
[a^{\dag}_{\ell\alpha}(x+1)a^{}_{\ell\alpha}(x) + 
  {\rm h. c.}],
  \\
  n_\ell(x) &=& a^{\dag}_{\ell\alpha}(x)a^{}_{\ell\alpha}(x)-1.
\end{eqnarray} 
The field $B_\ell(x)$ is the local kinetic energy, while $n_\ell(x)$
is the local electron density relative to half filling.  The two order
parameters differ in symmetry since $B_\ell(x)$ is
even and $n_\ell(x)$ is odd under a ${\cal Z}_2$ particle-hole
symmetry $a_{\ell\alpha}^{\vphantom\dagger}(x) \rightarrow
(-1)^{\ell+x} a_{\ell\alpha}^\dagger(x)$.  Using the usual
relations to rewrite $B_\ell$ and $n_\ell$ in terms of chiral
operators, bosonizing, and applying the semi-classical results (common
to both the SP and CDW phases) $\langle \theta_{\rho+} \rangle =
\langle \theta_{\sigma+} \rangle = \langle \varphi_{\sigma-} \rangle =
0$, one obtains
\begin{eqnarray}
  \langle B_\ell(x) \rangle & \sim & (-1)^{x+\ell}
  \langle \cos (\frac12 \varphi_{\rho-}) \rangle,
  \\
  \langle n_\ell(x) \rangle & \sim & (-1)^{x+\ell}
  \langle \sin (\frac12 \varphi_{\rho-}) \rangle.
\end{eqnarray}
Since $\langle \varphi_{\rho-} \rangle = 0,\pi$ in the SP and CDW phases,
respectively, we find $\langle B_\ell \rangle_{\rm SP} \sim
(-1)^{x+\ell}$, $\langle n_\ell \rangle_{\rm SP} = 0$ and $\langle
B_\ell \rangle_{\rm CDW} = 0$, $\langle n_\ell \rangle_{\rm CDW} \sim
(-1)^{x+\ell}$.  The SP phase thus breaks only the discrete
symmetry under translations by one lattice spacing (the translation in
the $y$-direction is, of course, equivalent to parity), while the CDW
phase breaks {\sl both} translational symmetry and the ${\cal Z}_2$
particle-hole symmetry.  These broken symmetries can be
depicted easily in the strong coupling limit, as shown in Fig. 11.

\begin{figure}[hbt]
\epsfxsize=8cm
\centerline{\epsfbox{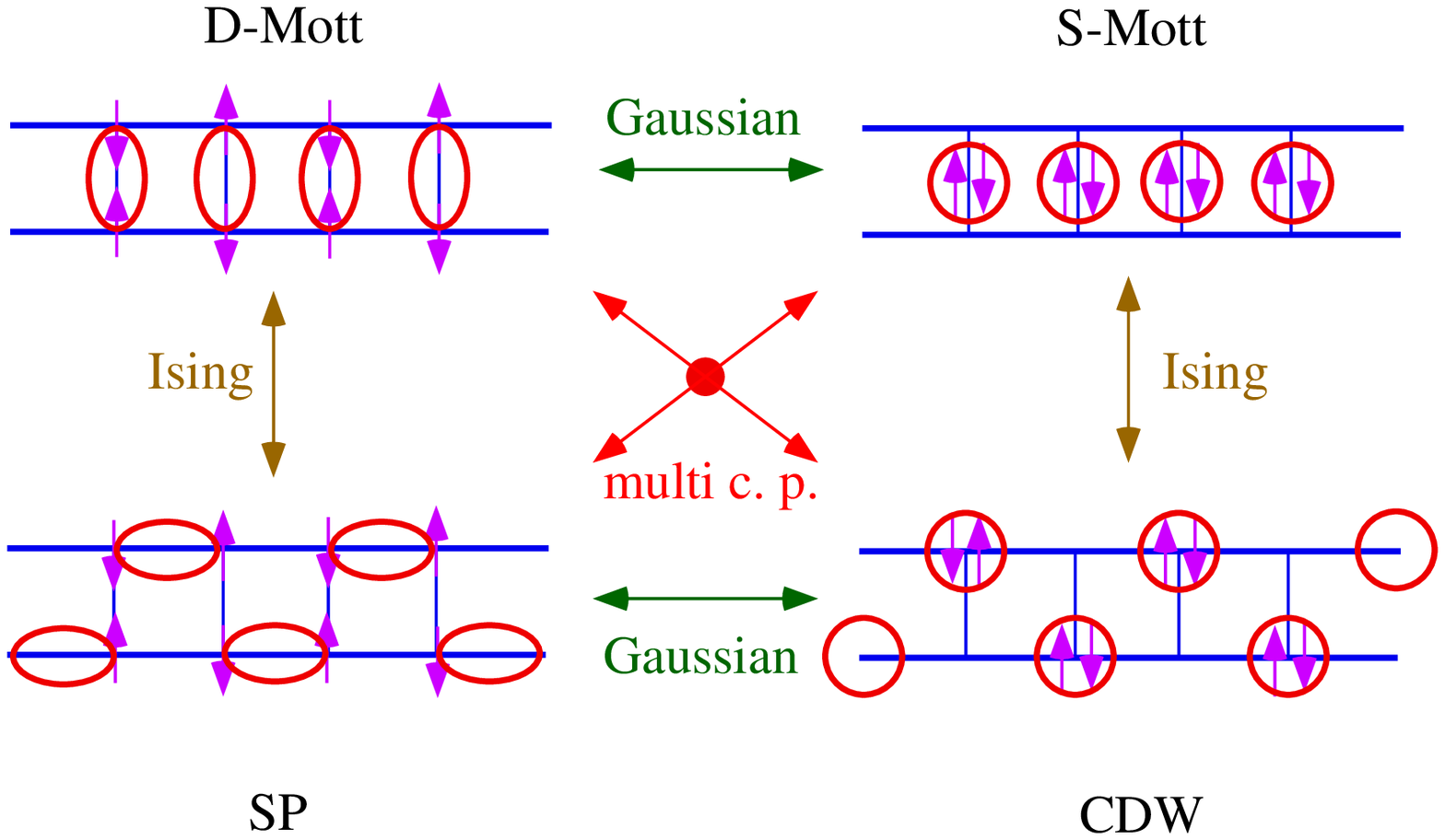}}
\vspace{5pt}
{\noindent FIG.  11: Schematic illustration of the four dominant
  phases, drawn for simplicity in the strong-coupling limit.  In the 
  D-Mott and S-Mott phases, neighboring rungs contain essentially
  decoupled pairs. Adjacent rungs are highly correlated in the
  SP and CDW phases, which furthermore break parity symmetry.}
\end{figure}

\subsection{SO(8) Symmetries and Degeneracies of the S-Mott, SP, and
  CDW Phases} 

Since the four dominant phases appear on essentially equal footing,
one might suspect that the S-Mott, SP, and CDW phases exhibit $SO(8)$
symmetries similar to that of the D-Mott phase.  We shall see that
this is indeed the case, but that the $SO(8)$ algebras are {\sl
  different} in each state.

Consider first the S-Mott.  In the previous subsection, it was shown
that the S-Mott is related to the D-Mott by a $\pi$ shift in
$\varphi_{\rho-}$.  It follows that if we define
\begin{equation}
  \theta_a^{\rm S} = \cases{ \theta_a & $a=1,2,3$ \cr
    \varphi_{\rho-}+\pi & $a=4$ \cr},
\end{equation}
and $\varphi_a^{\rm S} = \varphi_a$, the bosonized Hamiltonian in the
S-Mott phase takes the form of Eqs.~\ref{free_boson}-\ref{int_boson}
(with $\theta_a,\varphi_a$ replaced by $\theta_a^{\rm
  S},\varphi_a^{\rm S}$).  Consequently, a refermionization into the GN
form is again possible.  In particular, upon defining
\begin{equation}
  \eta_{\scriptscriptstyle PA}^{\rm S} = \cases{
    \eta_{\scriptscriptstyle PA} & $A=1\ldots 6$ \cr
    P\eta_{\scriptscriptstyle PA} & $A=7,8$ \cr},
  \label{eq:SMottMajoranas}
\end{equation}
the S-Mott Hamiltonian takes the ``canonical" 
GN form (Eq.~\ref{O8GrossNeveu}) in
terms of the $\eta_{\scriptscriptstyle PA}^{\rm S}$.  The sign changes in Eq.~\ref{eq:SMottMajoranas}\ imply that the
generators of the $SO(8)$ symmetry in the S-Mott phase 
(are different
from those of the D-Mott phase.  For instance, $G_{\rm
  S}^{\scriptscriptstyle 71} = G_{\scriptscriptstyle
  R}^{\scriptscriptstyle 71} - G_{\scriptscriptstyle
  L}^{\scriptscriptstyle 71}$, whose spatial integral is not an
$SO(8)$ generator in the D-Mott case.  However, since the Majorana
fermions for the two phases are equal for $A=1\ldots 6$, the D-Mott
and S-Mott do share a common $SO(6)$ subalgebra.

Similar constructions can be performed for the SP and CDW phases.
Recalling the duality of the previous subsection, we choose
\begin{equation}
  \theta_a^{\rm SP} = \cases{ \theta_a & $a=1,2$ \cr
    \varphi_{\sigma-} & $a=3$ \cr
    \varphi_{\rho-} & $a=4$ \cr}\;\;\; \varphi_a^{\rm SP} = \cases{
    \varphi_a & $a=1,2,4$ \cr \theta_{\sigma-} & $a=3$ \cr}
  \label{eq:SPphases}
\end{equation}
for the SP phase. Similarly for the CDW, we take
\begin{equation}
  \theta_a^{\rm CDW} = \cases{ \theta_a & $a=1,2$ \cr
    \varphi_{\sigma-} & $a=3$ \cr
    \varphi_{\rho-}+\pi & $a=4$ \cr}
  \label{eq:CDWphases}
\end{equation}
and $\varphi_a^{\rm CDW} = \varphi_a^{\rm SP}$.
As before, the GN form can be retained.  The appropriate Majorana
fermions in these cases are
\begin{equation}
  \eta_{\scriptscriptstyle PA}^{\rm SP} = \cases{
    \eta_{\scriptscriptstyle PA} & $A=1\ldots5,7,8$ \cr
    P\eta_{\scriptscriptstyle P6} & $A=6$ \cr}
\end{equation}
and
\begin{equation}
  \eta_{\scriptscriptstyle PA}^{\rm CDW} = \cases{
    \eta_{\scriptscriptstyle PA} & $A=1\ldots5$ \cr
    P\eta_{\scriptscriptstyle PA} & $A=6,7,8$ \cr}.
\end{equation}
Like the D-Mott and S-Mott, the SP and CDW phases share a common $SO(6)$ symmetry.
Moreover, the D-Mott and SP share an $SO(7)$ symmetry, as do the
S-Mott and CDW.

A final calculation is possible now that the appropriate bosonized
variables have been established.  In Sec.~\ref{sec:GN}\ and
App.~\ref{app:gauge}, the uniqueness of the D-Mott ground state was
established.  We also expect a unique ground state for the S-Mott
phase, but have yet to establish it.  In the SP and CDW phases,
discrete symmetries are broken, and one expects at least a two-fold
degeneracy in the thermodynamic limit.  Using the techniques applied
earlier (gauge equivalence of semiclassical solutions) to the D-Mott,
we can determine these degeneracies.  Details can be found in
App.~\ref{app:gauge}.  The result of such an analysis is that the
S-Mott indeed has a unique ground state, while the SP and CDW ground
states are each exactly two-fold degenerate.

\subsection{Full Set of $SO(5)$ Fixed Points}

Because of the relative simplicity of the $SO(5)$-invariant manifold
(5 coupling constants versus 9 for the general case), it is possible
to perform an exhaustive determination of the possible asymptotic
scaling trajectories under the RG.  To do so, we insert the power-law
ansatz of Eq.~\ref{power_law}\ into the $SO(5)$ RG equations,
Eqs.~\ref{eq:so5rg1}-\ref{eq:so5rg5}.  This set of five coupled
algebraic equations can be solved exactly, in contrast to the
corresponding set of nine $U(1)\times SU(2)$ equations, which have
proved intractable.  One finds fourteen solutions, delineated in
Table~3.  Five represent the states discussed so far: the gapless C2S2
and four dominant $SO(8)$-symmetric phases.

\vskip 0.1in
\begin{tabular*}{3.2in}{ccccc|@{\extracolsep{\fill}}c}
\hline \hline
48$g_{s}$& 48$g_{v}$&48$g_{t}$& 48$h_{s}$&48$h_{v}$&phase\\ \hline
0    &  0   & 0 & 0  &0   &C2S2\\ \hline \hline
-2   &  2  & 4  & 1 &-1  &D-Mott\\ \hline
-2   &  2  & 4  & -1 &1    &S-Mott\\ \hline
-2   &  -2  & 4 & -1 &-1  &SP\\ \hline
-2   &  -2  & 4 &1  &1     &CDW\\ \hline \hline
0&3&6&0&0&D-Mott $\leftrightarrow$ S-Mott \\ \hline
0&-3&6&0&0&SP $\leftrightarrow$ CDW \\ \hline
-(12/5)&0&(24/5)&0&-(6/5)&
D-Mott $\leftrightarrow$ SP \\ \hline
-(12/5)&0&(24/5)&0&(6/5)&
S-Mott $\leftrightarrow$ CDW \\ \hline
0&0&8&0&0&multi-critical\\ \hline \hline
-12 &0& 8& $\pm$ 6 &0 & $SO(5)\times SO(3)$ GN \\ \hline
-12&0&0&$\pm$6&0& $SO(5)$ WZW $\times$ \\
&&&&&$SO(3)$ GN\\ \hline \hline
\end{tabular*}
\vspace{.1in}
\\
{\noindent Table 3: Fourteen algebraic solutions of the $SO(5)$
RG equations}

Of the remainder, five represent critical points.  Consider first the
D-Mott$\leftrightarrow$S-Mott transition.  Taking the values in
Table~3, one finds that semi-classically the fields $\theta_{\rho+},
\theta_{\sigma+},$ and $\theta_{\sigma-}$ are pinned as in the D-Mott
and S-Mott states, but that neither the $\theta_{\rho-}$ nor the
$\varphi_{\rho-}$ field appears in the interaction Hamiltonian.  There
is thus a single gapless (central charge $c=1$) bosonic mode at the
critical point.  That this is indeed the critical point between the
D-Mott and S-Mott phases can be seen by perturbing slightly away from the
scaling trajectory.  If the perturbations are small, those 
terms involving
the gapped degrees of freedom will have negligible effect, and we need
only include the couplings involving the $\rho-$ fields.  As argued in
Sec.~\ref{sec:model}, $\cos n\theta_{\rho-}$ terms are not allowed by
translational invariance.  The low-energy Hamiltonian
density near the critical point (after integrating out the massive
fields) thus has the form
\begin{equation}
  {\cal H}_{\rm D\leftrightarrow S}^{\rm eff.} = {1 \over 8\pi} \left[
    (\partial_x \varphi_{\rho-})^2 +
    (\partial_x\theta_{\rho-})^2\right] - \lambda \cos 
\varphi_{\rho-}. 
\end{equation}
For $\lambda >0$, the semiclassical minimum occurs for
$\varphi_{\rho-} =0$, describing the D-Mott phase, while for
$\lambda<0$, the minimum shifts to $\varphi_{\rho-} = \pi$, yielding
the S-Mott phase.  We expect that the general form of this low-energy
{\sl critical} Hamiltonian will remain valid even in strong coupling,
though the Luttinger stiffness and velocity of the critical
$\varphi_{\rho-}$ mode will shift in this case.  What are the critical
properties of this transition?  The correlation length exponent is
determined by the scaling dimension of $\cos\varphi_{\rho-}$.  In a
general strong-coupling situation, this is a {\sl continuously
  variable} exponent.  In weak-coupling, however, it is determined.
In particular, refermionization implies $\cos\varphi_{\rho-} \sim
\psi_4^\dagger \tau^y \psi_4^{\vphantom\dagger}$, which acts as a
Dirac mass and has scaling dimension one.  In this limit then, the
correlation length $\xi \sim |\lambda|^{-\nu}$, with $\nu = 1$.  Both
in strong and weak coupling, the dynamical exponent $z=1$, as
determined by the quadratic bosonic kinetic energy.  This type of
$c=1$ continuously-variable critical point is known as a Gaussian
model, as shown in Fig. 11.  Of course, in neglecting the massive modes, we have thrown out
additional universal physics in the weak-coupling limit.  In
particular, these massive modes have a large $SO(6)$ symmetry, which
can be seen by rewriting the critical interaction Hamiltonian density 
using
Table~3 and Eqs.~\ref{eq:JvectorG}-\ref{eq:JtensorG},
\begin{equation}
  {\cal H}_{\rm D \leftrightarrow S}^{\rm int.}(\lambda=0) =
  g\sum_{\scriptscriptstyle A,B=1}^6 G_{\scriptscriptstyle
    R}^{\scriptscriptstyle AB} G_{\scriptscriptstyle
    L}^{\scriptscriptstyle AB}.
\end{equation}
The full weak-coupling critical symmetry is thus $U(1)_R \times U(1)_L
\times SO(6)$.  It may seem surprising that this critical point has
{\sl lower} symmetry than the massive dominant phases, which enjoy
$SO(8)$-invariance.  This is a result unique to the weak-coupling
limit.  With stronger interactions, corrections to the weak-coupling
scaling will break the $SO(8)$ symmetry, while leaving the $U(1)_R
\times U(1)_L$ critical symmetry (which results from truly
infinite-wavelength physics) intact.

Having understood the D-Mott$\leftrightarrow$S-Mott transition, it is
clear that the SP$\leftrightarrow$CDW transition is essentially
identical.  The Hamiltonian in this case differs only via the 
interchange of $\theta_{\sigma-}$ and $\varphi_{\sigma-}$,
which in any case are massive at this critical point.

The next two critical points are somewhat different.  For
concreteness, consider the D-Mott$\leftrightarrow$SP transition.
As before, three of the bosonic fields are massive, in this case
$\theta_{\rho+}, \theta_{\sigma+}, \theta_{\rho-}$.  These can be
integrated out, leaving the
$\sigma-$ fields critical.  However, here {\sl both}
$\theta_{\sigma-}$ and the dual $\varphi_{\sigma-}$ appear, so a
semi-classical analysis is not tenable.  Instead, we  refermionize
this single remaining bosonic field and its interactions {\sl after}
integrating out the three massive bosons (i.e. setting them to their
semi-classical minima).  The reduced Hamiltonian density
in this case is
\begin{equation}
  {\cal H}_{\rm D\leftrightarrow SP}^{\rm int.,eff.} = g i
  \eta_{\scriptscriptstyle R5} \eta_{\scriptscriptstyle L5} + 
\tilde\lambda
  i \eta_{\scriptscriptstyle R6} \eta_{\scriptscriptstyle L6}.
\end{equation}
Here $g$ is the finite coupling along the scaled RG trajectory, and
$\tilde\lambda$ is a deviation from the critical trajectory similar to
$\lambda$ in the Gaussian model above.  Since $g$ is non-zero, the
$\eta_{\scriptscriptstyle P5}$ Majorana fermion acquires a gap, and
only the single $\eta_{\scriptscriptstyle P6}$ Majorana fermion is
gapless at the critical point.  This is a central charge $c=1/2$
critical point, which uniquely identifies it as an Ising transition.
Indeed, the Ising nature of this transition is very physical, given
the discrete ${\cal Z}_2$ parity symmetry broken in the SP phase.
This also explains the duality between the D-Mott and SP phases found
earlier: this duality is nothing but the usual Kramers--Wannier
duality of the Ising model.  As before, we expect the Ising critical
behavior to be robust to corrections to the weak-coupling RG, so these
transitions should be in the same universality class even with strong
interactions.  In the weak-coupling limit, the massive degrees of
freedom again have higher symmetry, in this case including the
$\eta_{\scriptscriptstyle P5}$ Majorana fermion coming from the
$\sigma-$ fields.  The full weak-coupling critical theory is thus
${\cal Z}_2 \times SO(7)$, where the ${\cal Z}_2$ theory is the
conformally-invariant Ising model, as indicated in Fig. 11.  

Not surprisingly, the S-Mott$\leftrightarrow$CDW transition is also of
the Ising variety.  The ``multi-critical point'' in Table~3
describes the case when all four phases come
together at a point, i.e. when two transition lines cross.  It is
simply a direct product of the two critical theories above, i.e. a
Gaussian model and an Ising theory.  It is possible that these
theories actually become coupled if one re-introduces interactions
that were irrelevant at the non-interacting Fermi fixed point, but
we do not explore this possibility here.

The remaining four ``fixed points'' of the $SO(5)$-invariant RG
describe more exotic situations.  We have not observed any microscopic
Hamiltonians attracted to these phases, but some of these may perhaps
occur for some choices of bare interactions.  We suspect that these ``phases'' are unstable, and hence spurious for
physically relevant situations.  Nevertheless, we discuss them briefly
for completeness.  They are most easily understood by using the
representations in Eqs.~\ref{eq:JscalarG}-\ref{eq:IvectorG}.  The form of
the $SO(5)\times SO(3)$ case is then easily seen from the interaction
Hamiltonian density (taking the positive sign for $h_s$ for 
simplicity)

\begin{equation}
{\cal H}_{SO(5)\times SO(3) {\rm GN}}^{\rm int.} \sim g 
\left[
\sum_{\scriptscriptstyle A,B=1}^5 
G_{\scriptscriptstyle R}^{\scriptscriptstyle AB}
G_{\scriptscriptstyle L}^{\scriptscriptstyle AB} 
+ 3\hspace{-0.1cm} \sum_{\scriptscriptstyle A,B=6}^8 
G_{\scriptscriptstyle R}^{\scriptscriptstyle AB}
G_{\scriptscriptstyle L}^{\scriptscriptstyle AB}  
\right]\hspace{-0.05cm}.
\end{equation}

These interactions are precisely those of an $SO(5)\times SO(3)$ GN
model.  Both of the constituent GN models are massive, so this
represents another gapped phase.  The solution with the opposite sign
for $h_s$ can be converted into the same form by the canonical
transformation $\eta_{\scriptscriptstyle R6} \rightarrow -
\eta_{\scriptscriptstyle R6}$, so it is also a gapped phase of this
sort.  The remaining two phases can be understood similarly.  Note
that in these cases only the scalar interactions $g_s$ and $h_s$ are
non-zero.  This implies that the first five and last three Majoranas
are decoupled.  Furthermore, in this case since $g_v = g_t = 0$, the first
five Majorana fermions are non-interacting.  They comprise a gapless
$SO(5)$ Wess-Zumino-Witten (WZW) model with central charge $c=5/2$,
while the last three Majorana fermions combine to form an $SO(3)$ GN
model.  The interaction Hamiltonian density is thus simply
\begin{equation}
  {\cal H}_{SO(5) {\rm WZW} \times SO(3) {\rm GN}}^{\rm int.} \sim g 
  \sum_{\scriptscriptstyle  
    A,B=6}^8 G_{\scriptscriptstyle R}^{\scriptscriptstyle AB}
  G_{\scriptscriptstyle L}^{\scriptscriptstyle AB}.
\end{equation}

\section{Doping the D-Mott Phase}
\label{sec:doping}

In this Section we briefly consider the effects of doping away from
half-filling in the two-leg ladder.  In the weak coupling limit of
interest and with nearest neighbor hopping in the kinetic energy, the
ladder at half-filling was argued to scale onto the soluble
Gross-Neveu model which posseses an exact global $SO(8)$ symmetry.
Generally, doping away from half-filling will break down this large
symmetry, leaving only charge and spin conservation, with the much
smaller $U(1) \times SU(2)$ symmetry.  This can already be 
anticipated for the
non-interacting problem: When the Fermi energy moves away from zero
(half-filling) the Fermi velocities in the bonding and anti-bonding
bands will in general become unequal due to curvature in the
energy/wavevector dispersion.  For weak doping, however, this effect
is small.  Indeed, in the relativistic model derived in Section II
where the dispersion was linearized about the Fermi points, the small
curvature is ignored entirely.  In the following we focus on this very
low doping limit ($x=1-n \ll 1$), where the difference between the two
Fermi velocities can be safely ignored.  
We thus continue to employ the linearized relativistic model.
Nevertheless, as we shall see, even within this limit doping away from
half-filling breaks down the global $SO(8)$ symmetry of the 
Gross-Neveu
model, although in a rather straightforward manner.

To dope we consider adding a chemical potential term to the 
Gross-Neveu
Hamiltonian, $H$, with $H_\mu = H - \mu Q$, where $Q$ is
the {\it total} electron charge.  This charge can be written
\begin{equation}
Q = 2 \int\! dx\, (\psi^\dagger_{{\scriptscriptstyle R}1} 
\psi^{\vphantom\dagger}_{{\scriptscriptstyle R}1}
+\psi^\dagger_{{\scriptscriptstyle L}1} \psi^{\vphantom\dagger}_{{\scriptscriptstyle L}1}  ) = 
2 \int\! dx\, (G_{\scriptscriptstyle R}^{21} + 
G_{\scriptscriptstyle L}^{21}).
\end{equation}
Since $Q$ is a {\it global} $SO(8)$ generator, it commutes with the
full interacting Hamiltonian: $[Q,H]=0$.  Thus, even for $\mu \ne 0$
the states can still be labelled by $Q$, which, along with all the
generators $G^{{\scriptscriptstyle AB}}$ with $A,B=3\ldots 8$, remains
a good quantum number.  The $SO(8)$ multiplets will of course be split
by the presence of a non-zero $\mu$, lowering the energy of positively
charged excitations and raising the negatively charged ones.

The splitting of the $SO(8)$ multiplets can be conveniently visualized
in the large-N ``semiconductor" picture.  Of the four fermionic
particle/hole excitations of the fundamental GN fermions,
only the first one is charged and is shifted in energy.
Specifically, employing the semiclassical notation,
the excitations $(\pm 1,0,0,0)$ carry charge $Q=\pm 2$,
and are shifted by an energy $\Delta E_\mu = \mp 2 \mu $,
as depicted schematically in Fig.~12.  Provided this shift is smaller
than the energy gap, $-m < 2 \mu < m$, the {\it ground state}
remains unaltered:  The negative energy ``valence" bands
remain filled and the ``conduction" bands empty.    
In terms of electrons, the ladder remains at half-filling.

Similarly, the energies of the 16 kink excitations are
split into 8 with energy $m+\mu$ and 8 with energy $m-\mu$.  Of the 28
(two-fermion bound) states with energy $\sqrt{3} m$,
$16$ are neutral and unshifted by the chemical potential.
Of the others, 6 have charge $2$ and are shifted up in energy
by $2\mu$ and the other 6 down by $-2\mu$.

The situation is more
interesting when $2 \mu > m$.  In this case,
the energy of the states in the
``conduction" band for Cooper pairs drops below zero,
and the ground state will be radically altered.  In the large-N
limit the new ground state will consist of filling up
the negative energy states with a Fermi-sea
of Cooper pairs, as depicted in Fig. 12.  For $N=8$ the
pair excitations will {\it not} be describable in terms of free 
Fermions,
but one still anticipates the general picture of a conducting sea
of Cooper pairs to remain valid. 
Since the Fermions interact for finite $N$, this conducting sea will
be more correctly described as a Luttinger liquid.  In the limit
of very low doping, however, the Cooper-pairs will be very far apart
and well described in terms of hard-core Bosons or free-Fermions.
In this limit, the Luttinger liquid parameters should approach those
of Free fermions.  
It is probable that the
N=8 Gross-Neveu model remains integrable even in the doped case, since
the states can still be labelled by the same good quantum numbers,
so that exact statements about the doped Mott-insulator can be made.
In the following we are less ambitious, using known results
from integrability for the undoped case to infer the behavior in the 
very 
low doping limit.

\begin{figure}[hbt]
\epsfxsize=7cm
\centerline{\epsfbox{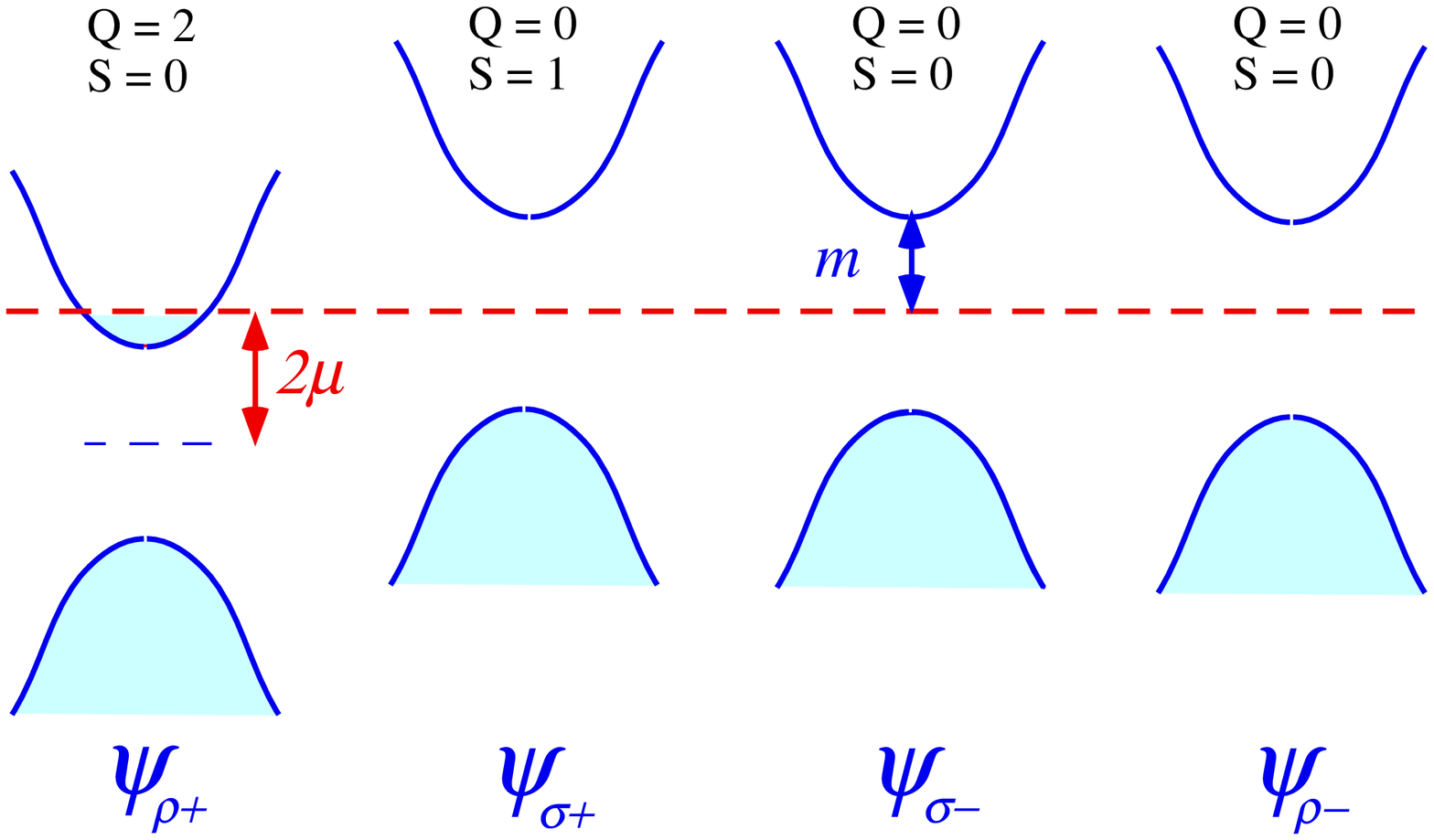}}
%\vspace{5pt}
{\noindent FIG. 12: Mean-field picture of the doped $SO(8)$ GN model. Only 
the first band (with $Q=2$) is shifted by the chemical potential. For
$2\mu > m$, Cooper pairs are added to the original ground state.}
\end{figure}

\subsection{Excitations with One pair present}

When $2 \mu > m$ the energy is lowered by adding Cooper
pairs to the system.  Here we consider first the case
$ 2 \mu = m + 0^+$, so that the concentration of pairs, denoted $x/2$,
is infinitesimal.  In this limit it is sufficient to
consider the properties in the presence of a {\it single}
Cooper pair.
The presence of even this one pair modifies
the spectrum of other excitations -- such as the spin or
single-particle gaps -- as we now briefly discuss.

Consider first the spin-gap, i.e. the energy of the lowest-lying spin
$s=1$ excitation.  In the undoped case, the lowest-lying triplet
states are the $\eta_{3,4,5}$ fundamental fermions, with momentum
$(\pi,\pi)$ and energy $m$.  As known from integrability, these
excitations interact via an {\it attractive} interaction with the
other GN fermions, including the single Cooper pair present due to
doping.  Indeed, the binding energy is known exactly and given by $E_b
= (2- \sqrt{3})m$.  With the single Cooper pair present, an
$s=1$ magnon can be added to the system into a bound state with the
Cooper pair, costing a reduced energy $m - E_b = (\sqrt{3} -1)m$.
Thus the spin-gap at infinitesimal doping $x=0^+$ is reduced from the
undoped value of $m$ down to $(\sqrt{3} -1)m$.  There are, of course,
also unbound $s=1$ excitations which can be created well away from the
Cooper pair, with energy $m$.  In fact, for $x \rightarrow 0$ the
energy $m$ spin excitations will dominate the spectral weight.  The
spectal weight for the lower energy $s=1$ bound states will presumably
vanish linearly with $x$.  It is worth emphasizing that the
discontinuity in the spin-gap at infinitesimal doping $x=0^+$ is a
general feature due to the presence of a magnon/Cooper-pair bound
state in the undoped Mott insulator, and is not an artifact of weak
coupling.  If such a bound state survives strong coupling, as
suggested by numerical RG on the two-leg ladder, a discontinuity
should be present.

It is also instructive to consider the energy gap for adding single 
electrons
in the presence of the single Cooper pair.  Adopting a convention
where $Q >0$ correspponds to ``hole" doping, consider the energy
to add a single electron with charge $-1$.  A single electron
can be created by adding a kink excitation,
for example an even kink with $(-1,-1,-1,-1)/2$ in the semiclassical 
notation.
When $\mu=0$ this costs an energy $m$, but is shifted {\it up}
in energy for non-zero chemical potential:  $E_1 = m + \mu$,
as depicted in Fig. 13.  When $2\mu = m + 0^+$ and the single 
Cooper pair
is added, the energy to add the electron can be lowered
from $E_1 = (3/2)m$ by binding the kink to the $(1,0,0,0)$
Cooper pair.  This forms a charge $Q=1$ hole state: an odd
kink with $(1,-1,-1,-1)/2$.  The associated binding energy equals $m$,
as follows directly from triality
(at $\mu=0$).  
Thus, at infinitesimal (hole) doping the energy to add an electron
drops by $m$, down to $m/2$.
As for the case of the spin-excitations, one expects
a continuum of unbound single electron excitations,
at energies above $3m/2$.

\subsection{Excitations with Many pairs}

For $2\mu > m$ the ``conduction band" for Cooper pairs will be
partially occupied.  In this case, one expects a continuum of low
energy particle/hole excitations created by exciting pairs across the
Fermi ``surface".  For the $SO(8)$ Gross-Neveu model the Cooper pairs
will presumably interact with one another, so that the semiconductor
picure of a non-interacting Fermi sea will not be quite correct.
Rather, the spinless gas of Cooper pairs will presumably form an
interacting Luttinger liquid.  In any event, one expects a continuum
of low energy excitations in the Cooper pair fluid, presumably with a
linear dispersion relation.  One might hope that the velocity of this
mode as a function of doping $x$ might be accessible from
integrability of the doped Gross-Neveu model.

It would also be very interesting to study the energy of the
spin-one excitations with {\it finite} doping.  A $s=1$ magnon
added to the system
will interact via an attractive interaction with the
sea of Cooper pairs.  For infinitesimal
doping ($x=0^+$) the corresponding spin-gap energy
was lowered due to the formation of a magnon/Cooper-pair
bound state.  With many pairs present, this energy will presumably
be further lowered, as depicted schematically in Fig.~12.

Finally, we consider briefly the spin one excitations at energies above threshhold.  These states would contribute to the spin-spectral function,
accessible via inelastic neutron scattering in the doped ladder.
Generally, we expect a continuum of states above threshhold, corresponding
for example to adding a magnon at $(\pi,\pi)$ and simulatenously
exciting multiple ``particle-hole" pairs in the (Cooper-pair)
Fermi-sea.  This continuum should contribute
to the spin-spectral function at any given momentum.  For example,
at momentum $(\pi,\pi)$, multiple particle-hole pairs with
zero net momentum will contribute spectral weight at all energies
above threshhold.  Due to this continuum of states, we do not expect
any delta-functions in the energy dependence of the spin-spectral 
function in the doped ladder.  

This expectation runs contrary to arguments put forward by Zhang for
the existence of a sharp $\pi-$resonance in the spin-spectral function
in the superconducting phase of models which exhibit an exact SO(5)
symmetry.  Zhang's argument has recently been applied to the doped
(power-law) superconducting phase of the two-leg ladder by Scalapino,
Zhang and Hanke.\cite{Scalapino97u} Below, we briefly reconsider
Zhang's argument for the sharp $\pi-$resonance, and show that in
addition to SO(5) symmetry, it relies on the existence of a {\it
  condensate} in the superconducting phase.  Being one-dimensional,
however, a true condensate does not exist in the ``superconducting"
phase of the two-legged ladder.  In our view, this invalidates the
argument for a sharp delta-function $\pi-$resonance in the doped ladders, even in the
presence of exact SO(5) symmetry.

Zhang's argument for the $\pi-$resonance rests on the fact that the 
$\pi$ operators, defined in Eq.~\ref{pi-op}, being global SO(5) (and SO(8))
generators, are exact eigen-operators
even with non-zero chemical potential:
\begin{equation}
[H_\mu , \Pi_a] = 2\mu \Pi_a   ,
\end{equation}
where $H_\mu = H -\mu Q$ and the subscript $a$ labels the three
components of the $\pi-$operators.  This implies that for any
eigenstate $H_\mu |E\rangle=E|E\rangle$ with energy $E$, the triplet
of states $\Pi_a |E\rangle$ are also exact eignestates, but with
energy $E + 2\mu$, provided they are non-vanishing.  Denote the exact
ground state of the doped ladder with $N$ Cooper-pairs as $|N\rangle$,
which satisfy $H_\mu |N\rangle=0$.  Adding an additional Cooper-pair
is accomplished with the operator ${\cal O}_1^+(x) = {\cal O}_s (x)
\psi^\dagger_1(x)$.  The zero momentum Fourier transform of this
operator, ${\cal O}_1^+(k=0)$,  creates a state with $N+1$
pairs, which can be decomposed as,
\begin{equation}
  {\cal O}_1^+(k=0) |N \rangle = c |N+1 \rangle + ...  ,
  \label{PI}
\end{equation}
where the dots denote excited states with $N+1$ pairs present.
Following Zhang, we can use the $\pi-$operators to rotate
Cooper-pairs at zero momentum into a triplet of magnons
at momentum $(\pi,\pi)$, since
\begin{equation}
[ \Pi_a , \psi^\dagger_1(x)] = \sqrt{2} \eta_a(x)  ,
\end{equation}
with $a=3,4,5$.  Acting with the $\Pi-$operator on
Eq.~\ref{PI}\ 
and using the above commutation relation and the fact that ${\cal
  O}_s$ commutes with $\Pi_a$, one obtains,
\begin{equation}
  {\cal O}_a (k=0) |N \rangle = {c \over
    \sqrt{2}} \Pi_a |N+1 
  \rangle + ...   , 
\end{equation}
where ${\cal O}_a(x) = {\cal O}_s(x) \eta_a(x)$.  
The left hand side is a spin $1$ triplet of states
with momentum $(\pi,\pi)$, built by adding a magnon
to the N-pair ground state.  Due to the SO(5) symmetry,
the states on the right side, 
$\Pi_a |N+1 \rangle$, are {\it exact} eigenstates with energy
$2\mu$.  As argued by Zhang,
the equality between the left and right sides implies
that the triplet of magnons will contribute a delta-peak
in the spin-spectral function at energy $2 \mu$ - the $\pi-$resonance.

However, this conclusion rests on the assumption of a non-vanishing overlap
between ${\cal O}_a |N \rangle$ and $\Pi_a |N+1\rangle$. 
But in the thermodynamic limit, the squared
overlap, $|c|^2$, is simply the (Bose) condensate density,
since \begin{equation}
  c = \langle N+1| {\cal O}_1^+(k=0) | N \rangle  .
\end{equation}
While non-zero
in a 2d superconductor, for the two-leg ladder the condensate
density is zero, and the argument for a delta-function $\pi-$resonance
is invalid.  The vanishing condensate density is a general
property of 1d systems which follows from the Mermin-Wagner theorem in 
the thermodynamic limit.  For finite $N$ at fixed pair density, we
expect $c$ to decay like an inverse power of the system length $L$.
 
\begin{figure}[hbt]
\epsfxsize=6cm
\centerline{\epsfbox{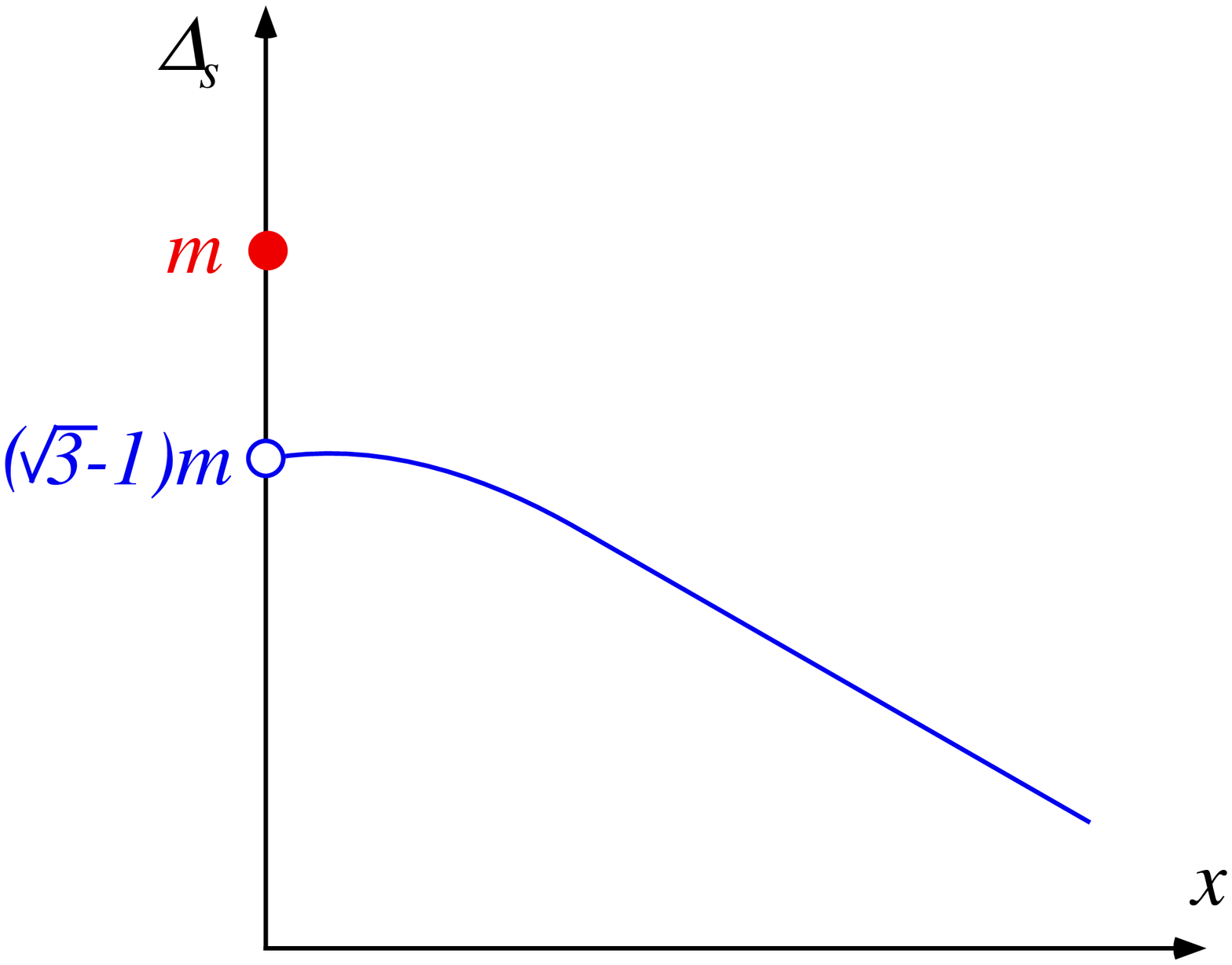}}
\vspace{5pt}
{\noindent FIG.  13: Spin-gap as a function of doping $x$. The
  spin-gap is discontinuous at $x=0$ due to the 
 formation of magnon-Cooper-pair bound states. }
\end{figure}

For this reason, we expect that a {\sl finite} length $SO(5)$-invariant 1D
model exhibits a $\delta(\omega-2\mu)$ peak in the spin spectral
function at momentum $(\pi,\pi)$ with weight (coefficient) decreasing
as some power $L^{-a}$.  Zhang has suggested\cite{Zhang:pc}\ that the
spin spectral function may have a corresponding algebraic singularity
{\sl in frequency} in an {\sl infinite} system.  To address the fate
of this finite-size peak in the thermodynamic limit, we consider now an
approximate calculation of the {\sl doped} spin-spectral
function in the infinite system.

To this end, we must determine the GN operator content of the
lattice spin operator.  Using the techniques of Sec.~\ref{sec:GN}, it
straightforward to show that the decomposition of $S^+_\ell(x)$
contains a term
\begin{equation}
  S^+_\ell(x) \sim (-1)^{\ell+x} {\cal O}_s(x)
  \psi_{\scriptscriptstyle 2R}^\dagger + \cdots
\end{equation}
Of course, many other operators are also present, but give either
negligible or identical contributions to the spectral function in the
regime of interest.  In the D-Mott phase, the string ${\cal O}_s$ has
negligible qualitative effects in correlation functions, since the
bosonic fields $\theta_a$ are all locked (i.e. only weakly
fluctuating) around $\theta_a=0$ in that case.  In the doped system,
however, there is an important modification to the $\theta_1$ field.
Since the derivative of this field is just the pair density
(Eq.~\ref{eq:PairWinding}), its average value has a mean slope
$\langle \theta_1(x)\rangle = -2\kappa_F x$, where $\kappa_F =
\pi(1-n)/2$ is the Fermi wavevector
for the sea of Cooper pairs; recall that $1-n$ is the concentration of holes in the
system.  Furthermore, there will be fluctuations of $\theta_1$ around
this mean value, corresponding to the density and phase waves of the
Cooper-pair fluid in the C1S0 state.

To account for both these effects, we redefine $\theta_1(x)
\rightarrow -2\kappa_F x + \theta_1(x)$, treating the shifted (zero mean)
$\theta_1$ field as a free Bose field, as appropriate for a free Fermi 
or Luttinger-liquid system.  The remaining three
($\theta_2,\theta_3,\theta_4$) fields remain locked, and we therefore
set these to zero inside the Jordan-Wigner string ${\cal O}_s$.  This
gives
\begin{equation}
  S^+_\ell(x) \sim e^{i(\pi-\kappa_F)x + i\pi\ell} e^{i\theta_1/2}
  \psi_{\scriptscriptstyle 2R}^\dagger . \label{eq:dopedspinrep}
\end{equation}
As carried out for the undoped case in Sec.~\ref{sec:correlators}, the
spin spectral function can be extracted from the analytically
continued Fourier transform of the imaginary time spin-spin
correlation function
\begin{equation}
  {\cal S}_{\ell\ell'}(x,\tau) \equiv \left\langle
    S^-_\ell(x,\tau) S^+_{\ell'}(0,0) \right\rangle.
\end{equation}
Using Eq.~\ref{eq:dopedspinrep}, one then finds
\begin{eqnarray}
  {\cal S}_{\ell\ell'}(x,\tau) & \sim & e^{-i(\pi-\kappa_F)x -
    i\pi(\ell-\ell')} \bigg\langle e^{-{i \over
        2}\left(\theta_1(x,\tau)-\theta_1(0,0)\right)} \nonumber \\
    & & \times \psi_{\scriptscriptstyle
      2R}^{\vphantom\dagger}(x,\tau)\psi_{\scriptscriptstyle
      2R}^\dagger(0,0)\bigg\rangle.
\end{eqnarray}

To proceed, we require a calculation of the above expectation value.
The simplest natural approximation, which will be our first attempt,
is to decouple the charge ($1$) and spin ($2$) sectors, calculating
the $\theta_1$ correlator as appropriate for a Luttinger liquid
(i.e. from a free Bose theory) and the $\psi_2$ correlator using the
``semiconductor'' free-fermion operators.  In particular, one finds
\begin{equation}
  \left\langle  e^{-{i \over
        2}\left(\theta_1(x,\tau)-\theta_1(0,0)\right)} \right\rangle
  \sim (x^2+\tau^2)^{-K/4}, \label{eq:Luttresult}
\end{equation}
where $K$ is the Luttinger parameter of the Cooper-pair fluid; $K=1$
corresponds to free fermions, as is appropriate for very low dopings.
Here we have set the Fermi velocity of the Cooper pair sea to one.
The fundamental fermion correlator is approximately
\begin{equation}
  \left\langle \psi_{\scriptscriptstyle
      2R}^{\vphantom\dagger}(x,\tau)\psi_{\scriptscriptstyle
      2R}^\dagger(0,0)\right\rangle_{\rm MF} \sim \int\! {{dp} \over {2\pi}} \, 
  e^{ipx -\epsilon_1(p)\tau} \Theta(\tau), \label{eq:FFresult}
\end{equation}
where $\Theta$ is the Heavyside step function.  To simplify
Eq.~\ref{eq:FFresult}, we have neglected to include the mean-field
``coherence factors''.  Because these are non-singular, their neglect
only modifies the final result by an overall smooth momentum-dependent
amplitude factor.  Multiplying the two terms in
Eqs.~\ref{eq:Luttresult}-\ref{eq:FFresult}, performing the Fourier
transform and analytically continuing to real frequencies gives the
spin spectral function
\begin{eqnarray}
  A_s^{\rm MF}(\pi-k,\pi;\omega) & \sim &\nonumber \\
  & & \hspace{-1.0in} {\rm Im}\, \int\! dx\, dp\, d\tau \, {e^{-ipx
      +\left(\omega - \epsilon_1(p+k-\kappa_F) + i\delta\right)\tau} \over
      {(x^2+\tau^2)^{K/4}}} \Theta(\tau),
\end{eqnarray}
where $\delta=0^+$ is a positive infinitesimal.  Singular behavior can
only arise from the large $x$,$\tau$ power-law behavior of the
denominator.  For large $x$, the oscillating exponential implies that
the integral is dominated by $p \approx 0$, so that the dispersion
$\epsilon_1$ can be linearized around this point.  Doing so, the $p$
and $x$ integrals can be readily performed.  Up to an overall constant
prefactor, one finds
\begin{equation}
  A_s^{\rm MF}(\pi-k,\pi;\omega) \sim {\rm Im}\, \int_0^\infty
  \!\! d\tau\, \tau^{-K/2} e^{\left(\omega-\epsilon_1(k-\kappa_F)+i\delta\right)\tau}.
\end{equation}
This integral can be related to a Gamma function by analytic
continuation.  Carrying this out carefully gives the final mean-field result
\begin{equation}
  A_s^{\rm MF}(\pi\!-\! k,\pi;\omega) \sim
  |\omega\!-\!\epsilon_1(k\!-\!\kappa_F)|^{-1+{K \over 2}}
  \Theta[\omega\!-\!\epsilon_1(k-\kappa_F)].
  \label{eq:MFssf}
\end{equation}

As suggested above, Eq.~\ref{eq:MFssf}\ indeed exhibits an algebraic
singularity.  For momentum $(\pi,\pi)$, $k=0$ above, and the
Fermi-level condition $\epsilon_1(\kappa)=2\mu$ for the
Cooper-pair fluid indeed implies the singularity is located at $\omega=2\mu$,
identifying it with the putative ``Pi resonance''.  Note, however,
that within this approximation identical ``resonances'' appear at {\sl 
  all} momenta, including a {\sl lower energy one} at $k=\kappa_F$.
Moreover, the resonance becomes more singular when the Luttinger parameter
$K$ decreases approaching a delta function as $K \rightarrow 0$, whereas the Pi-resonance should  
approach a delta function in the opposite limit of $K \rightarrow \infty$ where the Cooper-pair fluid develops off-diagonal long-ranged order.
Thus, it is unclear whether the above resonance for the 1d model
has any connection
with the two-dimensional Pi-resonance.

Moreover, further reflection on the nature of the mean-field approximation used
above, leads us to question the validity of the singular
behavior at finite frequency.
While it might well be correct for the $O(N=\infty)$ GN model, the
fundamental fermions, e.g. Cooper pairs and magnons, are {\sl strongly
interacting} for the $N=8$ case of interest, as evidenced, e.g. by
the $O(1)$ binding energy for the mass $\sqrt{3}m$ bound states and
the degeneracy of the fundamental fermion and kink excitations in the
D-Mott phase.  While interactions will not significantly modify the
$\theta_1$ correlator above (since the Cooper-pair fluid remains a
Luttinger liquid), they would appear to have a drastic effect upon the
$\psi_2$ Green's function.  In general, this Green's function
describes the propagation of a single massive injected particle into
and interacting with a Luttinger liquid.  Similar problems have been
extensively studied,\cite{Neto96}\ and one finds that the massive
particle will generally {\sl radiate} both energy and momentum into
the Luttinger liquid, decaying in the process.  From such decay
processes, we generally expect a finite lifetime and hence broadening
of the algebraic singularity above.  For large $N$, the interaction
and hence the broadening would be small, but we see no reason for this
to be the case for $N=8$.  Furthermore, one might naively expect that
the minimum energy singularity at $k=\kappa_F$ would survive, since it
is at the bottom of the $\psi_2$ band and thus naively has no states
to decay to.  The mean-field approximation, however, misses the
existence of bound states, including the Cooper pair-magnon bound state
which lies below the band minina at very low doping.
In general, we expect that even the $k=\kappa_F$
fundamental fermion can decay into this bound state (radiating
excitations in the Luttinger liquid) in the interacting 
system, washing out the algebraic singularity even here.

In summary, the above argument suggests that above threshhold the
spin-spectral function at finite doping will be smooth as a function
of energy, with no singularities.  Since this conclusion is based on a
number of physical arguments and approximations, we cannot rule out
the possibility of some high-energy singular structure.  Certainly
singular behavior at $\omega=2\mu$ would be a remarkable phenomenon.
On a much firmer standing is the spin-gap threshold energy, which is
presumably a universal function of doping $x$ for the Gross-Neveu
model.  One might hope that the precise functional form for this
energy gap is accessible via integrability.

\acknowledgements We are grateful to Anton Andreev, Natan Andrei,
David Gross, Victor Gurarie, Charlie Kane, Andreas Ludwig, Chetan
Nayak, Joe Polchinski, Hubert Saleur, Doug Scalapino and Shou-Cheng
Zhang for illuminating conversations.  This work has been supported by
the National Science Foundation under grant Nos. PHY94-07194,
DMR-9400142 and DMR-9528578.

\appendix
\section{RG equations}
\label{8RG}

For the weakly interacting two-leg ladder with particle-hole
symmetry at half-filling
there are nine marginal non-chiral four fermion interactions,
as discussed in detail in Section II.
The leading order renormalization group (RG) flow
equations for the corresponding nine interaction strengths are,
\begin{eqnarray}
{\dot b}^{\rho}_{11} &=& -(b^{\rho}_{12})^{2}
-\frac{3}{16}(b^{\sigma}_{12})^{2}
+4(u^{\rho}_{12})^{2}+\frac{3}{4}(u^{\sigma}_{12})^{2},
\\
{\dot b}^{\sigma}_{11}&=& -2b^{\rho}_{12}b^{\sigma}_{12}
-\frac12 (b^{\sigma}_{12})^{2}-(b^{\sigma}_{11})^{2}
-8u^{\rho}_{12}u^{\sigma}_{12}
\nonumber\\
&&-2(u^{\sigma}_{12})^{2},
\\ 
{\dot b}^{\rho}_{12}&=& -2b^{\rho}_{11}b^{\rho}_{12}
-\frac{3}{8} 
b^{\sigma}_{11}b^{\sigma}_{12}+2b^{\rho}_{12}f^{\rho}_{12}
+\frac{3}{8} b^{\sigma}_{12}f^{\sigma}_{12}
\nonumber\\
&&+16u^{\rho}_{12}u^{\rho}_{11},
\\
{\dot b}^{\sigma}_{12}&=& -2b^{\rho}_{11}b^{\sigma}_{12}
-2b^{\rho}_{12}b^{\sigma}_{11}-b^{\sigma}_{12}b^{\sigma}_{11}
+16 u^{\rho}_{11}u^{\sigma}_{12}
\nonumber\\
&&+2f^{\rho}_{12}b^{\sigma}_{12}
+2b^{\rho}_{12}f^{\sigma}_{12}-b^{\sigma}_{12}f^{\sigma}_{12},
\\ 
{\dot f}^{\rho}_{12}&=& (b^{\rho}_{12})^{2}
+\frac{3}{16}(b^{\sigma}_{12})^{2}+16(u^{\rho}_{11})^{2}
+4(u^{\rho}_{12})^{2}
\nonumber\\
&&+\frac{3}{4}(u^{\sigma}_{12})^{2},
\\
{\dot f}^{\sigma}_{12}&=& 2b^{\rho}_{12}b^{\sigma}_{12}
-\frac12 (b^{\sigma}_{12})^{2} -(f^{\sigma}_{12})^{2}
+8u^{\rho}_{12}u^{\sigma}_{12}
\nonumber\\
&&-2 (u^{\sigma}_{12})^{2},
\\
{\dot u}^{\rho}_{11} &=& 2b^{\rho}_{12}u^{\rho}_{12}
+4f^{\rho}_{12}u^{\rho}_{11}
+\frac{3}{8} b^{\sigma}_{12}u^{\sigma}_{12},
\\
{\dot u}^{\rho}_{12} &=& 2b^{\rho}_{11}u^{\rho}_{12}
-\frac{3}{8}b^{\sigma}_{11}u^{\sigma}_{12}
+4b^{\rho}_{12}u^{\rho}_{11}+2f^{\rho}_{12}u^{\rho}_{12}
\nonumber\\
&&+\frac{3}{8}f^{\sigma}_{12}u^{\sigma}_{12},
\\
{\dot u}^{\sigma}_{12} &=& -2b^{\sigma}_{11}u^{\rho}_{12}
+2b^{\rho}_{11}u^{\sigma}_{12}-b^{\sigma}_{11}u^{\sigma}_{12}
+4b^{\sigma}_{12}u^{\rho}_{11}
\nonumber\\
&&+2f^{\sigma}_{12}u^{\rho}_{12}
+2f^{\rho}_{12}u^{\sigma}_{12}-f^{\sigma}_{12}u^{\sigma}_{12}.
\end{eqnarray}
Here ${\dot g} \equiv 2\pi v dg/dl$ with $b = e^{dl}$ the 
dimensionless recaling length of the RG transformation.
The last three flow equations describe the renormalization of
momentum non-conserving Umklapp processes.

\section{Gauge Redundancy}
\label{app:gauge}

The Bosonized sine-Gordon form for the $SO(8)$ Gross-Neveu model
appears to have a highly degenerate ground state.  In terms of the
four Boson fields $\theta_a$,
the semiclassical ground states correspond to spatially uniform values
chosen to minimize
the potential $V(\theta) = -g \sum_{a \ne b} \cos(\theta_a) 
\cos(\theta_b)$.
Solutions include $\theta_a = 2\pi n_a$ as well as $\theta = 2\pi n_a 
+ \pi$
for {\it arbitrary} integers $n_a$.  
But as we shall see, in most situations these multiple solutions
actually correspond to the same {\it physical} state.
To see which solutions are physically equivalent, it is necessary
to relate the $\theta_a$ and their dual
fields $\varphi_a$ to the original Boson fields,
$\phi_{{\scriptscriptstyle P}i\alpha}$, introduced when the 
electron fermion operators
were bosonized.  Local gauge transformations, $\phi_{pi\alpha}
\rightarrow \phi_{pi\alpha} + 2\pi N_{pi\alpha}(x,\tau)$ for integer
$N_{{\scriptscriptstyle P}i\alpha}$ 
leave the electron operators invariant, and so do not
change the physical state. 
Thus, any shift in $\theta_a$ and $\varphi_a$
which corresponds to an {\it integer} shift
in $\phi_{{\scriptscriptstyle P}i\alpha}/2\pi$ is redundant, 
and leaves the physical state unchanged.

To establish whether or not
two different semiclassical solutions,
$\theta_a$ and $\theta_a^\prime$, are actually physically
equivalent we proceed as follows.
For the given (spatially constant) shift $\delta \theta_a = (\theta_a 
- \theta_a^\prime)/2\pi$,
we ask whether it is possible to choose appropriate
(spatially constant) shifts $\delta \varphi_a$ so that the chiral 
fields
$\phi_{{\scriptscriptstyle P}i\alpha}/2\pi$ are changed by integers.
The choice for $\delta \varphi_a$ is uncontrained,
since the full interacting Hamiltonian is invariant
under {\it arbitrary} spatially constant shifts in the
four $\varphi_a$ fields.  If it is possible, then the two 
semiclassical
solutions are physically equivalent.  For physically inequivalent
states, it will not be possible
to choose $\delta \varphi_a$ to give the required integer
shifts.

To implement the above procedure we need an expression relating
the bare chiral fields 
$\phi_{{\scriptscriptstyle P}i\alpha}$ to $\theta_a$ and 
$\varphi_a$.  This can be obtained from
\begin{eqnarray}
\phi_{{\scriptscriptstyle P}i\alpha}&=& \frac14 (
  \varphi_{\rho+}+\alpha \varphi_{\sigma+}
  -q\alpha\varphi_{\sigma-}-q\varphi_{\rho-})
  \nonumber\\
  &&+\frac14  P(
  \theta_{\rho+}+\alpha \theta_{\sigma+}
  -q\alpha \theta_{\sigma-}-q\theta_{\rho-}),
  \label{chiral}
\end{eqnarray}
where $q=(-1)^{i}=1, -1$ for bonding and anti-bonding bands
respectively.  For the D-Mott phase the relation between $\theta_{a},
\varphi_a$ and the fields $\theta_{\mu \pm}$, $\varphi_{\mu \pm}$ is
given explicitly 
in Eq.~\ref{mott_field}.  In the S-Mott phase, the eqivalence of
$\theta_4 = \varphi_{\rho-}$ is modified to be $\theta_4^S =
\varphi_{\rho-} + \pi$, but this $\pi$ difference does not effect the
{\it shifts} $\delta \theta_a$ between {\it different} semiclassical
states.  Thus the ground state degeneracies in the D-Mott and S-Mott
phases must necessarily be the same.  In subsection B.1 below we show
that both of these phases have unique ground states.  It is necessary
to consider the SP and CDW phases separately (in subsection B.2
below), since there is a non-trivial modification in the relation
between $\theta_{a}, \varphi_a$ and the fields $\theta_{\mu \pm}$,
$\varphi_{\mu \pm}$.  As we shall show, in these latter two phases the
ground state is {\it two-fold} degenerate - corresponding physically
to the spontaneous breaking of a discrete parity symmetry (see Section
VI).

\subsection{D-Mott and S-Mott phases}

In the D-Mott and S-Mott 
phases, shifts in the fields $\theta_a$ and $\varphi_a$
induce shifts in the chiral fields, 
$\delta \phi_{{\scriptscriptstyle P}i\alpha}$,
of the general form
\begin{equation}
\delta \phi^{}_{{\scriptscriptstyle P}i\alpha} = 
\frac14 P(\bbox{A}_{{\scriptscriptstyle P}})_{ab} 
\delta\theta_{b} +
\frac14 (\bbox{A}_{{\scriptscriptstyle P}})_{ab} 
\delta\varphi_{b},
\label{bare_O8}
\end{equation}
where $a=1\uparrow, 1\downarrow,
2\uparrow, 2\downarrow$ labels the band and spin indices and 
$b=1,\ldots,4$ labels the four flavors of the
sine-Gordon bosons.  Here and below, all shifts will be measured
in units of $2\pi$, so that for example $\delta \phi = (\phi - 
\phi^\prime)/2\pi$.
The matrices $\bbox{A}_{\scriptscriptstyle P}$ can be explicitly 
constructed by using Eqs.~\ref{chiral}, \ref{mott_field},
\begin{equation}
(\bbox{A}_{{\scriptscriptstyle P}})_{ab} = \left(
\begin{array}{cccc}
1&1&1&P\\
1&-1&-1&P\\
1&1&-1&-P\\
1&-1&1&-P
\end{array}\right).
\end{equation}
It will be convenient to separate out the two contributions coming
from 
the shifts in $\theta_a$ and $\phi_a$, respectively, by defining
\begin{equation}
\delta \Theta_{{\scriptscriptstyle P}a}=
P(\bbox{A}_{{\scriptscriptstyle P}})^{ab} \delta\theta_{b}, \qquad
\delta \Phi_{{\scriptscriptstyle P}a}=
(\bbox{A}_{{\scriptscriptstyle P}})^{ab} \delta\varphi_{b}. 
\label{theta}
\end{equation}

Comparing two semiclassical solutions, $\theta_a$ and 
$\theta_a^\prime$, determines the shifts 
$\delta \Theta_{{\scriptscriptstyle P}a}$.  These 
two solutions are physically equivalent, provided shifts $\delta 
\varphi_a$ can be chosen so that
the following eight constraint equations are satisfied,
\begin{equation}
\Theta_{{\scriptscriptstyle P}a} + 
\delta\Phi_{{\scriptscriptstyle P}a}  = 
4 N_{{\scriptscriptstyle P}a} ,
\label{rlgauge}
\end{equation}
with {\it integer} $N_{{\scriptscriptstyle P}a}$.  In this case, all 
eight shifts $\delta \phi_{{\scriptscriptstyle P}i\alpha}$ are integer, 
and the electron 
fields
are left unchanged.  

Since the four shifts, $\delta \varphi_a$, determine
{\it both} right and left vectors, 
$\delta \Phi_{{\scriptscriptstyle R}a}, \delta 
\Phi_{{\scriptscriptstyle L}a}$, 
these two vectors are not independent, and similarly
for the $\theta$ shifts.  Indeed one can see that, 
\begin{eqnarray}
\delta\Phi_{{\scriptscriptstyle R}a}  
&=& {1 \over 2} (\bbox{M})_{ab} \delta\Phi_{{\scriptscriptstyle L}b},
\\
\delta\Theta_{{\scriptscriptstyle R}a } 
&=& - {1 \over 2} (\bbox{M})_{ab} 
\delta\Theta_{{\scriptscriptstyle L}b},
\end{eqnarray}
with $\bbox{M} = 2 \bbox{A}^{}_{{\scriptscriptstyle R}} 
\bbox{A}_{{\scriptscriptstyle L}}^{-1}$ or,
\begin{eqnarray}
(\bbox{M})_{ab} 
&=& \left(
\begin{array}{cccc}
1&-1&1&1\\
-1&1&1&1\\
1&1&1&-1\\
1&1&-1&1
\end{array} \right).
\end{eqnarray}

We can now use the eight constraint equations to eliminate $\delta 
\Phi$
and arrive at four equations for $\delta \Theta$.
To this end, 
upon multiplying by $\bbox{M}$ on the left sector of 
Eq.~\ref{rlgauge}, 
one obtaines $\delta\Phi_{{\scriptscriptstyle R}a} 
-\delta\Theta_{{\scriptscriptstyle R}a}
=4(\bbox{M})_{ab} N_{{\scriptscriptstyle L}a}$. 
Upon combining with the right sector of Eq.~\ref{rlgauge} one obtains,
\begin{equation}
\delta\Theta_{{\scriptscriptstyle R}a}
= 2 N_{{\scriptscriptstyle R}a} 
- (\bbox{M})_{ab} N_{{\scriptscriptstyle L}b} . 
\label{key}
\end{equation}
Two semiclassical solutions (which determine 
$\delta \Theta_{{\scriptscriptstyle R}a}$)
are then physically equivalent provided these four equations
have solutions for {\it integer} $N_{{\scriptscriptstyle P}a}$.

All of the semiclassical solutions take the form
$\theta_{a}=2n_{a}\pi$ or $\theta_{a}=(2n_{a}+1)\pi$ with
arbitrary integers $n_a$.  It is straightforward to show
that for {\it any} two of these
solutions the difference $\delta \theta_a$
corresponds to $\delta \Theta_{{\scriptscriptstyle R}a}$ which are
either even integer for all $a=1,...,4$ or all odd integers.
When they are even integers, Eq.~\ref{key} can be solved
for integer $N_{{\scriptscriptstyle R}a}$ 
by taking $N_{{\scriptscriptstyle L}a}=0$.
For odd integer $\delta \Theta_{{\scriptscriptstyle R}a}$ 
a solution with integer
$N_{{\scriptscriptstyle R}a}$ is also possible by taking,
for example, $N_{{\scriptscriptstyle L}a}= \delta_{a1}$.

We have thereby established 
the physical equivalence between all of the semiclassical
solutions.  This implies that
the D-Mott and S-Mott ground states
are unique.

\subsection{SP and CDW phases}
In the SP and CDW phases, the relation between
$\theta_a, \varphi_a$ and $\theta_{\mu \pm}, \varphi_{\mu \pm}$
are changed, so the above conclusions are modified.
In particular, one has
\begin{equation}
\theta^{SP}_{3}= \varphi_{\sigma-} , \qquad
\varphi^{SP}_{3}=\theta_{\sigma-},
\end{equation}
in the SP phase and $\theta^{CDW}_3 = \theta^{SP}_3 +\pi$,
$\varphi^{CDW}_3 = \theta^{SP}_3$ in the CDW phase.  Because the Boson fields are defined
differently, the matrix $\bbox{A}_{\scriptscriptstyle P}$ 
which relates the two sets
of fields in Eq.~\ref{bare_O8} is modified.  The appropriate
matrix in this case, denoted 
$\bbox{\tilde{A}}_{\scriptscriptstyle P}$, is given by,
\begin{equation}
(\bbox{\tilde{A}}_{{\scriptscriptstyle P}})_{ab}= \left(
\begin{array}{cccc}
1&1&P&P\\
1&-1&-P&P\\
1&1&-P&-P\\
1&-1&P&-P
\end{array}\right).
\end{equation}
Notice that $\bbox{\tilde{A}}_{\scriptscriptstyle R}
 = \bbox{A}_{\scriptscriptstyle R}$, although the
left matrices
differ in the third column.
Similarly, the matrix $\bbox{M}$ is also modified, becoming
\begin{equation}
(\bbox{\tilde{M}})_{ab} = \left(
\begin{array}{cccc}
0&0&2&0\\
0&0&0&2\\
2&0&0&0\\
0&2&0&0
\end{array} \right).
\end{equation}

Physical equivalence between two semiclassical solutions 
for the SP or CDW phases is, once again, established by finding a 
solution
of Eq.~\ref{key} with integer $N_{{\scriptscriptstyle P}i\alpha}$, 
except with
$\bbox{\tilde{M}}$ replacing $\bbox{M}$.  As before,
the difference between any two of the semiclassical solutions
leads to either even integer or odd integer
$\delta \Theta_{{\scriptscriptstyle R}a}$.
For even integer $\delta \Theta_{{\scriptscriptstyle R}a}$
a solution is again possible by
taking $N_{{\scriptscriptstyle L}a}=0$ and choosing appropriate integers for 
$N_{{\scriptscriptstyle R}a}$.
However, a solution in the integers is {\it not} possible
for two semiclassical solutions differing by an
odd integer shift vector 
$\delta \Theta_{{\scriptscriptstyle R}a}$ (since 
$\bbox{\tilde{M}}_{ab}
N_{{\scriptscriptstyle L}b}$ is always odd).  Two such semiclassical
solutions would thus correspond to physically distinct
phases. 

The fact that the ground state is actually two-fold degenerate
can be established as follows.
Consider two specific semiclassical solutions,
$\theta_a^1 = 0$ and $\theta_a^2 = 2\pi \delta_{a1}$.  One can 
readily show that the shift vector, $\delta \Theta_a^{12}$, 
connecting these two states
is an odd integer vector, so that these two states are physically 
distinct.
Next consider an arbitrary third semiclassical solution, $\theta_a^3$.
If the relative shift vector between the first and third 
solutions, $\delta \Theta_a^{13}$ is even then the physical states 
are equivalent.  
If, on the other hand,
$\delta \Theta_a^{13}$ is an odd integer,
then $\delta \Theta_a^{23}$ is necessarily even,
and the second and third solutions describe the same physical state.
It is thus clear that there are only two physically distinct
ground states in the SP and CDW phases.  As discussed in Section VII
this two-fold degeneracy corresponds to a spontaneous breaking of
a discrete parity symmetry.

\section{Gamma Matrices and Spinor Representations}
\label{app:spinor}

In this appendix, we discuss some technical details of gamma matrices
and spinor representations of $SO(5)$.  In general, there are two
types of representations for $SO(N)$.  The first are tensors, which
transform like products of vectors. Irreducible representations are
then found by taking symmetric and anti-symmetric combinations (Young
tableaux). However, to describe how (complex) fermions transform under
rotations, the second representation, the spinor one, is necessary.
It has already been used in constructing invariants in
Sec.~\ref{sec:SO5}, but here we review the mathematics in somewhat more
technical detail, in order to allow the reader to perform
concrete calculations if he or she so desires.  To explain the spinor
representation, let us introduce a set of $N$ generalized Dirac
matrices which obey the Clifford algebra,
\begin{equation}
\{ \Gamma^{{\scriptscriptstyle A}}, 
\Gamma^{{\scriptscriptstyle B}} \} = 2 \delta_{{\scriptscriptstyle AB}},
  \label{Dirac_matrices}
\end{equation}
where $A,B=1, 2,\ldots,N$. We then construct the $N(N-1)/2$ 
generators defined as commutators between all pairs of these Dirac 
matrices,
\begin{equation}
\Gamma^{{\scriptscriptstyle AB}}=\frac{i}{4} 
[ \Gamma^{{\scriptscriptstyle A}}, \Gamma^{{\scriptscriptstyle B}}].
\end{equation}
It is easy to show that these generators satisfy the $SO(N)$ 
commutation relations
\begin{equation}
[\Gamma^{{\scriptscriptstyle AB}}, \Gamma^{{\scriptscriptstyle CD}}]
=i(\delta_{{\scriptscriptstyle AD}}\Gamma^{{\scriptscriptstyle BC}}
-\delta_{{\scriptscriptstyle AC}}\Gamma^{{\scriptscriptstyle BD}}
-\delta_{{\scriptscriptstyle BD}}\Gamma^{{\scriptscriptstyle AC}}
+\delta_{{\scriptscriptstyle BC}}\Gamma^{{\scriptscriptstyle AD}}) .
\label{SON_algebra}
\end{equation}
For $N=5$, we choose a specific set of matrices
to represent the $SO(5)$ group. The minimum dimension of a set of five
matrices which satisfy the Clifford algebra is $4\times 4$.  Our 
particular choice is
\begin{equation}
\Gamma^{1}\hspace{-0.1cm}=\hspace{-0.1cm}\left( \hspace{-.1cm}
\begin{array}{cc}
0 & i\sigma_{y}\\
-i\sigma_{y} & 0
\end{array} 
\hspace{-.1cm}\right)
\Gamma^{2}\hspace{-0.1cm}=\hspace{-0.1cm}\left( \hspace{-.1cm}
\begin{array}{cc}
0 & \sigma_{y}\\
\sigma_{y}&0
\end{array} 
\hspace{-.1cm}\right)
\Gamma^{3,4,5}\hspace{-0.1cm}=\hspace{-0.1cm}\left( \hspace{-.1cm}
\begin{array}{cc}
-\bbox{\sigma} & 0\\
0&-\bbox{\sigma}^{*}
\end{array} 
\hspace{-.1cm}\right).
\label{matrices}
\end{equation}

A useful property of the spinor representation is its ``reality''.
This means that the conjugate representation $-(\Gamma^{ab})^{*}$ also
obeys the algebra, and is equivalent under a unitary transformation to
the original representation.  This follows because it is always
possible to find a matrix $R$ which satisfies the properties
\begin{eqnarray}
R^{2} &=&-1, \qquad R^{-1}=R^{\dag}=R^{t}=-R,
\nonumber\\
R^{-1} \Gamma^{{\scriptscriptstyle AB}} 
R &=& -(\Gamma^{{\scriptscriptstyle AB}})^{*}, \qquad
R^{-1} \Gamma^{{\scriptscriptstyle A}} R 
= (\Gamma^{{\scriptscriptstyle A}})^{*}.
\end{eqnarray}
For $N=5$ with our particular choice of Dirac matrices in
Eq.~\ref{matrices}, the matrix $R$ is simply
\begin{equation}
R=\left( 
\begin{array}{cc}
0 & {\bf 1}\\
-{\bf 1}&0
\end{array} 
\right),
\end{equation}
where ${\bf 1}$ is the two by two identity matrix. The 
matrix $R$ is useful in constructing irreducible representations of 
$SO(5)$.

As we have seen in Sec.~\ref{sec:SO5}, these abstract matrices can be
elevated to physical operators by sandwiching them between two {\sl
spinors}.  The useful details are already given in the text of
Sec.~\ref{sec:SO5}.  Here we provide some
reasoning and motivation for the choice of spinor taken there.  For
convenience, we copy the spinor definition from
Eq.~\ref{Rabello_spinor}: 
\begin{equation}
  \Psi(\bbox{k}) = \left(
    \begin{array}{c}
      a_{\alpha}(\bbox{k})\\
      \phi_{\bbox{k}} a^{\dag}_{\alpha}(-\bbox{k}+\bbox{N})
    \end{array}
  \right),
\end{equation}
where $\bbox{N}=(\pi, \pi)$.
Here $\phi_{\bbox{k}}$ is a complex function
with absolute value one, chosen by Rabello et. al.\cite{Rabello97u}
to have D-wave symmetry in two-dimensions.  As discussed in Section VI, this
factor plays no role in the case of the two-leg ladder, and can be
set to unity.  At first blush, the particular
choice of spinor appears rather
arbitrary.
It is not, for several reasons.  At half filling, the system is
particle-hole symmetric.  For every hole excitation at momentum
$\bbox{k}$ created by $a(\bbox{k})$, there is a particle excitation
counterpart at momentum $\bbox{k}-\bbox{N}$ created by
$a^{\dag}(\bbox{k}-\bbox{N})$. Parity symmetry implies there is also a
particle excitation at the opposite momentum $-\bbox{k}+\bbox{N}$.
Because these excitations occur symmetrically, they are chosen as upper
and lower components in the spinor. The use of the parity symmetry is
not essential. However, it is rather convenient for later analysis in
weak coupling because, by such a construction, all components have the
same chirality, i.e. act on the same side of the Fermi surface. Since
the four-components spinor $\Psi(\bbox{k})$ contains
excitations at both $\bbox{k}$ and $-\bbox{k}+\bbox{N}$, the momentum
$\bbox{k}$ is only allowed to run over half of the Brillouin zone. 
The halved region in momentum space is also known as the folded
Brillouin zone (shown in Fig. 9). 
Finally, one would like the spinor to obey canonical
anti-commutation relations so that it annihilates or creates fermionic
excitations. This is the origin of the constraint on $\phi_{\bbox{k}}$:
direct calculation verifies that, provided
$|\phi_{\bbox{k}}|^{2}=1$, the canonical anti-commutation relation is
satisfied,
\begin{equation}
  \{ \Psi^{}_{a}(\bbox{k}), \Psi^{\dag}_{b}(\bbox{k}') \} = 
  (2\pi)^{d} \delta_{ab}\delta(\bbox{k}-\bbox{k}').
\end{equation}
Further straightforward algebra demonstrates that when the spinor
satisfies canonical anti-commutators, the currents in
Eqs.~\ref{eq:Jscalar}-\ref{eq:Ivector}\ satisfy appropriate Kac-Moody
algebras.  This exercise, which we do not reproduce here, verifies that these currents are indeed $SO(5)$
scalars, vectors, and tensors, as indicated in Sec.~\ref{sec:SO5}.

We conclude this appendix by obtaining expressions which
relate the 28 $SO(5)$ currents defined in Sec. VI, to the
28 $SO(8)$ currents, $G_{\scriptscriptstyle P}^{{\scriptscriptstyle AB}}$,
introduced in Sec. IV.
These relations can be determined by
bosonizing the $SO(5)$ currents,
rewriting in terms of the
GN bosons $\theta_a$ and $\varphi_a$ and the Klein factors,
refermionizing, and finally changing from Dirac to Majorana fermions.  For
example, 
\begin{eqnarray}
{\cal J}_{\scriptscriptstyle P}^{\scriptscriptstyle 21} & = & 
\frac{1}{2\pi} \partial_{x} \phi_{{\scriptscriptstyle P}\rho+} =
\psi^{\dag}_{{\scriptscriptstyle P}\rho+}
\psi^{}_{{\scriptscriptstyle P}\rho+} \nonumber \\
& = & i\eta_{{\scriptscriptstyle P2}}\eta_{{\scriptscriptstyle P1}}
= G^{\scriptscriptstyle 21}_{\scriptscriptstyle P}.
\end{eqnarray}
The general relations can be conveniently presented
in the following form:
\begin{equation}
G^{{\scriptscriptstyle AB}}_{{\scriptscriptstyle P}}= \left[
\begin{array}{cc}
\bbox{T}_{{\scriptscriptstyle P}}&
-\bbox{V}_{{\scriptscriptstyle P}}^{t}\\
\bbox{V}_{{\scriptscriptstyle P}}&
\bbox{S}_{{\scriptscriptstyle P}}\\
\end{array}
\right]_{AB} .
\end{equation}
The $5 \times 5$
anti-symmetric tensor matrix 
$\bbox{T}_{{\scriptscriptstyle P}}^{{\scriptscriptstyle AB}}
={\cal J}_{{\scriptscriptstyle P}}^{{\scriptscriptstyle AB}}$
for $A \neq B$ and it is zero for $A=B$. The $3 \times 5$ vector 
matrix is
\begin{equation}
\bbox{V}_{{\scriptscriptstyle P}}^{{\scriptscriptstyle CB}} 
=\frac12 \left(
\begin{array}{c}
{\cal J}_{{\scriptscriptstyle P}}^{{\scriptscriptstyle B}}\\
-{\rm Im}{\cal I}_{{\scriptscriptstyle P}}^{{\scriptscriptstyle B}}\\
P{\rm Re}{\cal I}_{{\scriptscriptstyle P}}^{{\scriptscriptstyle B}}\\
\end{array}
\right)_C .
\end{equation}
Finally, the $3 \times 3$ anti-symmetric scalar matrix is
\begin{equation}
\bbox{S}_{{\scriptscriptstyle P}}
=-\frac12 \left(
\begin{array}{ccc}
0&&\\
{\rm Re}{\cal I}_{{\scriptscriptstyle P}}&0&\\
P{\rm Im}{\cal I}_{{\scriptscriptstyle P}}&
P{\cal J}_{{\scriptscriptstyle P}}&0\\
\end{array}
\right).
\end{equation}

\section{$SO(5)$ currents in SU(2)$\times$U(1) and $SO(8)$ notation}
\label{app:so5currents}

In Section VI the most general set of $SO(5)$ invariant
interactions for the weak coupling
two-leg ladder were expressed as products
of right and left moving $SO(5)$ currents,
see Eq.~\ref{SO5-gen}.
Here we re-express these five interactions in terms of
charge and spin currents with lower $U(1)\times SU(2)$ symmetry,
which were introduced in Section II.
The products of $SO(5)$ scalar, vector, and tensor currents are
re-expressed as
\begin{eqnarray}
{\cal J}_{{\scriptscriptstyle R}} 
{\cal J}_{{\scriptscriptstyle L}} &=& (J_{{\scriptscriptstyle R}11}
-J_{{\scriptscriptstyle R}22}-2)
( J_{{\scriptscriptstyle L}11}-J_{{\scriptscriptstyle L}22}-2),
\\
{\cal J}^{{\scriptscriptstyle A}}_{{\scriptscriptstyle R}} 
{\cal J}^{{\scriptscriptstyle A}}_{{\scriptscriptstyle L}} &=& 
4 (\bbox{J}_{{\scriptscriptstyle R}11}
-\bbox{J}_{{\scriptscriptstyle R}22})
(\bbox{J}_{{\scriptscriptstyle L}11}-\bbox{J}_{{\scriptscriptstyle L}22})
\nonumber\\
&&+2(I^{\dag}_{{\scriptscriptstyle R}12}I^{}_{{\scriptscriptstyle L}21}
+I^{\dag}_{{\scriptscriptstyle L}21}I^{}_{{\scriptscriptstyle R}12}),
\\
{\cal J}^{{\scriptscriptstyle AB}}_{{\scriptscriptstyle R}} 
{\cal J}^{{\scriptscriptstyle AB}}_{{\scriptscriptstyle L}} &=& 
\frac12 (J_{{\scriptscriptstyle R}11}+J_{{\scriptscriptstyle R}22}-2)
( J_{{\scriptscriptstyle L}11}+J_{{\scriptscriptstyle L}22}-2)
\nonumber\\
&&+2(\bbox{J}_{{\scriptscriptstyle R}11}
+\bbox{J}_{{\scriptscriptstyle R}22})
(\bbox{J}_{{\scriptscriptstyle L}11}+\bbox{J}_{{\scriptscriptstyle L}22})
\nonumber\\
&&-4(\bbox{I}^{\dag}_{{\scriptscriptstyle R}12}
\bbox{I}^{}_{{\scriptscriptstyle L}21}
+\bbox{I}^{\dag}_{{\scriptscriptstyle L}21}
\bbox{I}^{}_{{\scriptscriptstyle R}12}).
\end{eqnarray}
Notice that these three interactions conserve the number of particles
in each band. The remaining two $SO(5)$ invariant interactions,
involving anomalous scalars and 
vectors, scatter particles from one band to 
the other. In terms of the $U(1)\times SU(2)$
charge and spin currents, they are
\begin{eqnarray}
{\cal I}^{}_{{\scriptscriptstyle R}}
{\cal I}^{}_{{\scriptscriptstyle L}} 
+{\cal I}^{\dag}_{{\scriptscriptstyle R}}
{\cal I}^{\dag}_{{\scriptscriptstyle L}}
&=& 4 (J_{{\scriptscriptstyle R}21}J_{{\scriptscriptstyle L}21}
+J_{{\scriptscriptstyle R}12}J_{{\scriptscriptstyle L}12}),
\\
{\cal I}^{{\scriptscriptstyle A}}_{{\scriptscriptstyle R}}
{\cal I}^{{\scriptscriptstyle A}}_{{\scriptscriptstyle L}}
+{\cal I}^{{\scriptscriptstyle A}\dag}_{{\scriptscriptstyle R}}
{\cal I}^{{\scriptscriptstyle A}\dag}_{{\scriptscriptstyle L}}
&=&16 (\bbox{J}_{{\scriptscriptstyle R}21}\bbox{J}_{{\scriptscriptstyle L}21}
+\bbox{J}_{{\scriptscriptstyle R}12}\bbox{J}_{{\scriptscriptstyle L}12})
\nonumber\\
&&-2(I^{\dag}_{{\scriptscriptstyle R}11}I^{}_{{\scriptscriptstyle L}22}
+I^{\dag}_{{\scriptscriptstyle R}22}I^{}_{{\scriptscriptstyle L}11}
\nonumber\\
&&+I^{\dag}_{{\scriptscriptstyle L}11}I^{}_{{\scriptscriptstyle R}22}
+I^{\dag}_{{\scriptscriptstyle L}22}I^{}_{{\scriptscriptstyle R}11}).
\end{eqnarray}

For a given set of five $SO(5)$ invariant interaction parameters,
these operator identities enable us to obtain
the corresponding values of the nine forward, backward
and Umklapp scattering amplitudes;
\begin{eqnarray}
b^{\rho}_{11}=& g_{s}+\frac12 g_{t}, \qquad
&b^{\sigma}_{11}=-4g_{v}-2g_{t},
\\
b^{\rho}_{12}=&4h_{s}, \qquad
&b^{\sigma}_{12}=-16h_{v},
\\
f^{\rho}_{12}=&-g_{s}+\frac12 g_{t}, \qquad
&f^{\sigma}_{12}=4g_{v}-2g_{t},
\\
u^{\rho}_{11}=&-2h_{v}, \qquad
&u^{\rho}_{12}=g_{v}, \qquad
u^{\sigma}_{12}=2g_{t}.
\end{eqnarray}
From these, and the nine RG flow equations in
Appendix A, one can obtain a closed set of five RG flow equations
for the five $SO(5)$ invariant coupling constants, given explicitly
in Appendix E.

It is also instructive to re-express the five $SO(5)$ invariant
interactions in terms of the $SO(8)$ currents - specifically the 28
$SO(8)$ generators $G^{{\scriptscriptstyle AB}} =
i\eta_{{\scriptscriptstyle A}}\eta_{{\scriptscriptstyle B}}$,
comprising the vector (fundamental) representation of $SO(8)$.  For
the first three $SO(5)$ interactions one finds,
\begin{eqnarray}
  {\cal J}_{{\scriptscriptstyle R}} {\cal J}_{{\scriptscriptstyle L}} 
&=& -4 G^{{\scriptscriptstyle 78}}_{{\scriptscriptstyle R}} 
G^{{\scriptscriptstyle 78}}_{{\scriptscriptstyle L}}, 
\label{eq:JscalarG}
\\
{\cal J}^{{\scriptscriptstyle A}}_{{\scriptscriptstyle R}} 
{\cal J}^{{\scriptscriptstyle A}}_{{\scriptscriptstyle L}} 
&=& 4\sum^{5}_{A=1} G^{{\scriptscriptstyle A6}}_{{\scriptscriptstyle R}}
G^{{\scriptscriptstyle A6}}_{{\scriptscriptstyle L}},
\label{eq:JvectorG}
\\
{\cal J}^{{\scriptscriptstyle AB}}_{{\scriptscriptstyle R}} 
{\cal J}^{{\scriptscriptstyle AB}}_{{\scriptscriptstyle L}} &=& 
 \sum^{5}_{A,B=1}G^{{\scriptscriptstyle AB}}_{{\scriptscriptstyle R}} 
G^{{\scriptscriptstyle AB}}_{{\scriptscriptstyle L}}. \label{eq:JtensorG}
\end{eqnarray}
As expected, these expressions show that
$G^{{\scriptscriptstyle 78}}, 
G^{{\scriptscriptstyle A6}}$ and $G^{{\scriptscriptstyle AB}}$ 
(for $A,B=1,..,5$) transform under $SO(5)$ rotations
as scalar, vector and 
(rank two) tensor, respectively.
The remaining two anomalous $SO(5)$ invariant interactions
can similarly be re-expressed as,
\begin{eqnarray}
{\cal I}^{}_{{\scriptscriptstyle R}}{\cal I}^{}_{{\scriptscriptstyle L}} 
+{\cal I}^{\dag}_{{\scriptscriptstyle R}}
{\cal I}^{\dag}_{{\scriptscriptstyle L}}
&=&8( G^{{\scriptscriptstyle 67}}_{{\scriptscriptstyle R}}
G^{{\scriptscriptstyle 67}}_{{\scriptscriptstyle L}}
+G^{{\scriptscriptstyle 68}}_{{\scriptscriptstyle R}} 
G^{{\scriptscriptstyle 68}}_{{\scriptscriptstyle L}}),
\label{eq:IscalarG}
\\
{\cal I}^{{\scriptscriptstyle A}}_{{\scriptscriptstyle R}}
{\cal I}^{{\scriptscriptstyle A}}_{{\scriptscriptstyle L}}
+{\cal I}^{{\scriptscriptstyle A}\dag}_{{\scriptscriptstyle R}}
{\cal I}^{{\scriptscriptstyle A}\dag}_{{\scriptscriptstyle L}}
  &=&-8 \sum^{5}_{A=1} 
(G^{{\scriptscriptstyle A7}}_{{\scriptscriptstyle R}} 
G^{{\scriptscriptstyle A7}}_{{\scriptscriptstyle L}}
+G^{{\scriptscriptstyle A8}}_{{\scriptscriptstyle R}} 
G^{{\scriptscriptstyle A8}}_{{\scriptscriptstyle L}}).
  \label{eq:IvectorG}
\end{eqnarray}
It is clear that $G^{{\scriptscriptstyle 67, 68}}$ and
$G^{{\scriptscriptstyle A7, A8}}$ transform as $SO(5)$ scalars and vectors, 
respectively.

\section{$SO(5)$ RG equations}
\label{app:so5rg}

For the weakly interacting two-leg ladder
at half-filling,
requiring $SO(5)$ symmetry reduces the
number of marginal four-fermion
interactions from nine down to five.
Due to symmetry, one expects
the RG flow equations to close in the
manifold of $SO(5)$ invariant models.
This closure can be demonstrated explicitly by combining
the expressions obtained in Appendix D which specify
the five dimensional $SO(5)$ invariant manifold,
with the general RG flow equations in Appendix A.
When re-expressed in terms of the $SO(5)$ couplings,
the nine RG flow equations are seen to be redundant - only
5 are independent.  Thus confined to the
$SO(5)$ invariant manifold, the set of independent
RG flow equations can be written,
\begin{eqnarray}
{\dot g}_{s}&=& -16 h_{s}^{2} -80h_{v}^{2}, \label{eq:so5rg1}
\\
{\dot g}_{v}&=& 8g_{v}g_{t}-32h_{s}h_{v},
\\
{\dot g}_{t}&=& 8g_{v}^{2}+6g_{t}^{2}+64h_{v}^{2},
\\
{\dot h}_{s}&=&-4g_{s}h_{s}-20g_{v}h_{v},
\\
{\dot h}_{v}&=&-4g_{v}h_{s}-4g_{s}h_{v}+8g_{t}h_{v}. \label{eq:so5rg5}
\end{eqnarray}

%\bibliography{pub}

\begin{references}
\bibitem{foot:LRO}
  In one dimension, strong quantum fluctuations prohibit true long-range 
  order.  Nevertheless, many 1d Mott insulators, including e.g. the
  half-filled 1d Hubbard model, develop instead power-law magnetic
  correlations and still maintain low-energy magnon modes. Such systems
  are thus close counterparts of higher-dimensional systems with broken
  spin-rotational invariance.
\end{references}

\end{multicols}
\end{document}